\documentstyle[12pt,epsfig]{ioplppt}    

\def\ifmath#1{\relax\ifmmode #1\else $#1$\fi}%
\def\leqsim{\mathbin{\;\raise1pt\hbox{$<$}\kern-8pt\lower3pt\hbox{$\sim$}\;}}
\def\geqsim{\mathbin{\;\raise1pt\hbox{$>$}\kern-8pt\lower3pt\hbox{$\sim$}\;}}
\def\GeV{\ifmmode \hbox{\rm Ge\kern -0.1em V}\else
                  \hbox{\mathrm{Ge\kern -0.1em V}}\fi}%
\def\MeV{\ifmmode \hbox{\rm Me\kern -0.1em V}\else
                  \hbox{\mathrm{Me\kern -0.1em V}}\fi}%
\def\keV{\ifmmode \hbox{\rm ke\kern -0.1em V}\else
                  \hbox{\mathrm{ke\kern -0.1em V}}\fi}%
\def\eV{\ifmmode \hbox{\rm e\kern -0.1em V}\else
                 \hbox{\mathrm{e\kern -0.1em V}}\fi}%

\newcommand{\CoM}        {centre-of-mass}
\newcommand {\ee}         {\mathrm{e}^+\mathrm{e}^-}
\newcommand {\eleft}      {\rm{e}_{\rm L}}
\newcommand {\eright}     {\rm{e}_{\rm R}}
\newcommand {\nel}        {\nu_{\rm e}} 
\newcommand {\nmu}        {\nu_{\mu}} 
\newcommand {\ntau}        {\nu_{\tau}} 
\newcommand {\MZ}      {m_{\mathrm{Z}}}
\newcommand {\MW}      {m_{\mathrm{W}}}
\newcommand {\MH}      {m_{\mathrm{H}}}
\newcommand {\Mt}      {m_{\mathrm{t}}}
\newcommand {\Mb}      {m_{\mathrm{b}}}
\newcommand {\GZ}      {\Gamma_{\mathrm{Z}}}
\newcommand {\GW}      {\Gamma_{\mathrm{W}}}
\newcommand {\ff}         {{\rm f}\overline{\rm f}}
\newcommand {\lept}         {\ell^+\ell^-}
\newcommand {\lepta}         {\ell\overline{\ell}}
\newcommand {\bb}    {{\mathrm b\overline{\mathrm b}}}
\newcommand {\cc}    {{\rm c\overline{\rm c}}}
\newcommand {\ttb}   {{\mathrm t\overline{\mathrm t}}}
\newcommand {\Gf}      {{\Gamma_{\ff}}}
\newcommand {\Gb}      {{\Gamma_{\bb}}}

\newcommand {\Gl}      {{\Gamma_{\lepta}}}
\newcommand {\Ge}      {{\Gamma_{\rm e\overline{\rm e}}}}
\newcommand {\Gmuon}      {{\Gamma_{\mu\overline{\mu}}}}
\newcommand {\Gtau}      {{\Gamma_{\tau\overline{\tau}}}}
\newcommand {\Ginv}       {\Gamma_{\mathrm{inv}}}

\newcommand {\Ghad}       {\Gamma_{\mathrm{had}}}
\newcommand {\Gnu}        {\Gamma_{\nu\overline{\nu}}}
\newcommand {\ptau} {\mbox{$\cal P_{\tau}$}}
\newcommand {\ALR} {\mbox{$A_{\rm {LR}}$}}
\newcommand {\Alrfff}   {\tilde{\rm {A}}_{FB}^{\rm f}}
\newcommand {\Alrfbb}   {\tilde{\rm {A}}_{FB}^{\rm b}}
\newcommand {\Abb}   {\ifmath{A_{\mathrm{FB}}^{\mathrm{b\bar{b}}}}}
\newcommand {\roots}      {\ifmath{\sqrt{s}}}
\newcommand {\rootsp}      {\ifmath{\sqrt{s'}}}
\newcommand {\cAe} {\mbox{$\cal A_{\rm e}$}}
\newcommand {\cAm} {\mbox{$\cal A_{\mu}$}}
\newcommand {\cAt} {\mbox{$\cal A_{\tau}$}}
\newcommand {\cAf} {\mbox{$\cal A_{\rm f}$}}

\newcommand {\cAl} {\mbox{$\cal A_{\ell}$}}
\newcommand {\cAb} {\mbox{$\cal A_{\rm b}$}}
\newcommand {\cAc} {\mbox{$\cal A_{\rm c}$}}

\newcommand {\Afbpol}     {A^{0,\,\ell}_{\rm {FB}}}
\newcommand {\chit}       {\chi^{2}}
\newcommand {\chisq}       {\chi^{2}}

\newcommand {\Ree}      {R_{\mathrm{e}}}
\newcommand {\Rmu}      {R_{\mu}}
\newcommand {\Rtau}      {R_{\tau}}
\newcommand {\Rl}      {R_{\ell}}
\newcommand {\GF}      {G_{\mathrm{F}}}
\newcommand {\thw}        {\theta_{\mathrm{W}}}
\newcommand {\alphamz}    {\alpha(\MZ)}
\newcommand {\alphasmz}    {\alpha_{\rm{s}}{\rm(\MZ)}}
\newcommand {\alphasmw}    {\alpha_{\rm{s}}{\rm(\MW)}}
\newcommand {\alfas}   {\alpha_s}

\newcommand  {\Zzero}   {\mbox{${\mathrm{Z}}$}}
\newcommand {\Ae}         {a_{\rm{e}}}
\newcommand {\Abcoup}     {a_{\rm{b}}}
\newcommand {\Accoup}     {a_{\rm{c}}}
\newcommand {\Af}         {a_{\rm{f}}}

\newcommand {\Alll}       {a_{\ell}}
\newcommand {\Vnu}        {v_{\nu}}
\newcommand {\Vmu}        {v_{\mu}}
\newcommand {\Vtau}       {v_{\tau}}
\newcommand {\Amu}        {a_{\mu}}
\newcommand {\Anu}        {a_{\nu}}
\newcommand {\Atau}       {a_{\tau}}
\newcommand {\Ve}         {v_{\rm{e}}}
\newcommand {\Vf}         {v_{\rm{f}}}
\newcommand {\Vq}         {v_{\rm{q}}}
\newcommand {\Vlll}       {v_{\ell}}
\newcommand {\Vb}         {v_{\rm{b}}}
\newcommand {\Vc}         {v_{\rm{c}}}
\newcommand {\ppb}        {p\bar{p}}
\newcommand {\qqb}        {q\bar{q}}
\newcommand {\ffb}        {\mathrm{f}\bar{\mathrm{f}}}
\newcommand {\ffbp}       {{\mathrm f\bar{f}^{'}}}
\newcommand {\fbp}        {{\mathrm \bar{f}^{'}}}
\newcommand {\udb}        {u\bar{d}}
\newcommand {\csb}        {c\bar{s}}
\newcommand {\qpqb}       {q + \bar{q}}
\newcommand {\qpqbll}     {\qqb\rightarrow\lept}
\newcommand {\qpqbd}      {q_{1} + \bar{q_{2}}}
\newcommand {\mumu}       {\mu^+\mu^-}
\newcommand {\tautau}     {\tau^+\tau^-}
\newcommand {\eeff}       {\ee\rightarrow\ff}

\newcommand {\eeee}       {\ee\rightarrow\ee}
\newcommand {\eemumu}     {\ee\rightarrow \mu^+\mu^-}

\newcommand {\eehad}      {\ee\rightarrow\mathrm{hadrons}}
\newcommand {\eeqq}       {\ee\rightarrow {\rm q}\overline{\rm q}}
\newcommand {\eeqqgluon}  {\ee\rightarrow {\rm q}\overline{\rm q}{\rm g}}
\newcommand {\eeqqg}      {\ee\rightarrow {\rm q}\overline{\rm q}(\gamma)}
\newcommand {\eell}       {\ee\rightarrow \lept}
\newcommand {\eeww}       {\ee\rightarrow\mathrm{W^+W^-}}
\newcommand {\Afb}     {A_{\mathrm{FB}}}
\newcommand {\Afbzl}     {A^{0,\,\ell}_{\rm {FB}}}
\newcommand {\Afbze}     {A^{0,\,{\rm e}}_{\rm {FB}}}
\newcommand {\Afbzf}     {A^{0,\,{\rm f}}_{\rm {FB}}}
\newcommand {\Afbzm}     {A^{0,\,\mu}_{\rm {FB}}}
\newcommand {\Afbzt}     {A^{0,\,\tau}_{\rm {FB}}}
\newcommand {\Afbzb}     {A^{0,\,{\rm b}}_{\rm {FB}}}
\newcommand {\Afbzc}     {A^{0,\,{\rm c}}_{\rm {FB}}}
\newcommand {\Afbl}     {A_{\mathrm{FB}}^{\ell}}
\newcommand {\Afbf}     {A_{\mathrm{FB}}^{\mathrm{f}}}
\newcommand {\Afbc}     {A_{\mathrm{FB}}^{\mathrm{c}}}
\newcommand {\Afbb}     {A_{\mathrm{FB}}^{\mathrm{b}}}
\newcommand {\Ztoqq}     {\mathrm{Z}\rightarrow q\overline{q}}

\newcommand {\Ztoll}     {\mathrm{Z}\rightarrow \lept}

\newcommand {\Zff}       {\mathrm{{Z}f\overline{f}}}
\newcommand {\Ztoff}     {\mathrm{{Z}\rightarrow f\overline{f}}}
\newcommand {\Ztonn}     {\mathrm{{Z}\rightarrow \nu\overline{\nu}}}
\newcommand {\sigmaf}  {\sigma_{\rm F}}
\newcommand {\sigmab}  {\sigma_{\rm B}}
\newcommand {\sigmal}  {\sigma_{\rm L}}
\newcommand {\sigmar}  {\sigma_{\rm R}}
\newcommand {\sigmalf}  {\sigma_{\rm L}^{\rm F}}
\newcommand {\sigmalb}  {\sigma_{\rm L}^{\rm B}}
\newcommand {\sigmarf}  {\sigma_{\rm R}^{\rm F}}
\newcommand {\sigmarb}  {\sigma_{\rm R}^{\rm B}}
\def\shad{\sigma_{\mathrm{h}}^{0}}
\newcommand {\Itf}         {t^3_{\rm f}}
\newcommand {\rhof}        {\rho_{\rm f}}
\newcommand {\Qf}          {q_{\rm f}}

\newcommand {\swsq}       {1-\MW^2/\MZ^2}
\newcommand {\swsqa}       {\sin^2\!\thw}
\newcommand {\swsqsq}      {\sin^4\!\thw}

\newcommand {\swsqeffff}   {\sin^2\!\theta_{\rm{eff}}^{\rm {f}}}
\newcommand {\swsqeffl}    {\sin^2\!\theta_{\rm{eff}}^{\rm {lept}}}

\newcommand {\cwsq}       {\cos^2\!\thw}
\newcommand {\swsqb}      {\sin^2\!\overline{\theta}_{\mathrm{W}}}

\newcommand {\mch}      {\multicolumn {2} {c}}
\newcommand {\mcha}      {\multicolumn {2} {c}}
\newcommand {\HI} {\mbox{$114 \le \MH~[{\rm {GeV}}] \le 1000$}}
\newcommand {\tI} {\mbox{$169.2 \le \Mt~[{\rm {GeV}}] \le 179.4$}}
\newcommand {\Rbb}        {{R_{\mathrm{b}}}}
\newcommand {\Rcc}        {{R_{\mathrm{c}}}}

\newcommand {\Rccz}       {{R^{0}_{\mathrm{c}}}}
\newcommand {\Rbbz}       {{R^{0}_{\mathrm{b}}}}
\newcommand {\Rqqz}       {{R^{0}_{\mathrm{q}}}}
\newcommand {\Gq}      {{\Gamma_{\rm q}}}
\newcommand {\stq}        {\tilde{\rm {t}}}
\newcommand {\chargino}   {\tilde{\chi}}
\newcommand {\MZS}     {{\rm{m_Z^2}}}
\newcommand {\MWS}     {\rm{m_W^2}}
\newcommand {\bbbar}        {{\rm{b}\bar{\rm{b}}}}
\newcommand {\ccbar}        {{\rm{c}\bar{\rm{c}}}}
\newcommand {\Ab}      {\rm{A_b}}
\newcommand {\Ac}      {\rm{A_c}}
\def\Dstar{\mbox{$\mathrm{ D^\star}$}}
\newcommand {\Dstarp} {{\rm D}^{*+}}

\newcommand {\Dstarpm} {\mbox{${\mathrm D}^{*\pm}$}}

\newcommand {\Dzero} {{\rm D}^0}
\newcommand {\Dplus} {{\rm D}^+}

\newcommand {\Ds}   {{\rm{D_s}}}
\newcommand {\Lc}   {{\rm{\Lambda_c}}}
\def\ra{\rightarrow}

\def\avQfb{{\langle \mbox{\rm Q}_{\mbox{\rm FB}} \rangle}}
\newcommand {\chibar}     {\bar{\chi}}
\newcommand {\eb}         {\epsilon_{b}}
\newcommand {\effc}         {\epsilon_{c}}
\newcommand {\effuds}         {\epsilon_{uds}}
\newcommand {\Rb}   {\ifmath{R_{\mathrm{b}}}}

\newcommand {\Rc}   {\ifmath{R_{\mathrm{c}}}}
\newcommand {\Cb} {{\cal C}_{\rm b}}
\newcommand{\lnu}   {\mbox{${\ell\overline{\nu}_{\ell}}$}}
\newcommand{\enu}   {\mbox{${e\overline{\nu}_{e}}$}}
\newcommand{\mnu}   {\mbox{${\mu\overline{\nu}_{\mu}}$}}
\newcommand{\tnu}   {\mbox{${\tau\overline{\nu}_{\tau}}$}}
\newcommand{\Wtolnu} {\mbox{$\mathrm{W}\rightarrow\ell\overline{\nu}_{\ell}$}}
\newcommand{\Wtoenu}{\mbox{$\mathrm{W}\rightarrow\mathrm{e\overline{\nu}_{e}}$}}
\newcommand{\Wtomnu}   {\mbox{$\mathrm{W}\rightarrow\mu\overline{\nu}_{\mu}$}}
\newcommand{\Wtotnu}   {\mbox{$\mathrm{W}\rightarrow\tau\overline{\nu}_{\tau}$}}
\newcommand{\Wtoqqb} {\mbox{$\mathrm{W}\rightarrow q_{i}\overline{q}_{j}$}}

\newcommand {\mtrans}      {m_{\mathrm{trans}}}
\newcommand {\dalfahad}   {\Delta\alpha^{(5)}_{\rm had}}
\newcommand {\Qwcs}         {Q_{\rm W}({\rm Cs})}
\newcommand {\eone}       {\epsilon_{1}}
\newcommand {\etwo}       {\epsilon_{2}}
\newcommand {\ethree}     {\epsilon_{3}}
\newcommand {\Drho}  {\Delta\rho}
\newcommand {\Dkap}  {\Delta\kappa}
\newcommand{\Dr}{\Delta{r}}
\newcommand{\Drw}{\Delta{r}_{\mathrm{w}}}
\newcommand{\Drrem}{\Delta{r}_{\mathrm{remainder}}}
\newcommand {\seff}       {\sin^2\!\theta_{eff}}
\newcommand {\szt}     {\rm{s_0^2}}
\newcommand {\czt}     {\rm{c_0^2}}
\newcommand {\roottwo}    {\sqrt{2}}
\newcommand {\NTC}      {{\rm{N_{TC}}}}

\newcommand {\NTF}      {{\rm{N_{TF}}}}
\begin{document}
\title{Precision Electroweak Tests of the Standard Model}
\author{Peter B Renton\dag}
\address{\dag\ Denys Wilkinson Building, Keble Road, Oxford OX1 3RH 

e-mail p.renton1@physics.ox.ac.uk}


\hspace*{3.0cm} {\it to be published in Reports on Progress in Physics}

\begin{abstract}

 The present status of precision electroweak data is reviewed.
These data include measurements of $\eeff$, taken at the Z resonance at LEP, 
which are used to determine the mass and width of the $\Zzero$ boson. 
In addition, measurements
have also been made of the forward-backward asymmetries for leptons
and heavy quarks, and also the final state polarisation of the
$\tau$-lepton. At SLAC, where the electron beam was polarised,
measurements were made of the left-right polarised asymmetry, $\ALR$, 
and the left-right forward-backward asymmetries for b and c quarks.

 The mass, $\MW$, and width, $\GW$, of the W boson have been measured
at the Tevatron and at LEP, and the mass of the top quark, $\Mt$, has been
measured at the Tevatron. These data, plus other electroweak data, are
used in global electroweak fits in which various Standard  Model
parameters are determined.
A comparison is made between the results of the direct measurements
of $\MW$ and $\Mt$ with the indirect results coming 
from electroweak radiative corrections. 
Using all precision electroweak data fits are also made to determine limits on
the mass of the Higgs boson, $\MH$. The influence on these limits of specific
measurements, particularly those which are somewhat inconsistent with the
Standard Model, is explored. The data are also analysed in terms of the 
quasi model-independent $\epsilon$ variables.

 Finally, the impact on the electroweak fits of the improvements in the 
determination of the W-boson and top-quark masses, expected from the 
Tevatron Run 2, is examined.

\end{abstract}

%
%

\section{Introduction\label{sec-intro}}

 The Standard Model (SM) has not come from a sudden inspirational flash
of brilliance, be it in the bath or elsewhere! Instead it is a compact
summary of experimental facts and theoretical ideas. It has taken the
painstaking work of many thousands of researchers, both experimenters
and theorists, over several decades to achieve the form of the model that we
know today. Apart from a few possible discrepancies, which are discussed
in detail below, the model is consistent with a huge amount of precisely
measured physical quantities. The various pieces of data which have, from
time to time, shown conflict with the SM have, on analysing more (or better
quality) data, returned to the fold of compatibility with the SM.

 Yet the Standard Model is almost certainly wrong! It is clearly incomplete,
as it does not include the force of gravity. It also suffers from severe
theoretical problems when the higher-order perturbative corrections to the
Higgs boson mass are computed. These corrections would most naturally be
infinite in the SM, and can only be rendered finite by rather inelegant means.
The introduction of {\it Supersymmetry}, in which a supersymmetric partner
for each SM particle is introduced, can cure the potential divergences 
associated with the Higgs particle. However, the cost is a large increase
in the number of {\it parameters} in the model. As yet, there is no direct
evidence for Supersymmetry.

 This review is organised as follows. In section \ref{sec-SM} a brief review
of the Standard Model is given, together with a discussion of
electroweak radiative corrections. The properties and results of 
the Z boson are described in section \ref{sec-Zdata}, and those of 
the W boson in section \ref{sec-wmass}.
In section \ref{sec-alfaem}, the running of the electromagnetic coupling 
constant $\alpha$(s) is discussed, and in section \ref{sec-sm_other}
other electroweak measurements are briefly described.
In section \ref{sec-sm_tests} the constraints and tests of the validity
of the Standard Model are examined, and in  section \ref{sec-future} the
future prospects for electroweak measurements are assessed. There is
a brief summary in section \ref{sec-summary}.

\section{The Standard Model\label{sec-SM}}

\subsection{\bf The building blocks\label{sec-bblocks}}

 In the development of the Standard Model experimental discovery and
theoretical insights have gone hand in hand\footnote{For more details and
a bibliography see, for example,~\cite{pbrbook}}. Experimentally, the discovery
of the electron, then the nucleus and its proton and neutron constituents,
was the starting point. Studies on radioactive nuclei led to the 
recognition that, in addition to
{\it electromagnetic} interactions, there were also {\it weak} interactions,
and to the hypothesis of the feebly interacting neutrino. The experimental
discovery of a more massive lepton, the muon, was (and still is) a puzzle. 
The third charged lepton, the $\tau$, completes the known spectrum.
In addition, it was experimentally established that the neutrinos 
associated with the electron and the muon have different interactions. 
Our current picture is
that we have three {\it generations of leptons}: the electron, muon and tau, 
and their corresponding neutrinos. These  neutrinos are light on
the scale of the charged leptons, and there is an increasing body of 
experimental evidence, from neutrino-oscillation experiments, indicating
that one or more may have non-zero mass. 

 The discovery of the pion, followed by the kaon and many other meson and
baryon states, manifested the {\it strong interaction}. 
The study of the lowest-lying meson and baryon multiplets led to the
hypothesis of non-integrally charged up(u), down(d) and 
strange(s) {\it quarks}. Initially the quark-model was regarded by many
as a convenient classification scheme for hadronic states, without the
quarks necessarily having any physical significance. The extension of this 
static quark idea to a more physical dynamic interpretation 
came as a result of studies of deep-inelastic scattering off nucleons, 
using both charged lepton and neutrino beams. At high enough energies, 
the scattering was consistent with being off the fractionally charged 
up and down quark constituents of the nucleon. 
These results also suggested, by energy-momentum conservation, that
there was a further fundamental constituent of the nucleon, the {\it gluon}, 
which did not participate directly in these interactions. 
Studies of the process $\eeqq$ had shown that hadronic jets were formed along 
the directions of the quark and antiquarks, giving a two jet topology in the 
bulk of the events. The study of three-jet events 
showed the existence of the process $\eeqqgluon$, in which the gluon
produced a jet of hadrons similar in properties to those 
produced by quarks and antiquarks.

Various experimental observations
involving quarks, together with the symmetry principles of the wave-functions
of the lightest baryons, led to the concept of {\it colour}; that is, each
quark comes in three distinct types, or colours. This, plus further work 
on these 
strong interactions and the discovery of charm(c), bottom(b) and 
top(t) quarks, led to the formulation of the 
theory of {\it Quantum Chromodynamics (QCD)}, which attempts to
describe the strongly interacting behaviour of quarks, antiquarks and gluons.

 Pioneering work, post Second World War, showed that the potentially infinite 
quantities in {\it Quantum Electrodynamics (QED)}, such as the mass and
charge of the electron, could be controlled through the procedure 
of {\it renormalisation}. This renewed attempts to produce a combined
description of electromagnetic and weak interactions in terms
of a relativistic quantum field theory. The existing theory of
weak interactions at that time was the V-A theory, which grew out of 
Fermi's theory. In the V-A theory the fermions have only left-handed couplings.
These ideas could be extended to include a virtual massive charged vector 
boson W, with a role analogous to that of the photon in QED, but the resulting 
theory was not renormalisable.
A crucial piece of experimental input was the discovery 
of {\it neutral currents} (NC)
in neutrino-nucleon deep inelastic scattering. Prior to 
this, {\it charged current} (CC)
interactions had been found, which could be described by the exchange of a 
virtual charged W boson, leading to a final state containing a charged 
lepton. In the neutral current interactions there was no final state charged
lepton,
and this indicated the exchange of a massive neutral boson, the Z boson.

 Further work has also shown that the six quarks can be grouped into three
{\it doublets} u-d, c-s and t-b. This grouping is similar to that of the
three left-handed lepton doublets, each of which consists of a neutrino and its
corresponding charged lepton. There is an underlying SU(2) symmetry, and the
quantum number corresponding to these doublets is called
{\it weak isospin}; in analogy with the strong isospin successfully 
developed in understanding strong interactions. The leptons and quarks 
also have right-handed singlet states corresponding to a U(1) symmetry, 
with quantum number {\it weak hypercharge}.

 The {\it unified} electroweak 
theory which evolved, and which was later shown to be renormalisable,
described the interactions of the spin-$\frac{1}{2}$ fermions (leptons 
and quarks) and the spin-1 gauge bosons (photon, Z and W$^{+-}$).
To describe the neutral current interactions, the physical photon (A$^{0}$) 
and Z boson (Z$^{0}$) fields
are written as a linear combination of the neutral gauge 
bosons W$^{0}_{3}$ and B, which correspond to the SU(2) and U(1) groups
respectively, as follows 
\begin{eqnarray}\label{weakang}
Z^{0} = \cos\theta_{W}W^{0}_{3} - \sin\theta_{W}B^{0} \nonumber\\
A^{0} = \sin\theta_{W}W^{0}_{3} + \cos\theta_{W}B^{0},
\end{eqnarray} 
where $\theta_{W}$ is the {\it Weinberg} or {\it weak mixing angle}.

 This theory, due mainly to Glashow, Salam and Weinberg,
 unified the weak and electromagnetic forces, but with just
these spin-$\frac{1}{2}$ fermions and spin-1 bosons all the 
particles in this theory are massless. The W and Z bosons have masses
of about $\MW \simeq $80 GeV and $\MZ \simeq$ 91 GeV respectively, 
whereas the direct experimental
limit on the photon mass is m$_{\gamma} \leq$ 2 10$^{-16}$ eV~\cite{pdg2001}, 
and is consistent with being massless.
A massless gauge boson is described by only two spin states, whereas there
are three for the massive case.
A possible remedy was provided by the introduction of a complex weak doublet 
of scalar 
bosons, the {\it Higgs doublet}, which gives mass to the fermions and 
the massive
gauge bosons. Through the process of spontaneous local symmetry breaking  
three members of the Higgs doublet gets `eaten' to form the longitudinal spin
states of the massive gauge bosons.
This leaves one massive physical neutral scalar particle,
the {\it Higgs boson} H. This {\it Higgs mechanism} gives the relationship 
that $\MW$ = $\MZ$$\cos\theta_{W}$, or $\swsqa = \swsq$, between the
masses of the W and Z bosons.  These massive gauge bosons were discovered in 
high energy proton-antiproton collisions at CERN. 
The mass of the Higgs boson is not predicted
by the theory, and so far it has not been discovered. There is, however,
a possible signal, with a significance of about 2 standard deviations, from the
combined LEP data, at a mass of 115 GeV~\cite{gianotti}.

 The fermion masses are accounted for by introducing a fermion-Higgs
field with {\it Yukawa couplings}. This does the job, but there is no
predictive power on the fermion masses since each mass corresponds to
an arbitrary coupling parameter, and must therefore be specified from the
experimentally measured value.

 For the strong interaction sector of the SM,
invariance under {\it local colour transformations} leads to the theory
of QCD and the group SU(3)$_{\rm c}$. The gauge bosons are the eight massless
gluons. 

\subsection{\bf Electroweak interactions\label{sec-eweak}}

 The charged fermions have both left-handed and right-handed states. The 
electromagnetic interaction depends only on the charge of the fermion and
does not distinguish between these states. However, the weak interaction 
does depend on the {\it handedness} of the state. 
For example, for the first
generation of leptons, the left-handed electron state $\eleft$ and the
neutrino $\nu_{\rm e}$ form a weak isospin doublet, 
whereas the right-handed
state $\eright$ is a singlet. In addition to 
the left-handed {\it weak isospin} t, there is
a further quantum number, the {\it weak hypercharge} y; defined such that the
charge q = t$_{3}$ + y/2.
The properties of the fundamental fermions are given 
in table \ref{tab-fermions}\footnote{Direct evidence for the last remaining
building block, the $\tau$ neutrino, has recently been presented by
the DONUT Collaboration~\cite{donut}, which uses emulsion detectors in an intense
neutrino beam at Fermilab, in which the electron and muon neutrino fluxes
are heavily suppressed.}. 
Note that the classification for neutrinos is
that of the {\it Minimal} Standard Model, in which the neutrinos are massless.

\begin{table}
\caption{Quantum numbers for the fermions in the Standard Model,
         where q is the charge, and $t$
and $t_{3}$ denote the weak isospin and its third component.
\label{tab-fermions}} 

\begin{indented}
\lineup
\item[] 
\begin{tabular}{cccccc} 
\br
 \multicolumn{3}{c}{\bf Fermion Generation}  &  
 \multicolumn{3}{c}{\bf Quantum Numbers} \\
1 & 2 & 3 & $q$ &  $t$ & $t_{3}$          \\ \\ 
\mr
  \multicolumn{3}{c}{\bf Leptons}  \\  \\
    $ \left( \begin{array}{c} \nel \\ e  \end{array} \right)_{L} $
&   $ \left( \begin{array}{c} \nmu \\  \mu \end{array} \right)_{L} $
&   $ \left( \begin{array}{r} \ntau \\ \tau \end{array} \right)_{L} $

&  $\begin{array}{c} 0 \\ -1 \end{array}$
&  $\frac{1}{2}$
&  $\begin{array}{c} +\frac{1}{2} \\ -\frac{1}{2} \end{array}$ \\

$e_{R}$ & $\mu_{R}$ & $\tau_{R}$ & $-1$ & $0$ &$0$ \\
\\
  \multicolumn{3}{c}{\bf Quarks}  \\  \\
     $ \left( \begin{array}{c} u \\ d  \end{array} \right)_{L} $
&    $ \left( \begin{array}{c} c \\ s  \end{array} \right)_{L} $
&    $ \left( \begin{array}{c} t \\ b  \end{array} \right)_{L} $

&  $\begin{array}{c} +\frac{2}{3} \\ -\frac{1}{3} \end{array}$
&  $\frac{1}{2}$
&  $\begin{array}{c} +\frac{1}{2} \\ -\frac{1}{2} \end{array}$ \\

$u_{R}$  & $c_{R}$ & $t_{R}$  & $+\frac{2}{3}$ &$0$ & $0$ \\
$d_{R}$  & $s_{R}$ & $b_{R}$  & $-\frac{1}{3}$ &$0$ & $0$ \\
\br
\end{tabular}
\end{indented}
\end{table}

 The electroweak theory is based on the symmetry of invariance under local
gauge transformations. That is, the physical equations are invariant
under a phase change applied independently at each space-time point.
The group corresponding to weak isospin is SU(2)$_{\rm L}$, whereas that
for weak hypercharge is U(1)$_{\rm y}$. In the electroweak theory there
is invariance under SU(2)$_{\rm L}\otimes$U(1)$_{\rm y}$.
There are constants g and g$^{'}$
corresponding to the SU(2)$_{\rm L}$ and U(1)$_{\rm y}$ groups. These are
related to the electron charge e and $\sin\thw$ by
\begin{eqnarray}\label{eandg}
e = g\sin\thw = g^{'}\cos\thw .
\end{eqnarray} 

\subsection{\bf Electroweak couplings of SM particles\label{sec-couplings}}

\begin{figure}[htbp]
\vspace*{13pt}
\vspace*{-2.0cm}
\begin{center}
\mbox{
\epsfig{file=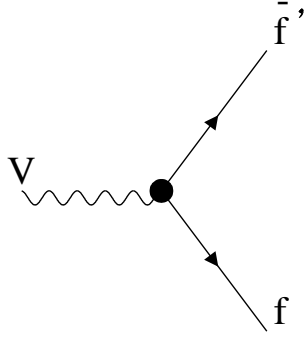,height=7cm}}
\end{center}
\vspace*{-1.0cm}
\caption
{The interaction vertex of a Vector Boson V (W or Z) with a fermion-antifermion
pair.}
\label{fig-vffbar}
\end{figure}

 The interaction vertices involving a Vector Boson (V=W,Z) and a 
fermion-antifermion pair (see fig.\ref{fig-vffbar}) are of great importance.
These can be classified as {\it charged} or {\it neutral} currents,
depending on the nature of the Vector Boson.

 {\it Charged current } interactions involve the W$^{+-}$ bosons, and
the interaction Lagrangian for the W$\ffbp$  vertex has the form
\begin{eqnarray}\label{lag-cc} 
{\cal L}_{CC} = -\frac{g}{2\sqrt{2}}W_{\mu}^{-}j^{\mu}_{CC},
\end{eqnarray}
with
\begin{eqnarray}\label{lag-ccj}
j^{\mu}_{CC} = \bar{f^{'}}\gamma^{\mu}(1-\gamma^{5})f,
\end{eqnarray}
where (f,$\fbp$) = ($\ell^{-}$,$\bar{\nu_{\ell}}$) for leptons
and ($q_{\frac{1}{3}}$,$\bar{q_{\frac{2}{3}}}$) for quarks,
with $q_{\frac{1}{3}}$ = (d,s,b) and $q_{\frac{2}{3}}$ = (u,c,t).
This form corresponds to pure left-handed coupling of the W to fermions.

 The {\it neutral current } Lagrangian for the Z$\ff$ vertex can be
written as
\begin{eqnarray}\label{lag-nc}
 {\cal L}_{NC}^{Z} = -\frac{g}{2\cos\thw}Z_{\mu}j^{\mu}_{NC},
\end{eqnarray}
with
\begin{eqnarray}\label{lag-ncj}
 j^{\mu}_{NC} = \bar{f}\gamma^{\mu}( \Vf-\Af\gamma^{5})f 
             = \bar{f}\gamma^{\mu}[C_{L}^{f}(1-\gamma^{5}) +
C_{R}^{f}(1+\gamma^{5}) ]f.
\end{eqnarray}

Here $\Vf$ and $\Af$\footnote{The alternative symbols g$_{Vf}$ (or g$_{V}^{f}$)
and g$_{Af}$ (or g$_{A}^{f}$) are also used.} are the vector- and axial-vector
couplings. There are both 
left-handed (C$_{L}^{f} = \ell_{f} = (\Vf+\Af)/2$) and
right-handed (C$_{R}^{f}$ = r$_{f}$ = ($\Vf-\Af)/2$) terms. 
In the SM, at tree-level,
\begin{eqnarray}\label{coup-nc}
 \Vf = \Itf - 2\Qf\swsqa,  \hspace*{1.0cm}  \Af =  \Itf,
\end{eqnarray}
where t$_{f}^{3}$ is the third component of the weak-isospin 
(see tab.\ref{tab-fermions}). 

 The photon-fermion coupling, for a fermion of charge q$_{f}$, has the form 
\begin{equation}\label{lag-gam}
 {\cal L}_{NC}^{\gamma} = -eq_{f}A_{\mu}j^{\mu}_{\gamma} \nonumber , \\
\end{equation}
with 
\begin{equation}\label{lag-gam1}
j^{\mu}_{\gamma} = \bar{f}\gamma^{\mu}f,
\end{equation}
that is, the left- and right-handed couplings are equal, and there is only
a vector term. For the case of polarised fermions, the left- and
right-handed fermion components, i.e. (1-$\gamma^{5}$) and (1+$\gamma^{5}$), 
are used. More details and specific Born-level calculations in the SM
can be found in \cite{pbrbook}.

 Many of the electroweak interactions discussed later involve the
scattering of two fermions i.e. f$_{1}$f$_{2} \rightarrow$ f$_{3}$f$_{4}$.
The matrix element for this process has the form
\begin{eqnarray}\label{matelem}
 {\cal M}_{fi} \propto  j_{12}^{\mu} (P_{V})_{\mu\nu} j_{34}^{\nu},
\end{eqnarray}
with  j$_{12}^{\mu}$ = 
$\bar{f_{2}}\gamma^{\mu}(\epsilon_{V}-\epsilon_{A}\gamma^{5})f_{1}$ etc. 
The couplings $\epsilon_{V}$ and $\epsilon_{A}$ depend on the form of the 
interaction, as discussed above. The vector-boson {\it propagator} P$_{V}$ has
the form 1/s for a photon, and the relativistic Breit-Wigner resonance
form
\begin{eqnarray}\label{bwig}
P_{V} \propto \frac{1}{ s-M_{V}^{2} + is\Gamma_{V}/M_{V} }
\end{eqnarray}
for V=W or Z, where s is the centre-of-mass energy squared. For the case of
a neutral current interaction involving charged particles, both $\gamma$
and Z exchange are possible, so there is a $\gamma$-Z interference term.
This latter term has an energy dependence proportional to the 
difference (s-$\MZ^{2}$), and so
changes sign in going from below to above the Z-pole. The
fermion decay f$_{1} \rightarrow$ f$_{2}$f$_{3}$f$_{4}$ has a CC matrix
element similar in form to that in eqn.(\ref{matelem}),  
with a W-boson far from its mass shell except for the decay of the top-quark.

 The Higgs boson is a neutral scalar particle in the SM, of (unknown)
mass $\MH$. In the kinematic regime explored at LEP the main
decays considered are H $\rightarrow \ff$. The coupling is proportional
to the fermion mass m$_{f}$, so  H $\rightarrow \bb$ 
and H $\rightarrow \tautau$ are the dominant decay modes for $\MH$ up to
about 120 GeV. For higher masses the decays  H $\rightarrow$ WW and ZZ
become increasingly important.

\subsection{\bf Parameters and predictions of the SM\label{sec-params}}

In the Standard Model there is invariance under the 
combination SU(3)$_{\rm c}\otimes$SU(2)$_{\rm L}\otimes$U(1)$_{\rm y}$.
The Lagrangian of the Standard Model can be completely specified, but exact
solutions have so far not been obtained. Instead, low-order perturbation theory
is used. The infinities arising in the renormalisation process are rendered
harmless by expressing the results of these perturbative calculations in 
terms of precisely known physical quantities.

The QCD part of the SM, corresponding to the SU(3)$_{\rm c}$ group, has
only one parameter: the gauge coupling constant g$_{\rm s}$. This constant
is not specified by the theory. 
Higher order QCD effects introduce a dependence on the momentum scale Q in 
the QCD coupling parameter $\alpha_{\rm S}$. The value of $\alpha_{\rm S}$(Q)
decreases as the scale Q increases. It is usual to specify the value at
the scale Q = $\MZ$; thus giving $\alpha_{\rm S}$($\MZ$).

 The electroweak sector is more complicated. There are constants g and g$^{'}$
corresponding to the SU(2)$_{\rm L}$ and U(1)$_{\rm y}$ groups. 
The Higgs potential term in ${\cal L}_{\rm SM}$ is
\begin{eqnarray}\label{higgspot}
V(\phi) = \mu^{2}\mid\phi\mid^{2} + \lambda^{2}\mid\phi\mid^{4},
\end{eqnarray}
with parameters $\mu$ ($\mu^{2} < $ 0) and $\lambda$. Alternatively these
can be cast in terms of the vacuum expectation value ({\it vev}), $v$, and the 
mass of the Higgs boson, $\MH$.

 There are also parameters associated with the masses and mixing angles 
of the leptons and quarks. For the lepton sector, in the case of 
massless neutrinos, there are three parameters corresponding to the 
charged lepton masses. For the quark sector there are parameters for
the quark masses, and there are four mixing angles needed to describe the
CC transitions between quark states. The set of these fermion
parameters is referred to as $\{\rm{m_{f}}\}$.
This gives in total 17 parameters for the electroweak sector,
and all but two of these (g and g$^{'}$) are associated 
with the Higgs field. The number of parameters is of course larger if the
neutrino sector also has mass. Thus the Higgs can solve the problem of 
giving mass to the particles, but only at the expense of a large number 
of parameters.

 In addition, the SM embodies some explicit assumptions which do not
directly appear as parameters. The hypothesis of {\it lepton universality}
is assumed. That is, the overall couplings of the leptons are the same, and
any differences arise only from the lepton masses.
The same is also the case with the quark sector. It is thus important to
test these hypotheses experimentally.

 In addition to the aesthetic problem that the Higgs mechanism introduces 
a large number of parameters, there is a further, and more fundamental,
problem with the Higgs boson. The mass of the scalar Higgs boson has
contributions from loop diagrams which involve integration over the
four-momentum of the loop. The resulting Higgs mass depends on the upper limit
of this integration. In the SM there is no upper scale to fix this limit.
The SM does not include the force of {\it gravity}, which
is another serious deficiency.
Using the {\it Planck scale} of $\simeq$ 10$^{19}$ GeV as an upper limit
could resolve the problem, but only at the expense of having to {\it fine-tune}
cancellations to many significant figures. The problem of the large difference
between the {\it electroweak scale} and that where new physics, beyond the SM,
enters is known as the {\it hierachy or naturalness problem}. 

 A possible solution to the hierachy problem is the introduction of a
further symmetry of nature called {\it supersymmetry}. This corresponds
to space-time transformations which change fermions into bosons, and
{\it vice versa}. The introduction of supersymmetry (or SUSY) essentially
doubles the number of fundamental particles. There is a scalar super-partner
for each fermion and a spin-$\frac{1}{2}$ super-partner
for each spin-1 particle. In the calculation of the Higgs mass, the loop
contributions from the s-particles have the opposite sign to those of
their corresponding particles, and thus tend to cancel the divergent
loop contributions. However, none of these super-particles (or s-particles)
have been observed, so supersymmetry, and this cancellation, cannot be exact. 
For the cancellation to work, the s-particles cannot be too massive; and that
sets an upper limit on the s-particle masses in the TeV range. 

 The introduction of supersymmetry into the SM, the simplest version of
which is the {\it Minimal Supersymmetric Standard Model (MSSM)}, brings
with it many (more than 100) extra prameters. There is, without as yet
any experimental guidance, a choice among many possibilities as to how
supersymmetry is broken. The phenomenology is thus complicated and is
outside the scope of this article. 

\subsection{\bf Electroweak Radiative Corrections\label{sec-radcor}}

 In the electroweak sector, as discussed above, there are the SM parameters
g and g$^{'}$, or e and $\swsqa$, as well as the Higgs mass $\MH$ 
and {\it vev} $v$. Alternatively, these can be expressed in terms of the 
electromagnetic coupling constant $\alpha$, the gauge boson masses $\MZ$ and
$\MW$, and also $\MH$. In addition, there are the fermion parameters 
$\{\rm{m_{f}}\}$, and the QCD parameter $\alpha_{\rm s}$.

 For SM predictions at the scale $\MZ$, the QED coupling, $\alpha$, is needed 
at this scale. Although $\alpha(0)$, at scale q$^2$ = 0, is precisely 
known experimentally, there is, as discussed in section \ref{sec-alfaem},
some significant uncertainty on $\alphamz$. The
value of the Fermi coupling constant, $\GF$, is accurately determined from
measurements of $\mu$ decays~\cite{pdg2001}. 
Although $\GF$ is specified directly in terms
of $\MW$ at the Born-level, the introduction of higher-order loops means
that, in the calculation of $\MW$ in terms of $\GF$, a dependence on other
SM parameters enters the computation. 

 The relationship between the neutral and charged weak couplings
fixes the ratio of the W and Z boson masses, namely
\begin{equation}  \label{eq:rho}
  \rho = \frac{\MW^2}{\MZ^2\cwsq}.
\end{equation}
This $\rho$ parameter  is determined by the Higgs structure of the theory.
In the Minimal Standard Model, where there are only Higgs doublets,
$\rho$ is unity. Electroweak radiative corrections 
lead to $\rho$ = 1 + $\Delta\rho$.

 Electroweak corrections modify the tree-level relationships such that
\begin{equation}
  \label{eq:GFmod}
  \GF = \frac{\pi\alpha}{\sqrt{2}\MW^2\swsqa}\frac{1}{(1-\Dr)},
\end{equation}
where $\swsqa =\swsq$. The quantity $\Dr$, which is zero at tree-level, 
is given by
\begin{equation}
  \label{eq:Dr}
  \Dr = \Delta\alpha + \Drw.
\end{equation}
The term $\Delta\alpha$ controls the running of $\alpha$(s), and is given by
\begin{equation}\label{alfa1}
 \alpha (s)  = \frac{ \alpha(0)}{ 1 - \Delta \alpha(s) }
   = \frac{ \alpha(0)}{ 1 - \Delta \alpha_{lept}(s) -
\Delta \alpha_{top}(s) -  \Delta\alpha^{(5)}_{had}(s) },
\end{equation}
where $\alpha$(0) = 1/137.036. At LEP/SLD energy scales this becomes 
$\alphamz \simeq$ 1/129. 
The dominant term in $\Drw$ is given by $\Delta\rho$, defined above:
\begin{equation}
  \label{eq:drw}
  \Drw = -\frac{\cwsq}{\swsqa}\Delta\rho + \Drrem.
\end{equation}

\begin{figure}[htp]
\vspace*{13pt}
\vspace*{-1.0cm}
 \begin{center}
   \mbox{\epsfig{file=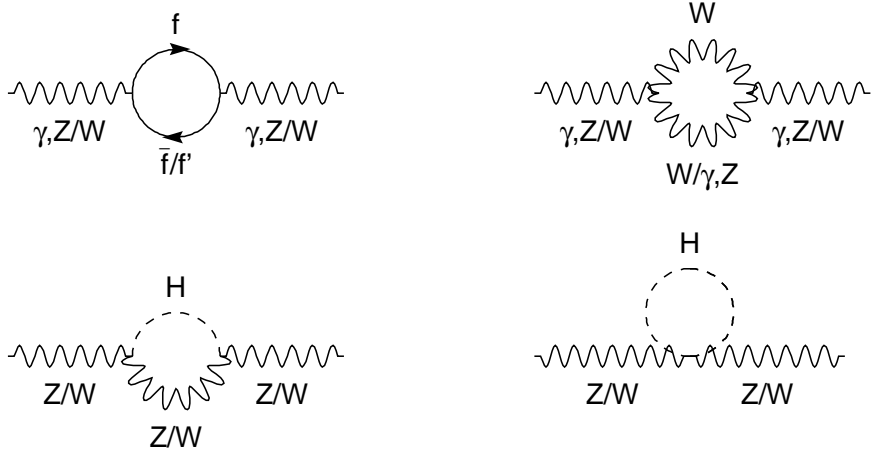,height=7cm}}
\end{center}
\vspace*{-0.5cm}
 \caption{Electroweak loop corrections to the gauge boson propagators.
}
\label{fig-props}
\end{figure}

Electroweak corrections enter through loop diagrams to the vector boson
propagators, see fig.\ref{fig-props}, and through vertex corrections.
The leading-order contributions to $\Drw$ depend on the masses of the
top quark, $\Mt$ and Higgs boson, $\MH$:
\begin{eqnarray}  \label{eq:drwtop}
  \Dr^{\mathrm{t}} & = & 
 -\frac{3\GF\MW^2}{8\sqrt{2}\pi^2}\frac{\Mt^2}{\MW^2}\frac{\cwsq}{\swsqa} + 
 \cdots\\
  \Dr^{\mathrm{H}} & = &
     \frac{11}{3}\frac{\GF\MW^2}{8\sqrt{2}\pi^2}\left(\ln\frac{\MH^2}{\MW^2}
-\frac{5}{6}\right) + \cdots . \label{eq:drwhiggs}
\end{eqnarray} 

The leading contributions are thus $\Dr^{\mathrm{t}} \propto \Mt^{2}$ and
$\Dr^{\mathrm{H}} \propto \ln(\MH)$.
The largest effect is from the top-quark mass, $\Mt$.
This is because of the large difference in mass 
between $\Mt$ ($\simeq$ 175 GeV) and that of the 
bottom-quark, $\Mb$ ($\simeq$ 5 GeV). 

 The large amount of electroweak data, accumulated from $\ee$ colliders working
at the Z peak and available in the early Nineties, was sufficiently
accurate to make an estimate of the top-quark mass from electroweak fits,
within the context of the SM. For example, the data available in Summer
1994~\cite{ichep94}, around the time of the top-quark discovery,
gave $\Mt$ = 173 $^{+12}_{-13}$ GeV for $\MH$ = 300 GeV, with additional
uncertainties of -20 and +18 GeV if the Higgs mass is varied from 60 to
1000 GeV. This SM prediction was in good 
agreement with the directly measured value of $\Mt \simeq$ 175 GeV.
This represents a triumph for the Standard Model. 
The current situation, and also the sensitivity to the Higgs boson mass,
is discussed in section \ref{sec-sm_tests}.

 A more detailed discussion of  electroweak radiative corrections, as well
as other theoretical issues of the SM, can be found in~\cite{okun,PCPreport}.

\section{The Z boson\label{sec-Zdata}}

 The data on the Z boson discussed in this review come from
the $\ee$ colliders at LEP in CERN, Geneva, Switzerland, and from the
Stanford Linear Collider (SLC), in California, USA. Most of the 
Z boson data are, or are very close to being, final. 
The data used here are mainly those presented at the
EPS 2001 Conference in Budapest, Hungary, in July 2001~\cite{EWWG01},
in which a full list of references to individual results can be found.

\begin{figure}[htp]
\vspace*{13pt}
\vspace*{-1.0cm}
 \begin{center}
   \mbox{
     \includegraphics[width=0.45\linewidth]{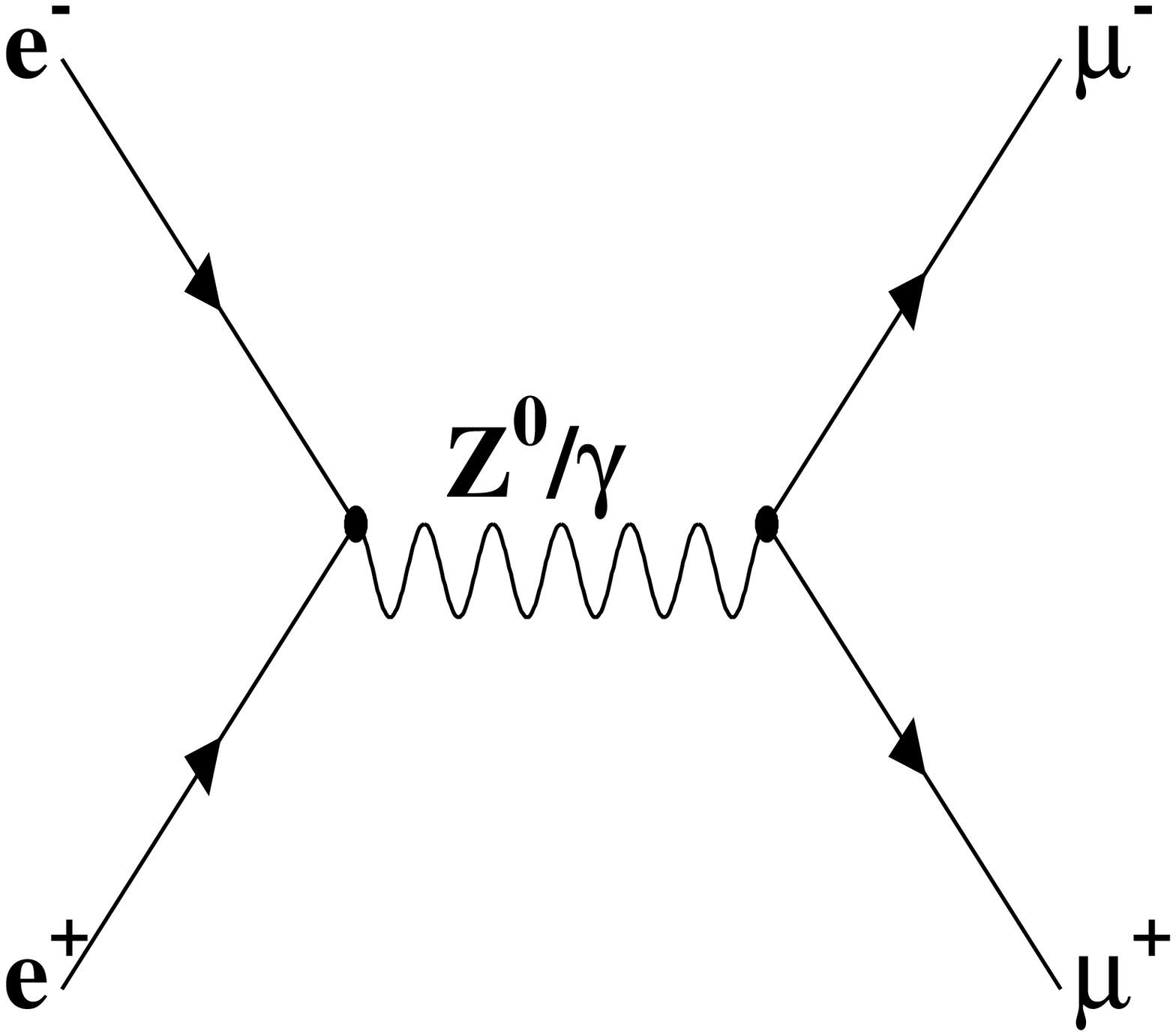}
     \includegraphics[width=0.45\linewidth]{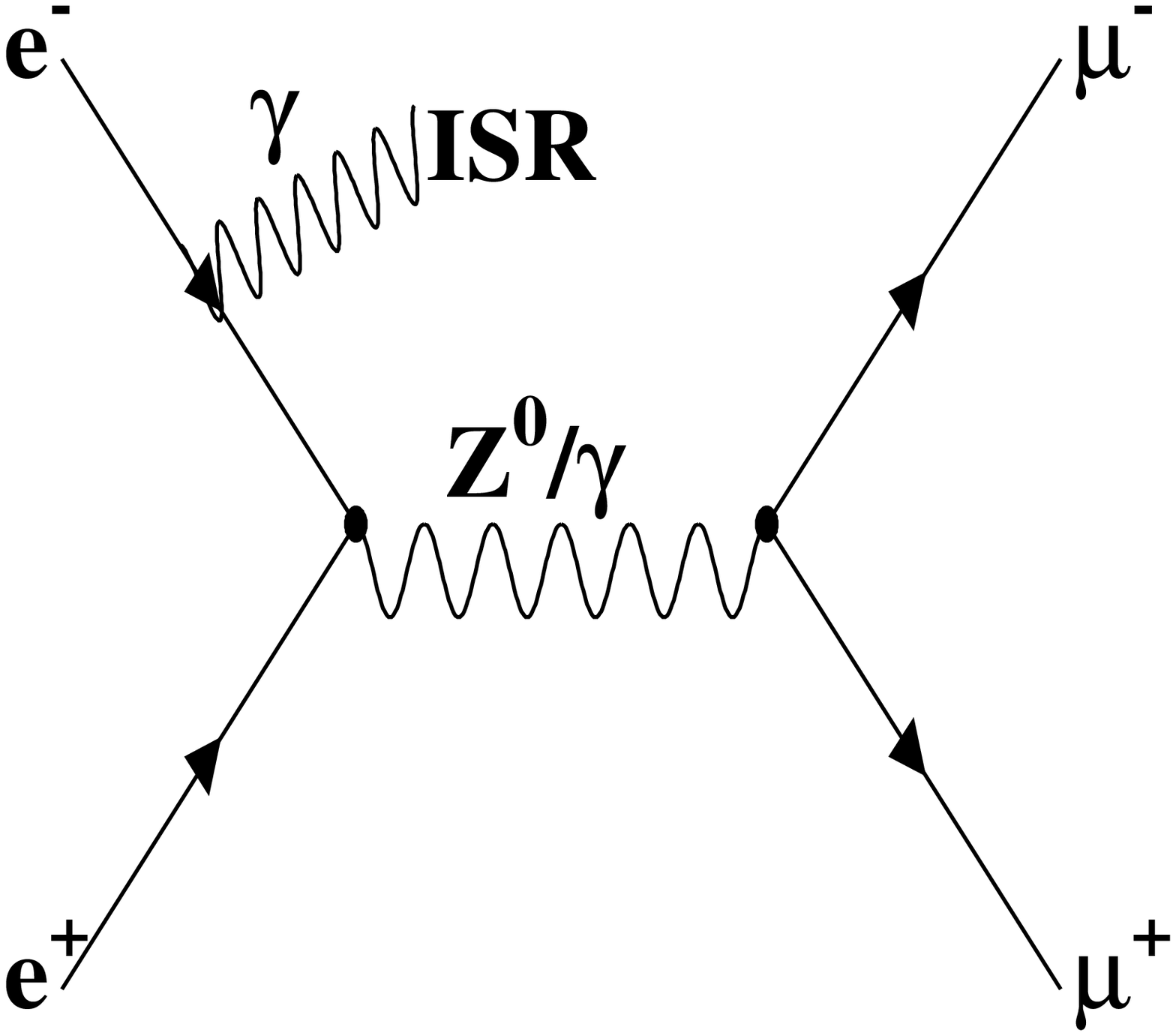}
     }
\end{center}
 \caption{Feynman diagrams for the lowest-order process $\eemumu$ and
for the case where there is initial state radiation of a photon.
}
\label{fig-mumu}
\end{figure}

 The lowest-order Feynman diagrams for the process $\eemumu$, and also
where there is an initial state radiation of a photon, are shown
in fig.~\ref{fig-mumu}.

\subsection{\bf Z boson variables\label{sec-zvar}}

 The Z boson decays to $\ffb$ and the 
$\Zff$ vertex can be described by effective vector ($\Vf$) and
axial-vector ($\Af$) coupling constants~\footnote{ These are in general 
both complex and energy dependent. In practice,
the values are extracted at the Z pole with the imaginary parts taken from
the SM; see ~\cite{PCPreport} for a detailed discussion.}
such that, to a very good approximation,
the Born-level formulae are retained. It is useful to define the 
{\it polarisation parameter}
\begin{equation}\label{eqn-z1}
\cAf = \frac{2\Vf\Af}{(\Vf^2 + \Af^2 )} 
     = \frac{ \ell_{f}^{2} - r_{f}^{2}} { \ell_{f}^{2} + r_{f}^{2} } \ \ .
\end{equation}
 The cross-section for $\eeff$, close to the $\Zzero$ pole, may be written in
terms of the Z mass, $\MZ$, and total width, $\GZ$, as
\begin{equation}\label{eqn-zbwig}
  \sigma_{f}(s) = \frac{\sigma_{f}^{0}}{(1+\delta_{\rm QED})}
  \frac{ s \GZ^{2} }
   { (s-\MZ^{2})^{2} + s^{2}\GZ^{2}/\MZ^{2} } \hspace*{0.5cm}
 + \sigma_{\gamma} + \sigma_{\gamma/\Zzero} \ ,
\end{equation}
where s is the square of the $\CoM$ energy. The QED 
factor $\delta_{\rm QED}$ = 3$\alphamz$/4$\pi$ cancels the corresponding
factor in the $\Ge$ partial width (see below). The 
contributions $\sigma_{\gamma}$ and $\sigma_{\gamma/\Zzero}$
from $\gamma$-exchange and $\gamma$-Z interference are small 
($\leqsim$1$\%$) in the region around the Z-pole. 
The Breit-Wigner form has an s-dependent width, which takes into
account higher-order effects~\cite{Bardin88}.
 $\sigma_{f}^{0}$ is
the {\it pole cross-section}, which is defined as
\begin{equation}
\sigma_{f}^{0} =  \frac{12\pi{ \Ge\Gf}}{\MZ^{2}\GZ^{2}}.
\end{equation}
 Here, $\Gf$ is the partial width for $\Ztoff$, which in turn can be
written as \footnote{All the formulae here are for m$_{f}$=0. In practice,
finite mass terms are taken into account.} 
\begin{equation}\label{eqn-zgf}
    \Gf = \frac{\GF \MZ^{3} }{ 6\pi\sqrt{2}}  ( \Vf^{2} + \Af^{2} )
    f_{QCD}  f_{QED} \ \ .
\end{equation}
 The final-state QED correction factor 
is f$_{QED}$ = 1 + 3$\alphamz$Q$_{f}^{2}$/4$\pi$,
whereas f$_{QCD}$ is unity for leptons 
and f$_{QCD}$ = 3(1 + c$_{q}\alphasmz/\pi$ + ..) for quarks, 
with c$_{q}\simeq$ 1.

 The formula for the cross-section $\sigma_{f}$(s) is an effective Born-level
formula and must be convoluted with QED radiative corrections before comparing
with experiment. The cross-sections which are measured in practice are 
for $\eehad$ and $\eell$, for $\ell$ = e,$\mu$,$\tau$. 
The hadronic width $\Ghad$ is the sum 
of $\Gamma_{q}$
for q=u,d,s,c,b. The total Z width, assuming lepton universality, 
is $\GZ = \Ghad + 3\Gl + \Ginv$, where $\Ginv = 3\Gnu$ in the SM.
The results below are expressed in terms of the ratio of the hadronic
to leptonic partial widths R$_{\ell}$ ($\ell$ = e,$\mu$,$\tau$), where
\begin{equation}
               \Rl = \frac{\Ghad}{\Gl}. 
\end{equation}

 The distribution of the angle $\theta$ of the outgoing fermion f, with
respect to the incident e$^{-}$ direction is given, at Born-level, by
\begin{equation}
     \frac{d\sigma}{d\cos\theta} \propto 1 + \cos^2\theta + 
     \frac{8}{3}\Afb \cos\theta.
\end{equation}
The {\it forward-backward asymmetry} is defined 
as $\Afbf = (\sigmaf - \sigmab)/
(\sigmaf + \sigmab)$, where 
 $\sigmaf$~($\sigmab$) are the cross-sections in 
the forward (backward) directions. The value measured at the Z-pole, the
{\it pole asymmetry}, is defined as
\begin{equation}\label{eqn-z2}
\Afbzf = \frac{3}{4}\cAe\cAf \ .
\end{equation}
 Since $\cAf$
depends on the ratio $\Vf$/$\Af$, a measurement of $\Afbzf$ depends on
both $\Ve/\Ae$ and $\Vf/\Af$. The effective couplings can also be written as
\begin{equation}\label{eqn-z3}
     \Af = \Itf \sqrt{\rhof} \ , \ \ \ \ \ 
 \frac{\Vf}{\Af} = 1 - 4|\Qf| \swsqeffff \ ,
\end{equation}
where $\Itf$ is the third component of the weak isospin.
The mixing angle defined for {\it leptons} ($\swsqeffl$) is 
used for reference.
Those defined for quarks have  small shifts, due to SM plus any
new physics~\cite{ORR}.

\subsection{\bf LEP lineshape scans and beam energy determination\label{sec-scans}}

 The LEP $\ee$ Collider took data at the Z-pole between 1989 and 
1995 ({\it LEP 1 phase}). There
were lineshape scans, in which data were taken at a series of energy settings
within $\pm$ 3 GeV of the Z-pole, in all years except 1992 and 1994. The early
scans had seven energy points, whereas those in 1993 and 1995
had three points. In the two latter scans, the two off-peak energy points 
were situated at about 1.8 GeV below and above the $\Zzero$ peak, and are
referred to as {\it peak-2} and {\it peak+2} respectively. 
The choice of these energies optimised the precision
on the Z width. In both these scans the
total off-peak luminosity was about 20 pb$^{-1}$, significantly higher
than in the 1991 and earlier scans. 

Good accuracy on the LEP beam energy E$_{LEP}$ is crucial in the
determination of $\MZ$ and $\GZ$. A precise energy calibration 
measurement (E$_{LEP}^{pol}$, with $\delta$E$_{cms} \leq$ 0.8 MeV) of 
the average circulating beam energy
can be made using the technique of {\it resonant depolarisation}.
This technique was available, but only at one energy point, for the 1991
scan. For the 1993 and 1995 scans it was  available at all the energy points.
For the two latter scans, most of the energy calibration 
measurements were performed after physics
data-taking at the end of fills. For the 1993 scan, about one-third of 
the off-peak fills (representing about 40$\%$ of the recorded off-peak 
luminosity) was calibrated.
For the 1995 scan about 70$\%$ of the 
recorded off-peak luminosity was calibrated.
 
 Although resonant  depolarisation gives a very precise energy value 
at the time of measurement, the LEP energy varies with time, due to the 
Earth tides and 
other effects such as the temperature of the dipole magnets.
Since the RF frequency, and thus the orbit length, is fixed, stresses in the
local rock structure result in changes to the position of the beams in the
quadrupole magnets. This changes the beam energy, since the effective dipole
field changes. These energy changes can be tracked accurately using 
measurements of the horizontal beam orbit positions, x$_{orb}$. 
For the 1993 scan a {\it model} for E$_{LEP}$ was developed, based 
on x$_{orb}$, together with correction
terms from the magnetic dipole fields and temperatures, the RF cavity
voltages, as well as other factors~\cite{LEPEN93}. 

 In the analysis of the 1993 scan a term was included for the increase 
in the energy 
during a fill, based on observations of an NMR device in a reference 
dipole magnet (outside the LEP ring), which showed an upward drift in the 
field together with occasional jumps of a few MeV. To understand better 
these effects in 1995, two NMR devices were inserted in 
dipole magnets in the LEP
ring. Furthermore, in four physics fills energy calibration measurements
were made at both the beginning and end of fills to measure the energy
change. The NMR devices showed a significant amount of noise, which was
largely anti-correlated in the two devices which were on opposite sides
of the LEP ring, and which was much smaller
between midnight and 5am. The main cause of this noise was traced to  
local electric trains, the earth currents from which travelled along the 
LEP beam pipe. When averaged over several fills
these NMR devices showed an increase in energy during a fill  of typically 
5 MeV (much larger than the rise estimated in ~\cite{LEPEN93}). The magnitude 
of the rise was confirmed using the beginning and end of fill
energy calibrations, as
well as studies of fills in machine development periods in 1993 and 1995
in which energy calibration measurements were made over periods of several
hours. 

 An improved model, including this rise term, was 
developed~\cite{LEPEN95}, both for the 1995 scan and also for a revised
analysis of the 1993 scan, and also the 1994 peak data. For each energy and year
a single normalisation 
parameter was used and the rms values of E$_{LEP}^{model}$ -  E$_{LEP}^{pol}$ 
were determined.  These rms values (typically a few MeV) are of importance 
in determining the energy error on those fills with no polarisation 
measurement, since it is assumed that these follow the same distribution.
The error for the 1995 scan from this effect is smaller than for 1993 since
a much higher fraction of the off-peak luminosity was collected in 
calibrated fills.

 The analysis of the energy values and errors for the years 1993 to
1995 includes estimates of the correlations both
between  energy points and between years ~\cite{LEPEN95}.
The correlations
between off-peak energy points are important because error components which are
highly correlated do not contribute significantly to $\delta\GZ$.
The contributions of the LEP energy uncertainties to the errors 
on $\MZ$ and $\GZ$ are $\delta\MZ$(LEP) $\simeq \pm$ 1.7 MeV 
and $\delta\GZ$(LEP) $\simeq \pm$ 1.2 MeV.

The uncertainty of the LEP cms energy spread ($\simeq$ 55 $\pm$ 1 MeV) also 
gives rise to an error on $\GZ$, amounting to $\pm$ 0.2 MeV, significantly
improved with respect to ~\cite{LEPEN93}. There are also smaller effects
on some of the other Z parameters, as detailed below.

\subsection{\bf LEP data on cross-sections and lepton 
asymmetries\label{sec-lsafb}}
 
 Between 1989 and 1995 a total of 15.5 million hadronic $\Zzero$ decays
and 1.7 million leptonic decays were recorded by the four LEP
experiments (ALEPH, DELPHI, L3 and OPAL). 
These events have distinctive topologies; for the {\it hadronic}
events there is a large visible energy and a large hadron multiplicity,
whereas for the {\it leptonic} events there are two, approximately 
back-to-back, high energy leptons.
Each experiment measures the total cross-section for
$\eehad$ and $\eell$ ($\ell$ = e,$\mu$,$\tau$), and the forward-backward
asymmetries for $\eell$, at each energy point. 

The data taking periods can be separated into two phases. In the first phase,
up to 1992, the energy determination was rather imprecise.
The second phase consisted of data from the 1993 and 1995 scans,
and from the $\simeq$ 60 pb$^{-1}$ on-peak data in 1994. Details of
the analysis methods of the four experiments can be found
in ~\cite{ALEPHLS,DELPHILS,L3LS,OPALLS}.

As discussed below, fits are made to the entire
dataset by each experiment, taking into account the correlations in systematic
errors arising for experimental effects such as detection efficiencies, the
LEP energy uncertainties and also theoretical uncertainties.

 To match these impressive statistics the systematic errors need to be well
understood. This is indeed the case. The experimental error on the
{\it luminosity}, which is determined from the $\eeee$ cross-section at small
angles, is determined to better than 0.1\% by each of the
four experiments. This requires knowledge of both the absolute and relative
positions of the detectors at the 10-20 $\mu$m level; an impressive 
achievement.

 The theoretical error on the luminosity has improved significantly since the
1994 ICHEP in Glasgow~\cite{ichep94}, when the error 
was $\delta\cal L/\cal L$(theory) = 0.25\%. 
More recent calculations, using BHLUMI 4.04
~\cite{Jadach,Ward}, include $\cal O$($\alpha^{2}$L$^{2}$) terms (where L
denotes the leading log term), as well as improved treatment of the $\gamma$-Z
interference contributions. The estimated theory error is
$\delta\cal L/\cal L$(theory) = 0.06\%. 
Note, however, that this error
is common between the LEP experiments\footnote{In fact for the OPAL experiment
only 0.054\% is common, as the effects of light fermion
pairs~\cite{Montagna} are also included, which reduces the uncertainty.} , 
and is comparable to that from
the combined experimental component, which is also about 0.06\%. 

\begin{figure}[htbp]
\vspace*{13pt}
\begin{center}
\mbox{
\epsfig{file=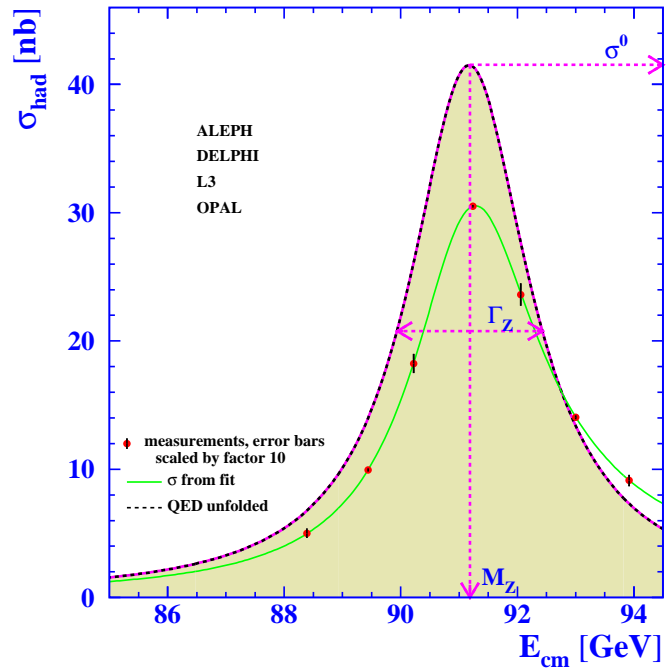,height=9cm}}
\end{center}
\caption
{Average LEP hadronic cross-sections, as a function of centre-of-mass energy.
The errors have been multiplied by a factor 10 for clarity.
The shaded area shows the cross-section, deconvoluted for the effects of QED,
which defines the Z parameters discussed in the text.}
\label{fig-had_xs}
\end{figure}

\begin{figure}[htbp]
\vspace*{13pt}
\begin{center}
\mbox{
\epsfig{file=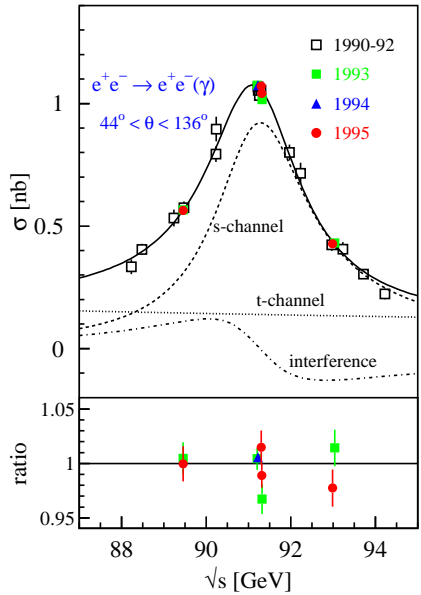,height=9cm}}
\end{center}
\caption
{Cross-section for Bhabha scattering, as a function of cms energy, from the 
L3 Collaboration, showing the s,t and s-t interference components.}
\label{fig-bhabha_xs}
\end{figure}

 The event selection efficiency for $\sigma$($\eehad$) is known to the
very accurate precision of $\leqsim$ 0.1\%. 
The averages of the measurements of the hadronic cross-sections, as a function
of centre-of-mass energy, are shown in fig.~\ref{fig-had_xs}. The 
importance of the effects of initial state QED radiative effects can
be seen, as the cross-section deconvoluted for these effects is also shown.  
The lepton cross-section efficiencies are somewhat 
less well  determined (0.1-0.7\%), and the systematic errors on these 
measurements are roughly comparable to the statistical errors.
The errors on $\Afbl$ are mainly statistical in nature. However, for the
reaction $\eeee$ there is a large t-channel contribution, and this is subtracted
in order to obtain the s-channel cross-sections and asymmetries; 
see, for example, 
fig.~\ref{fig-bhabha_xs}. 
The theoretical errors estimated for this subtraction~\cite{Beenakker}, which
are again common between the experiments, amount
to 0.025 nb on $\shad$, 0.024 on $\Ree$ and 0.0014 on $\Afbze$. The 
correlation between these, and
also the correlations induced because of the uncertainty in $\MZ$ in making
these subtractions, are all taken into account.

\begin{figure}[htbp]
\vspace*{13pt}
\begin{center}
\mbox{
\epsfig{file=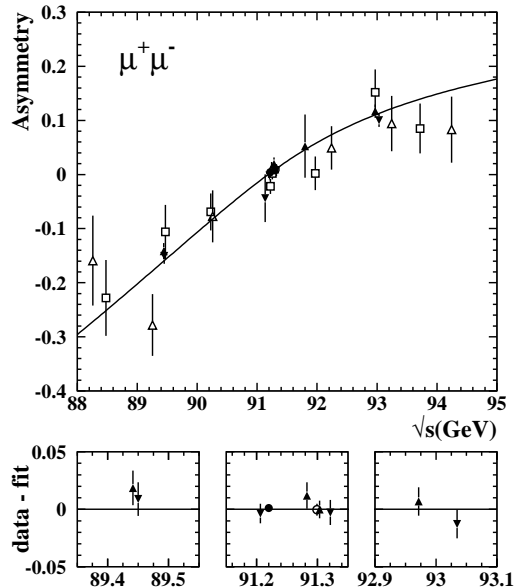,height=9cm}}
\end{center}
\caption
{Forward-backward asymmetry for $\eemumu$, as a function of cms energy, 
from the OPAL experiment.}
\label{fig-opal_mumu}
\end{figure}

 The forward-backward asymmetries for leptons have also been measured as
a function of $\roots$. An example of the results is shown 
in fig.~\ref{fig-opal_mumu}. 

\subsection{\bf Combining the LEP lineshape and asymmetries\label{sec-lep_comb}}

 The combination of the data from the four LEP experiments should, ideally,
take place at the level of the measured {\it observable} quantities: the
cross-sections and asymmetries. However, due to the very complicated nature
of the correlations, this has not been attempted. Instead,
for the purposes of combining the LEP data, each experiment provides the 
results of a (so-called) {\it model-independent} fit to their 
cross-section and asymmetry data in terms 
of nine variables. These are chosen to have small experimental correlations 
and are  $\MZ$, $\GZ$, $\shad$, $\Rl$ and $\Afbzl$ ($\ell$ = e,$\mu$,$\tau$). 
These so-called {\it pseudo-observables} or {\it POs}, as defined 
in ~\cite{PCPreport}, have been shown\footnote{The combination procedure
and justification are described in detail in ~\cite{ep_lsafb}.} 
to represent the results from each experiment, which consist of
around 200 cross-section and asymmetry measurements, to an excellent degree
of precision. 
 The results of a 9 parameter fit to the combined LEP data are given
in table~\ref{tab-nine}. 

The combination takes into account errors which are common between 
the experiments. These are shown in table~\ref{tab-common}. Those
arising from the uncertainty on the LEP energy determination and the
beam energy spread, the t-channel subtraction
and the theoretical uncertainty on the luminosity have already been
discussed. More details can be found in ~\cite{ep_lsafb}, where the
full correlation matrices are also given.

 The remaining error in table~\ref{tab-common}, the `theory' error, covers
the effects of QED uncertainties and differences in the precision
electroweak programs (TOPAZ0~\cite{TOPAZ0}, ZFITTER~\cite{ZFITTER} and
MIZA~\cite{MIZA}), used to extract the {\it  pseudo-observables}.
 
 Uncertainties in the QED corrections, both from the effects of uncalculated
higher-order terms and from initial state fermion pair radiation, have been
evaluated.  Initial state radiation corrections to ${\cal O}(\alpha^3)$
~\cite{Jadach91,Skrzypek92,Montagna97} (and see 
also ~\cite{Yennie61,Kuraev85,Berends88,Kniehl88,Jadach92,Jadach99,Jadach99a}
for earlier work) are included, and a
comparison of the predictions of TOPAZ0 and ZFITTER~\cite{TOPAZ0,ZFITTER}
has been made to evaluate the fermion pair uncertainties. 
The resulting uncertainties are estimated to be 0.3 and 0.2 MeV 
on $\MZ$ and $\GZ$ respectively, and 0.02\% on $\shad$.

  A comparison of the theoretical predictions for the cross-sections 
of the programs TOPAZ0 and ZFITTER, for a given set of SM input
parameters, has been performed~\cite{TOPAZ0,ZFITTER,HEP99_BARDIN}.  
These differences have been transformed into differences in the 
fitted {\it POs}. The uncertainties estimated in
this way are 0.1 MeV on both $\MZ$ and $\GZ$,
0.001nb on $\shad$, 0.004 on $\Rl$  and 0.0001 on $\Afbzl$. 

The largest uncertainty arising from the parameterisation used in
extracting the {\it POs} 
is from the $\gamma$-Z interference term for the $\eeqq$ channel, which is 
fixed to its SM value. Changing the Higgs mass from 100 and 1000 GeV,
gives a change of +0.23 MeV on $\MZ$. The uncertainty 
in $\alphamz$ (see section~\ref{sec-alfaem}) leads to a negligible 
change in $\MZ$. The effects on the other {\it POs} are
also negligible. 

 If {\it lepton universality} is imposed (evidence for this is discussed below),
then there are five variables. The results of the fit to the combined LEP
data are given in table~\ref{tab-five}. 
 It can be seen from tables~\ref{tab-nine} and ~\ref{tab-five} that 
the Z mass and width are
determined to $\delta\MZ$= 2.1 MeV and $\delta\GZ$= 2.3 MeV respectively.
These are impressive accuracies.

 As a cross-check of the LEP energy determination, the values of $\MZ$
for three separate periods of data taking, namely the early data up to 1992,
and the 1993 and 1995 energy scans, have been determined.
The values measured for these three periods are
$\MZ$ = 91.1904 $\pm$ 0.0065, $\MZ$ = 91.1882 $\pm$ 0.0033 
and $\MZ$ = 91.1866 $\pm$ 0.0024 GeV respectively,
giving confidence in the LEP energy determination. 

\begin{table}[htb]
\caption{Results and correlation matrix of the 9 parameter fit to the 
LEP data. The $\chit$/df of the average is 33/27, a probability 
of 21\%.}\label{tab-nine}
\footnotesize\rm

\begin{tabular}{lllrrrrrrrrr}
\br
quantity& value & error & $\MZ$ 
& $\GZ$ & $\shad$ & $\Ree$ & $\Rmu$ & $\Rtau$& $\Afbze$& $\Afbzm$& $\Afbzt$ \\
\mr
$\MZ$(GeV)  & 91.1876  & 0.0021 & 
  1.000 & -0.024 & -0.044 & 0.078 & 0.000 & 0.002 & -0.014 & 0.046 & 0.035 \\
$\GZ$(GeV)  &  2.4952  & 0.0023 & 
      &   1.00 &  -0.297 & -0.011 & 0.008 & 0.006 &  0.007 & 0.002 & 0.001 \\
$\shad$(nb) & 41.541   & 0.037  & 
      &        &  1.00 & 0.105 & 0.131 & 0.092 & 0.001 & 0.003 & 0.002 \\
$\Ree$       & 20.804   & 0.050  &
      &        &       &  1.00 &  0.069 & 0.046 & -0.371 & 0.020 & 0.013 \\
$\Rmu$       & 20.785   & 0.033  &
      &        &       &    & 1.00 & 0.069 & 0.001 & 0.012 & -0.003 \\
$\Rtau$        & 20.764   & 0.045  &
       &     &      &   &  & 1.00 & 0.003 & 0.001 & 0.009 \\
$\Afbze$     & 0.0145   & 0.0025 &  &  &   &  &  &  & 1.00 & -0.024 & -0.020 \\
$\Afbzm$     & 0.0169   & 0.0013 &  &  &  &  &  &  &  & 1.00& 0.046  \\
$\Afbzt$     & 0.0188   & 0.0017 &  &  &  &  &  &  &  &     & 1.00  \\
\br
\end{tabular}
\end{table}

\begin{table}[htb]
\caption{Total and common systematic error components for the 9 parameters.
}\label{tab-common}

\begin{indented}
\lineup
\item[]
\begin{tabular}{lrrrrr}
\br
quantity& total error & LEP energy & t-chann. & luminosity & theory \\
\mr
$\delta\MZ$(MeV)  &  2.1   &  1.7   & -      & -     & 0.3 \\
$\delta\GZ$(MeV)  &  2.3   &  1.2   & -      & -     & 0.2 \\
$\delta\shad$(nb) &  0.037 &  0.011 & -      & 0.025 & 0.008 \\
$\delta\Ree$      &  0.050 & 0.013  & 0.024  & -     & 0.004 \\
$\delta\Rmu$      & 0.033  &  -     &  -     & -     & 0.004 \\ 
$\delta\Rtau$     & 0.045  &  -     &  -     & -     & 0.004 \\ 
$\delta\Afbze$    & 0.0025 & 0.0004 & 0.0014 & -     & 0.0001 \\
$\delta\Afbzm$    & 0.0013 & 0.0003 &  -     & -     & 0.0001 \\
$\delta\Afbzt$    & 0.0017 & 0.0003 &  -     & -     & 0.0001 \\
\br
\end{tabular}
\end{indented}
\end{table}

\begin{table}[t]
\caption{ 
Results and correlation matrix of the 5 parameter fit to the
LEP data.
The $\chit$/df of the average is 37/31, a probability of 23\%.
}\label{tab-five}

\begin{indented}
\lineup
\item[]
\begin{tabular}{lllrrrrr}
\br
quantity& value & error & $\MZ$
& $\GZ$ & $\shad$ & $\Rl$ &$\Afbzl$  \\
\mr
$\MZ$(GeV)  & 91.1875  & 0.0021  &  1.00 & -0.023 & -0.045 & 0.033 &  0.055 \\
$\GZ$(GeV)  &  2.4952  & 0.0023  &       &   1.00 & -0.297 & 0.004 &  0.003 \\
$\shad$(nb) & 41.540   & 0.037   &       &        &  1.00 & 0.183  &  0.006 \\
$\Rl$       & 20.767   & 0.025   &       &        &       &  1.00 & -0.056 \\
$\Afbzl$    & 0.01714  & 0.00095 &       &        &       &       &  1.00  \\
 \br
\end{tabular}
\end{indented}
\end{table}

 Within the context of the Standard Model, the measurement of $\Rl$ can be
used to extract a value of the QCD coupling constant. The result
is $\alphasmz$ = 0.122 $\pm$ 0.004,
where the central value is for $\MH$ = 100 GeV. The value of $\alphasmz$
would increase by 0.003 if $\MH$ = 1000 GeV was used.

 Other quantities can be derived from these 9 or 5 parameter fits. Some of these
are given in table~\ref{tab-derived}.
The results for partial widths can also be transformed into $\Zzero$
branching ratios, giving $\Ztoqq$ = 69.911 $\pm$ 0.056 $\%$,
$\Ztoll$ = 10.0898 $\pm$ 0.0069 $\%$ and $\Ztonn$ = 20.000 $\pm$ 0.055 $\%$.
In addition the ratio $\Ginv$/$\Gl$= 5.942 $\pm$ 0.016 can be extracted.
When this is combined with the SM ratio $\Gnu$/$\Gl$ = 1.9912 $\pm$ 0.0012, 
this gives the number of light neutrinos: 
\begin{equation}
 {\rm N}_{\nu} = 2.9841 \pm 0.0083,
\end{equation}
which is 1.9 standard deviations from the SM value N$_{\nu}$ = 3.
The direct experimental verification that there are just three light
neutrinos is one of the most important results from LEP.

 The difference in the invisible widths between the measured and SM 
values ($\Ginv^{SM}$ = 501.7$^{+0.1}_{-0.9}$ MeV)
gives $\Delta\GZ^{\rm inv}$ = -2.7 $\pm$ 1.6 MeV, to be attributed
to possible non-standard contributions, {\it i.e.} not from $\Ztonn$.
This can be converted into a limit 
$\Delta\GZ^{\rm inv} < $ 2.0 MeV at the 95$\%$ c.l., where the limit is
calculated allowing for only positive values of $\Delta\GZ$. This can be
used to set limits on, for example, the pair production cross-sections
of `invisible' supersymmetric particles.

\begin{table}[t]
\caption{Quantities derived from the 9 and 5 parameter fits. The lepton 
partial width $\Gl$ is defined to be $\Ge$ for the case of lepton
universality. } 
\label{tab-derived}

\begin{indented}
\lineup
\item[] 
\begin{tabular}{cccc}
\br 
\mcha{Without Lepton Universality} & \mch{With Lepton Universality}  \\
\mr 
 $\Ge   (\MeV)$    &$  83.92 \pm 0.12$ & $\Gl  (\MeV)$ & $ 83.984 \pm 0.086$ \\
 $\Gmuon  (\MeV)$  &$  83.99 \pm 0.18$   & $\Ghad (\MeV)$ & $ 1744.4 \pm 2.0$ \\
 $\Gtau (\MeV)$&$  84.08 \pm 0.22$     & $\Ginv (\MeV)$ & $ 499.0 \pm 1.5 $ \\
\br 
\end{tabular}
\end{indented}
\end{table}

\subsection{\bf $\tau$ polarisation\label{sec-taupol}}

 The outgoing fermions in the $\ee$ annihilation are generally polarised.
However, this polarisation can only be measured in the case of 
the $\tau$-lepton, which
decays via $\tau^{-} \rightarrow \nu_{\tau}$W$^{*}$, with the virtual W$^{*}$
decaying to $\ell^{-}\bar\nu_{\ell}$ or $\qqb^{'}$, the latter 
leading to a variety of possible hadronic states. 
The $\tau$ polarisation ($\ptau$) is determined from studies of the decay
distributions of the $\tau$ leptons produced in $\Zzero$ decays. It is
defined as:  

\begin{equation}
\ptau = \frac{\sigma_R - \sigma_L}{\sigma_R + \sigma_L} \, ,
\end{equation}
where $\sigma_R$ and $\sigma_L$ are the $\tau$-pair cross-sections for the
production of a right-handed and left-handed $\tau^-$ respectively.
 
The angular distribution of $\ptau$, as a function of the
angle $\theta$ between the $e^-$ and the $\tau^-$, for $\roots = \MZ$,
is given by:
\begin{equation}
\label{eqn-taupol}
\ptau(\cos\theta)=
 - \frac{\cAt ( 1 +\cos{^2}\theta) + 2\cAe\cos\theta }
  { 1 +\cos{^2}\theta + 2\cAt\cAe\cos\theta },
\end{equation}
with $\cAe$ and $\cAt$ defined in equation~(\ref{eqn-z1}).
In equation~(\ref{eqn-taupol}) the small corrections for the effects of 
photon exchange, $\gamma-\Zzero$ interference
and electromagnetic radiative corrections for initial
and final state radiation are neglected. All of these effects 
are taken into account in the experimental analyses.

When averaged over all production angles $\cal {P}_{\tau}$ gives a
measurement of $\cAt$. Measurements of $\cal {P}_{\tau}(\cos\theta)$ provide
nearly independent determinations of both $\cAt$ and
$\cAe$, thus allowing a test of the universality of
the couplings of the $\Zzero$ to
$\rm ee$ and $\tau\tau$.

\begin{table}[htbp]
\caption[]{LEP results for $\cAt$ and $\cAe$ from the $\tau$ polarisation.
}
\label{tab-tau1}

\begin{indented}
\lineup
\item[] 
\begin{tabular}{lcc}
\br
expt.         &  $\cAt$                     &           $\cAe$  \\
\br
ALEPH         & $0.1451\pm0.0052\pm0.0029$  & $0.1504\pm0.0068\pm0.0008$  \\
DELPHI        & $0.1359\pm0.0079\pm0.0055$  & $0.1382\pm0.0116\pm0.0005$  \\
L3            & $0.1476\pm0.0088\pm0.0062$  & $0.1678\pm0.0127\pm0.0030$  \\
OPAL          & $0.1456\pm0.0076\pm0.0057$  & $0.1454\pm0.0108\pm0.0036$  \\
\br
LEP Average   & $0.1439\pm0.0035\pm0.0026$  & $0.1498\pm0.0048\pm0.0009$  \\
\br
\end{tabular}
\end{indented}
\end{table}
  
\begin{figure}[htbp]
\vspace*{13pt}
\begin{center}
\mbox{
\epsfig{file=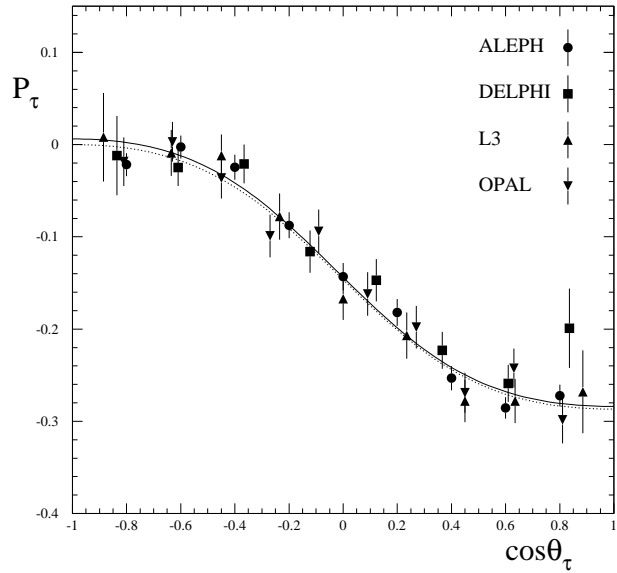,height=9cm}}
\end{center}
\caption
{Measurements of the $\tau$ polarisation as a function of polar angle.}
\label{fig-dptau_dcosth}
\end{figure}

Each of the LEP experiments has made separate $\ptau$ measurements using 
the five
$\tau$ decay modes e$\nu \overline{\nu}$, $\mu\nu \overline{\nu}$,
$\pi\nu$, $\rho\nu$ and
$a_{1}\nu$.
The $\rho\nu$ and $\pi\nu$ are the most sensitive channels, contributing
weights of about $40\%$ each in the average.
In addition, DELPHI and L3 have used an inclusive hadronic analysis.
The LEP combination is made on the results from each experiment already
averaged over the $\tau$ decay modes. 
The data are shown in fig.~\ref{fig-dptau_dcosth}.

Table~\ref{tab-tau1} shows the results
for $\cAt$ and $\cAe$ obtained by the four
experiments
and their combination.
  The LEP results are combined taking into account the (small) common
systematics from ISR, branching ratio uncertainties, hadronic
modelling, the theoretical uncertainties (using ZFITTER) and the correlations
in the extraction of $\cAe$ and $\cAt$ (typically 3$\%$). The combined
results are
\begin{equation}
\cAe = 0.1498 \pm 0.0049 \hspace*{2.0cm} \cAt = 0.1439 \pm 0.0043 \ \ ,
\end{equation}
with a $\chit$/df = 3.9/6. The correlation coefficient is 0.012.
The systematic components of these errors are
0.0009 and 0.0026, the one for $\cAe$ being much smaller as it is an
asymmetry measurement where many systematic effects cancel.
These values are compatible and, assuming lepton universality, can be 
combined to give 
\begin{equation}\label{eqn-ptau}
\cAl = 0.1465 \pm 0.0033 \ \ ,
\end{equation}
with a $\chit$/df = 4.7/7 and with a systematic error component of 0.0015.

\subsection{\bf Measurement of $\ALR$\label{sec-alr} at the SLC}

 The parameter $\cAe$ can be extracted directly if the incident 
electron beam is longitudinally polarised, by measuring the cross-sections 
for left-handed and right-handed incident beams. 
 The high values of longitudinal polarisation (P$_{e}\simeq$ 70-80 \%)
achieved at the SLC have allowed the SLD experiment to make an extremely
precise measurement of
\begin{equation}
\ALR = \frac{\sigmal - \sigmar} { \sigmal + \sigmar} = \cAe \ \ ,
\end{equation}
 where $\sigmal$($\sigmar$) is the total cross-section 
for a left-(right-)handed polarised incident electron beam. 
After the introduction of `strained lattice'  GaAs photocathodes in 1994
the average polarisation was between 73\% and 77\%. The polarisation was
measured by detecting beam electrons scattered by photons from a circularly
polarised laser, using a precision Compton polarimeter. Two further,
less precise, polarimeters have been used for verification. The estimated
error on the electron polarisation is about 0.5\%; much smaller than the
statistical error on the $\ALR$ measurement of 1.3\%. In the SLD detector
no final state selection is required, except that $\ee$ final states and 
those from non-resonant backgrounds are removed. The measurements 
essentially involve determining the ratios of the numbers of events detected 
with different polarisation settings, and are thus very insensitive to detailed 
knowledge of the detector acceptances and efficiencies. 

Combining all the data from 1992-8 (550K hadronic $\Zzero$ events)
gives~\cite{SLDALR}
\begin{equation}
\cAe = 0.1514 \pm 0.0022. 
\end{equation}

 Additional information can be obtained by measuring the 
{\it left-right-forward-backward asymmetry} for a specific fermion f:
\begin{equation}\label{eqn-alrfb}
\Alrfff = \frac{(\sigmalf - \sigmalb) - (\sigmarf - \sigmarb)} 
{ \sigmal + \sigmar} = \frac{3}{4}\cAf \ \ ,
\end{equation}
where $\sigmalf$($\sigmalb$) and $\sigmarf$($\sigmarb$) are the
forward(backward) cross-sections for fermion f for 
left- and right-handed polarised beams respectively. 
These measurements for leptons give~\cite{SLDosaka} 
$\cAe$ = 0.1554 $\pm$ 0.0060, $\cAm$ = 0.142 $\pm$ 0.015
and $\cAt$ = 0.136 $\pm$ 0.015.
This $\cAe$ value, when combined with that from $\ALR$, gives 
$\cAe$ = 0.1516 $\pm$ 0.0021, and has correlations of 0.038 and
0.033 with $\cAm$ and $\cAt$ respectively. The correlation between
$\cAm$ and $\cAt$ is 0.007. Assuming lepton universality, all these
results can be combined to give
\begin{equation}\label{eqn-alr}
\cAe = 0.1513 \pm 0.0021, \hspace*{1.0cm} \swsqeffl = 0.23098 \pm 0.00026.
\end{equation}
The SLD measurement of $\cAe$ is the single most precise 
determination, and the error is mostly statistics dominated.
The SLD $\cAe$ result is compatible with the less precise value  
from $\tau$-polarisation at the 0.3$\sigma$ level.
Assuming lepton universality, the SLD result for $\cAl$ is compatible
at the 1.3$\sigma$ level with the
value from $\tau$-polarisation. It is also compatible with the
result from $\Afbpol$ of $\cAl$ = 0.1512 $\pm$ 0.0042 to better
than 0.1$\sigma$.

\subsection{\bf Lepton universality\label{sec-leptuniv}}

\begin{figure}[t]
\vspace*{-1.2cm}
\vspace*{13pt}
\begin{center}
\mbox{
\epsfig{file=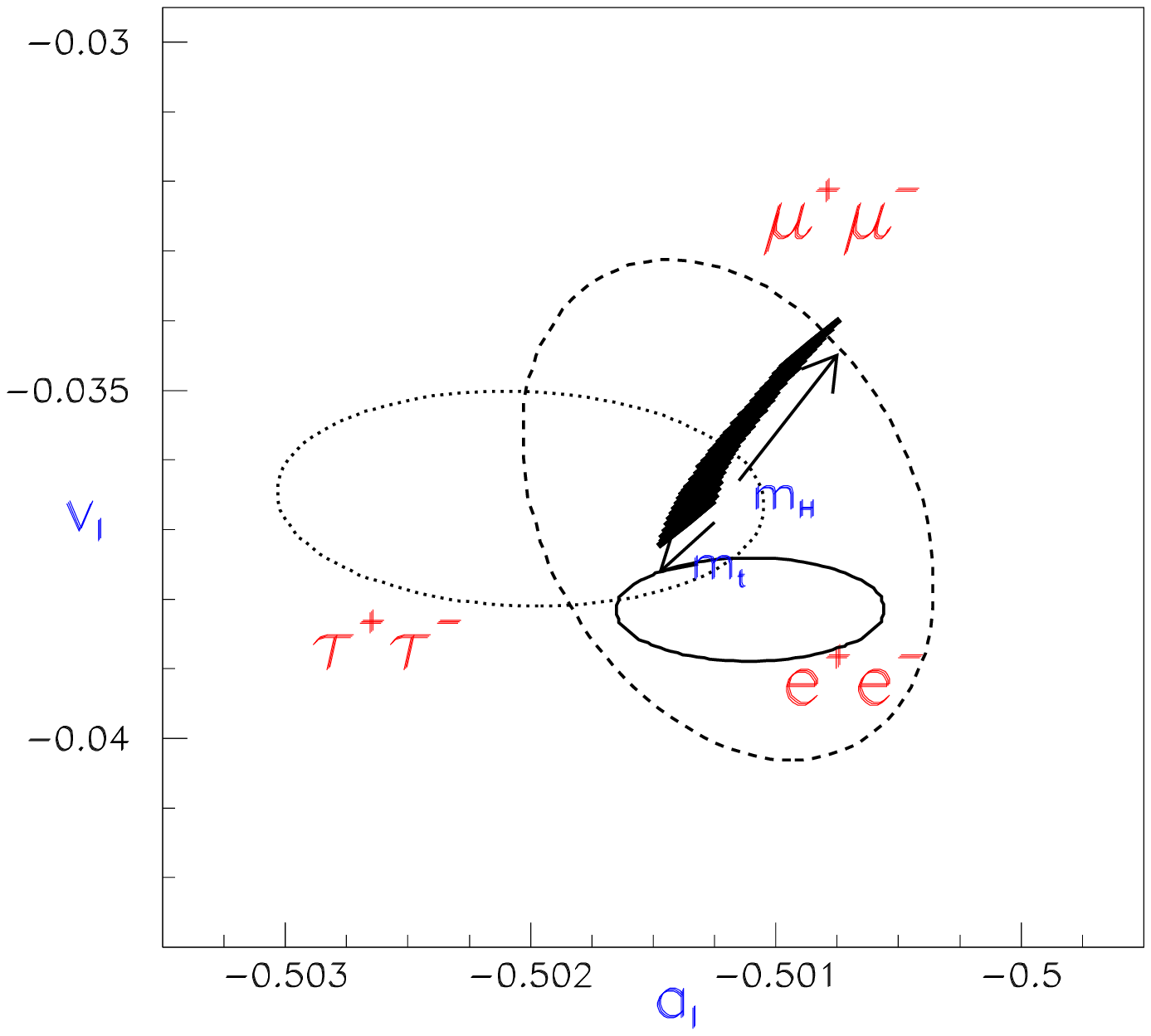,height=11cm}}
\end{center}
\caption {
 Contours of 70\% probability in the $\Vlll$-$\Alll$ plane from
  LEP and SLD measurements. The solid region
  corresponds to the Standard Model prediction for $\tI$ GeV
  and  $\HI$ GeV. The arrows point in
  the direction of increasing values of $\Mt$ and $\MH$.
}
\label{fig-leptuniv}
\end{figure}

  The data from the leptonic partial decay widths, forward-backward asymmetries,
$\tau$-polarisation ($\cAt$ and $\cAe$) and the SLD 
measurements ($\cAe$,$\cAm$ and $\cAt$) have been used to fit to $\Vlll$ 
and $\Alll$ ($\ell$=e,$\mu,\tau$), and thus
to test {\it lepton universality}.
The results are shown in table~\ref{tab-coupl} and fig.~\ref{fig-leptuniv}.
The correlations between the fitted values are rather small, the largest
being 0.38 between $\Amu$ and $\Atau$ and -0.29 between $\Vmu$ and $\Amu$;
the others are~$\leqsim$~0.15. The results for $\Vmu$ and $\Amu$ are the
least precise because only measurements of the forward-backward asymmetry
contribute. For $\Vtau$ and $\Atau$ the $\tau$-polarisation also contributes
and for $\Ve$ and $\Ae$ there are also contributions from the initial state
particles to the forward-backward asymmetries (see eqn.~\ref{eqn-z2}).

  The magnitudes of any differences in the couplings can be quantified by
fitting in terms of $\Vlll = \Ve + \Delta\Vlll$,  
$\Alll = \Ae + \Delta\Alll$ ($\ell=\mu,\tau$),
giving $\Delta\Vmu$ = 0.0014 $\pm$ 0.0024, $\Delta\Vtau$ = 0.0016 $\pm$ 0.0011,
$\Delta\Amu$ = -0.00009 $\pm$ 0.00067 
and $\Delta\Atau$ = -0.00093 $\pm$ 0.00076. Thus $\Delta\Vtau$ and 
$\Delta\Atau$ are 1.5 and 1.2 standard deviations respectively away
from zero.

\begin{table}[htbp]
\caption[]{
Values of the lepton vector and axial-vector couplings from LEP data
alone and with the addition of the SLD measurement of $\ALR$, without
and with the assumption of lepton universality.}
\label{tab-coupl}
\begin{indented}
\lineup
\item[] 
\begin{tabular}{ccc}
\br
&\multicolumn{2}{c}{Without Lepton Universality:} \\
 & LEP & LEP+SLD\\
\mr
$\Ve$    & $-0.0378   \pm 0.0011  $ & $-0.03816  \pm 0.00047 $ \\
$\Vmu$   & $-0.0376   \pm 0.0031  $ & $-0.0367  \pm 0.0023 $ \\
$\Vtau$  & $-0.0368   \pm 0.0011  $ & $-0.0366  \pm 0.0010 $ \\
$\Ae$    & $-0.50112  \pm 0.00035 $ & $-0.50111 \pm 0.00035$ \\
$\Amu$   & $-0.50115  \pm 0.00056 $ & $-0.50120 \pm 0.00054$ \\
$\Atau$  & $-0.50204  \pm 0.00064 $& $-0.50204  \pm 0.00064$ \\
\br
&\multicolumn{2}{c}{With Lepton Universality:   } \\
 & LEP & LEP+SLD\\
\mr
$\Vlll$    & $-0.03736  \pm 0.00066 $& $-0.03783  \pm 0.00041 $\\
$\Alll$    & $-0.50126 \pm 0.00026  $& $-0.50123 \pm 0.00026$\\
$\Vnu=\Anu$     & $+0.50068  \pm 0.00075 $& $+0.50068  \pm 0.00075 $\\
\br
\end{tabular}
\end{indented}
\end{table}

Thus the data are reasonably consistent with the universality hypothesis. 
The signs in fig.~\ref{fig-leptuniv} are plotted 
taking $\Ae <$ 0. Using this convention (this is justified from $\nu$-electron
scattering results~\cite{Winter95}), the signs of all couplings are uniquely
determined from LEP data alone. Note that the values of the lepton 
forward-backward
asymmetries away from the Z-pole vary as -(s-$\MZ^{2}$)$\Alll$. This term
also leads to a change in the sign around the Z-pole; see for example
fig.~\ref{fig-opal_mumu}.

\begin{figure}[t]
\vspace*{13pt}
\begin{center}
\mbox{
\epsfig{file=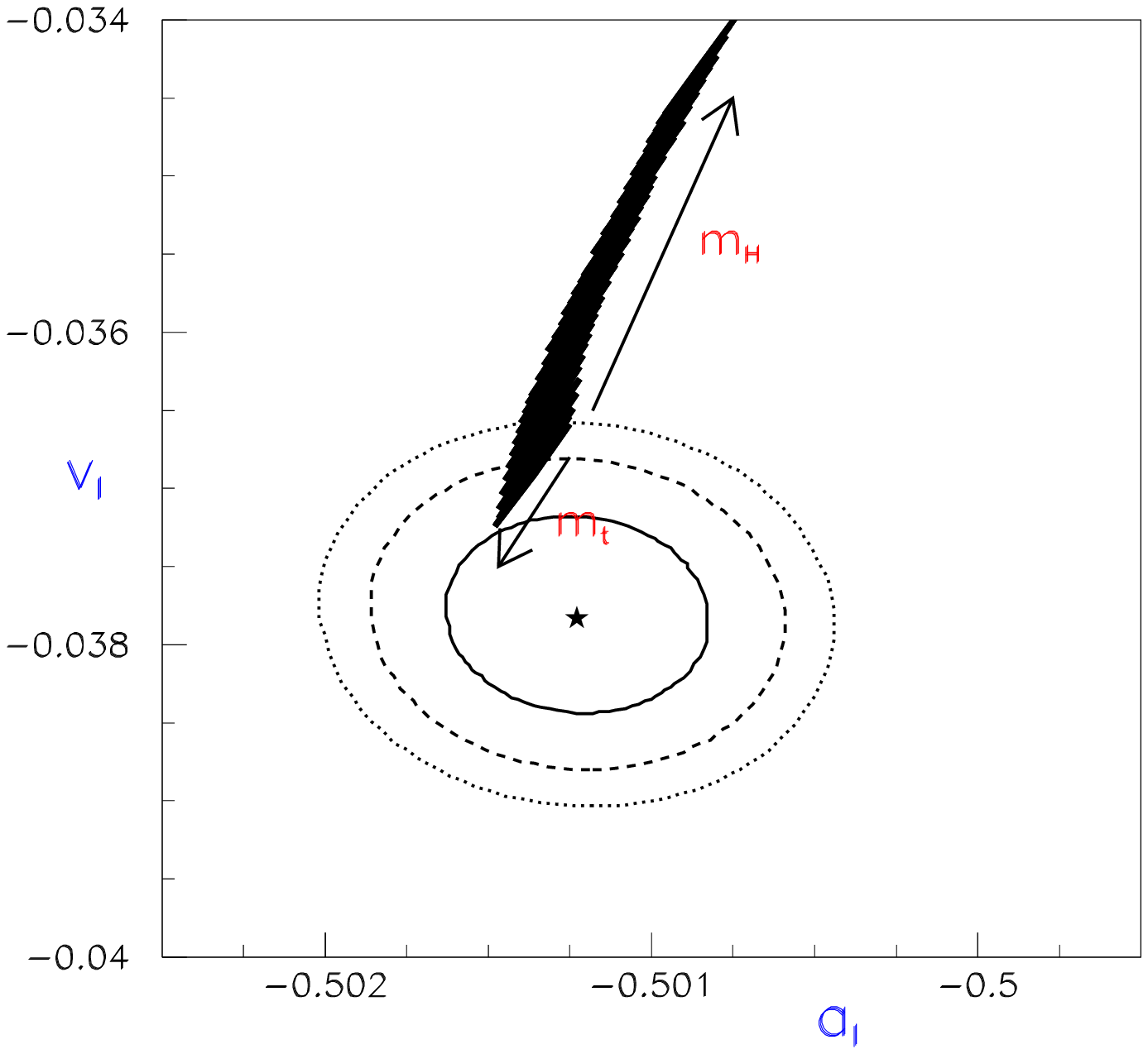,height=11cm}}
\end{center}
\caption {
 Contours of 70\%, 95\% and 99\%  probability in the $\Vlll$-$\Alll$ plane from
  LEP and SLD measurements. The solid region
  corresponds to the Standard Model prediction for $\tI$ GeV
  and  $\HI$ GeV. The arrows point in
  the direction of increasing values of $\Mt$ and $\MH$.
}
\label{fig-leptuniv1}
\end{figure}

 The results of a fit in which lepton universality is imposed are 
given in table~\ref{tab-coupl}  and fig.~\ref{fig-leptuniv1}. 
The value of the neutrino coupling comes essentially from $\Ginv$.
The value of $\Alll$ is different to the Born-level 
value (t$_{f}^{3}$ = -1/2; see eqn.\ref{eqn-z3}) by 4.7 standard deviations; 
indicating
sizeable electroweak corrections. 
It can be seen that the results are consistent with SM expectations,
provided the Higgs boson is relatively light.

\subsection{\bf Heavy Flavour results\label{sec-hflav}}

 It is of intrinsic interest to extract the $\Zzero$ couplings to individual
quark flavours, in contrast to the results described in 
section ~\ref{sec-lsafb}, which are
summed over 5 flavours. The quantities measured are the $\Zzero$ partial width 
ratios~\footnote{The symbols $\Rbbz$ and $\Rccz$ denote
specifically the ratios of $\Zzero$ partial widths.}
\begin{equation}
 \Rqqz = \frac{\Gq}{\Ghad} \hspace*{1.0cm} ( {\rm q= b,c}),
\end{equation}
the Z-pole forward-backward asymmetries for b and c 
quarks~\footnote{Some measurements on s-quarks have been made by the SLD,
DELPHI and OPAL Collaborations, either tagging s-quarks or 
assuming b and s quarks have the same couplings. 
These measurements are much less precise than those
for b and c quarks, but are however all compatible with SM predictions.},
$\Afbzb$ and $\Afbzc$, and the
direct measurements of $\Ab$ and $\Ac$ by SLD,
obtained by measuring $\Alrfff$ (see eqn.~\ref{eqn-alrfb}), for b and c quarks, 
with a polarised beam. 
Note that the propagator effects for the t-quark and Higgs,
as well as QCD effects, largely cancel in the ratio $\Rqqz$. However, 
for b-quarks, there are
significant SM vertex corrections from tWb couplings. These are essentially
independent of $\MH$ and lead to a decrease of $\Gb$ with increasing $\Mt$,
rather than an increase as for the other quark partial-widths. Furthermore,
$\Rbbz$ is sensitive to physics beyond the SM (e.g. from 
light $\stq$,$\chargino$ SUSY particles).

 Extracting relatively pure samples of events corresponding to individual quark
flavours is far from easy. Measurements exist for both c and b quarks, which
can be separated from light (u,d,s) quarks, and from each other, using their
characteristic properties (see table~\ref{tab-hquark}).

\begin{table}[htbp]
\caption[]{
Some properties of B hadrons and D mesons. 
}\label{tab-hquark}

\begin{indented}
\lineup
\item[] 
\begin{tabular}{lrrr}\br
{quantity} & {B}& D$^{+}$  &  D$^{0}$ \\
\br
lifetime (ps) & 1.6 & 1.0 & 0.4 \\
$<$ x$_{E}$ = E$_{had}$/E$_{beam}>$  & 0.7 & 0.5 & 0.5 \\
decay charged multiplicity & 5.5 & 2.2 & 2.2 \\
\br
\end{tabular}
\end{indented}
\end{table}

The  main selection criteria ({\it tags}) are as follows:-

\begin{itemize}

\item {\bf c-quarks:} D,D$^{*}$ mesons plus lifetime and lepton tags. 
The harder momentum fraction in direct c decay, compared to b$\rightarrow$c,
is also used. For $\Afbc$, the D/D$^{*}$ charges and the lepton charges, in
semileptonic decays, are used to distinguish c from $\bar{c}$. 

\item {\bf b-quarks:} lifetime, mass and lepton tags. The mass tag exploits
the fact that the b-quark decay products have relatively large invariant
masses. For $\Afbb$, the lepton charge is used, evaluating the contributions 
from b$\rightarrow\ell$ and b$\rightarrow$c$\rightarrow\ell$,
b$\rightarrow\bar{c}\rightarrow\ell$. Also used is the jet-charge for 
a specific hemisphere with respect to the thrust-axis,
Q$_{hemi}$= $\sum \mid $p$^{i}_{\mid\mid} \mid ^{\kappa}$   Q$_{i}$/
                  $\sum \mid $p$^{i}_{\mid\mid} \mid ^{\kappa}$, 
where p$^{i}_{\mid\mid}$ is the momentum component  of a hadron,
with charge Q$_{i}$, parallel to the thrust axis.
The power $\kappa$ is optimised for sensitivity.
The charge difference between the forward and backward
hemispheres, Q$_{F}$-Q$_{B}$, is related to the required asymmetry.
The sum, Q$_{F}$+Q$_{B}$, is sensitive to any bias and to
the charge resolution.      
For $\Rbb$, and to a lesser extent for $\Rcc$, the most accurate
results are from double-tag methods, as discussed below.

\end{itemize}

 The main background in the tagged b(c) quark sample is from c(b)quarks.
This means that the value of $\Rbbz$ is correlated to that of  $\Rccz$. It
is usual practice to give $\Rbbz$ at the SM value $\Rccz$ = 0.172.

 The main systematic uncertainties arise from:-
\begin{itemize}
\item[i)] the fraction of D$^{*}$, D$^{+}$, D$_{s}$, $\Lambda_{c}$ etc 
in $\ccbar$ events (particularly important for $\Rbb$) 
\item[ii)] b and c hadron lifetimes
\item[iii)] charm decay modes
\item[iv)] fraction of gluon-splitting g $\rightarrow \ccbar , \bbbar$ in 
hadronic Z events; the values used are the measured 
fractions g$_\ccbar$ = (2.96 $\pm$ 0.38)\% and g$_\bbbar$ = (0.25 $\pm$ 0.05)\%
\item[v)] semi-leptonic branching ratios and decay models 
\item[vi)] light quark fragmentation models
\item[vii)] correlations between hemispheres for double-tags.
\end{itemize}

\subsection{\bf Measurement of $\Rbb$\label{sec-rbb}}   

 The most accurate measurements of $\Rbb$ all employ a double-tag method. This 
involves determining the jet axis of the event ({\it thrust-axis}) and then
employing lifetime, mass, leptonic or other b-tags to each hemisphere to 
determine the number of hemispheres N$_t$, with a tag, and the number of 
events N$_{tt}$, with two tags. For a sample 
of N$_{had}$ hadronic $\Zzero$ decays one has
\newline
\begin{equation}
\label{eqn-rb}
   \frac{N_t}{2N_{\rm{had}}} = \eb \Rb 
                        + \effc  \Rc +
                        \effuds ( 1 - \Rb - \Rc ) ,\\ 
\end{equation}
\begin{equation}
\label{eqn-rb1}
   \frac{N_{tt}}{N_{\rm{had}}} = { \Cb} { ( \eb)^2 \Rb  }
                +   ( \effc)^2 \Rc +
                          (\effuds)^2 ( 1 - \Rb - \Rc ) ,
\end{equation}
where $\eb$, $\effc$ and $\effuds$ are the tagging efficiencies per 
hemisphere for b, c and light-quark events, and $\Cb \ne 1$
accounts for the fact that the tagging efficiencies between the 
hemispheres may be correlated. In practice,  $\eb\gg\effc\gg\effuds$, 
$\Cb \approx 1$, and the correlations for the other flavours are 
neglected. These equations can be solved to give \Rb\ and 
$\eb$ which, neglecting the c and uds backgrounds and the 
correlations, are approximately given by:
\begin{equation}
\label{eqn-rb2}
\eb \approx 2 N_{tt} / N_t  , \\
\end{equation}
\begin{equation}
\label{eqn-rb3}
\Rb \approx N_t^2 / (4N_{tt}N_{\rm{had}}).
\end{equation}
The double-tagging method has the advantage that the tagging efficiency 
is determined directly from the data, reducing the systematic error of the
measurement. The residual background of other flavours in the sample,
and the evaluation of the correlation between the tagging efficiencies
in the two hemispheres of the event, are the main sources of systematic
uncertainty in such an analysis. The use of powerful vertex detectors at
LEP has led to excellent b-tagging efficiencies. For example, DELPHI
achieves 30\%, with a 1.5\% background. Due, at least in part, 
to the closer proximity to
the interaction point an even better performance (50\% with a 2\% background)
is achieved by SLD.
The single/double-tag method has been extended by ALEPH and DELPHI to
multi-tags. This not only improves the statistical accuracy, but also 
reduces the systematic uncertainty due to hemisphere correlations 
and charm contamination.

\begin{figure}[t]
\vspace*{13pt}
\begin{center}
\mbox{
\epsfig{file=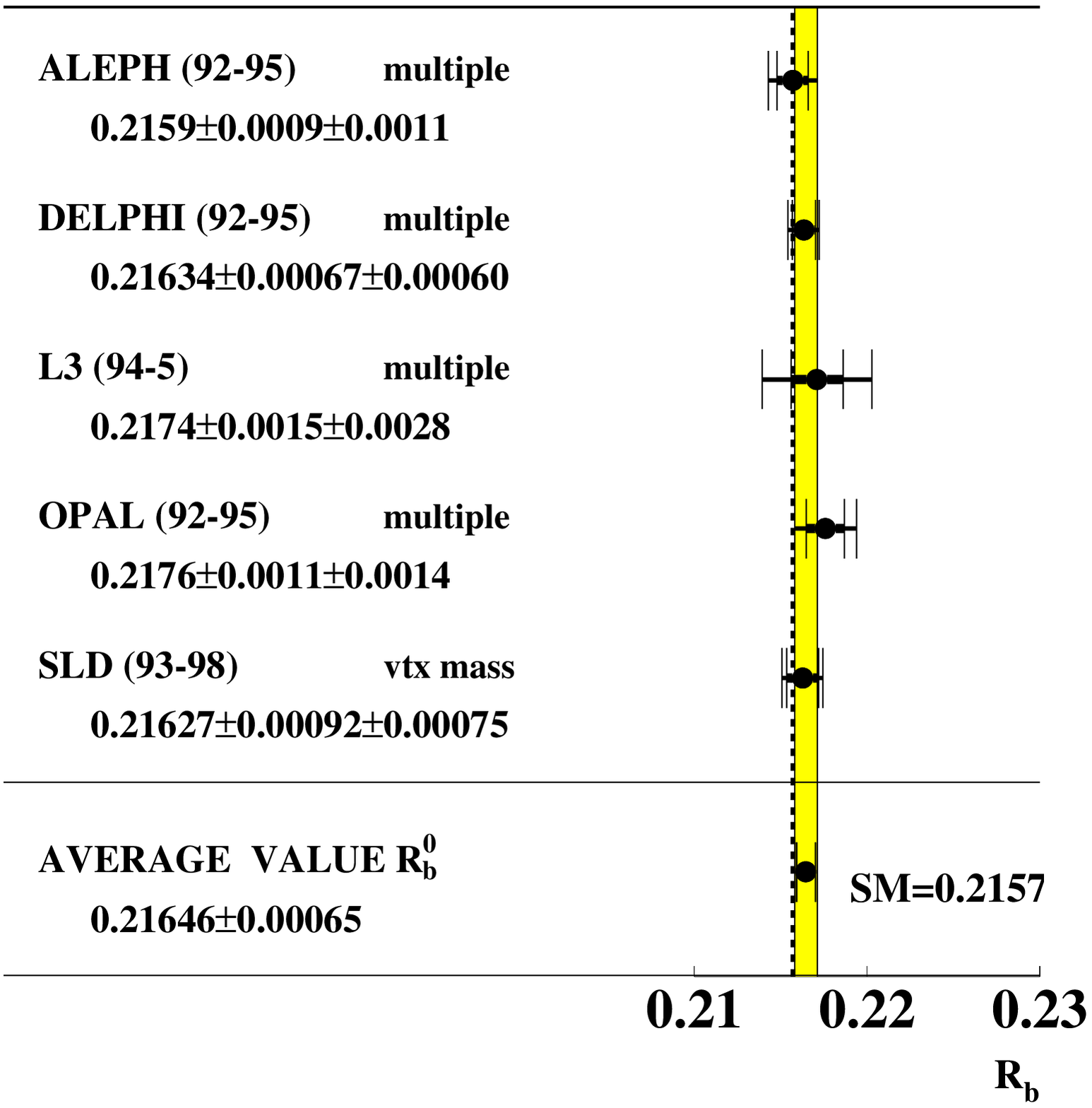,height=9cm}} 
\end{center}
\caption{
Results on $\Rbb$, for $\Rcc$ = 0.172, together with the 
average value for $\Rbbz$, 
from a fit to all the LEP and SLC heavy flavour data. 
The SM prediction for $\Mt$ = 174.3 GeV and $\MH$ = 150 GeV is 
shown as a dashed line.
}
\label{fig-rbb}
\end{figure}

 The results for $\Rb$ are shown in fig.~\ref{fig-rbb}.
 The combined LEP/SLD value of $\Rbbz =$ 0.21646 $\pm$ 0.00065 
thus has a relative precision of about 0.3\%. The
average value, when interpreted in terms of the SM, gives a value 
of $\Mt$ = 155 $^{+19 +1}_{-21 -0}$ GeV, where the central value
is for $\MH$ = 150 GeV and the second error corresponds to the range $\HI$,
and the constraint $\alphasmz$ = 0.118 $\pm$ 0.002 is used.
This is consistent with the direct determination~\cite{topmass}.
The combined statistical error of all the $\Rb$ measurements is 0.00043 and that
from the internal experimental systematics (track resolution, detection 
efficiencies of leptons etc) is  0.00029. The error due to common 
systematics is about 0.00039. The largest common systematic errors are
from uncertainties on gluon splitting into b and c quark pairs (0.00022),
QCD effects in hemisphere correlations (0.00018) and the branching 
ratio D $\rightarrow$ neutrals (0.00014).
In total, more than 20 possible sources of systematic error 
to $\Rbb$ are considered.
 
 At the time of the Summer Conferences in 1995, the average value 
was $\Rbbz$ = 0.2205 $\pm$ 0.0016 (for $\Rccz$ = 0.172)~\cite{pbr_lp95},
more than three standard deviations above the SM value.
The measured value was in the direction expected from light SUSY particles~
($\stq$,$\chargino$). However, SUSY particles in the mass range suggested
by the excess in  $\Rbbz$ have since been excluded by searches at LEP 2. 
The subsequent change in the $\Rbbz$ average  
is due to a combination of much improved statistics, purer tagging methods
and changes of some of the heavy flavour input parameters needed in the 
analysis.

\subsection{\bf Measurement of $\Rcc$\label{sec-rcc}}

 Tagging charm quarks with high efficiency and purity is unfortunately 
difficult. The cleanest tag is to use the decay sequence 
c$\rightarrow\Dstarp\rightarrow\Dzero\pi^{+}(\Dzero\rightarrow$K$^{-}\pi^{+}$),
 but this tags only about 0.5\% of c-decays, 
so is statistically limited. Other $\Dzero$ modes, which are somewhat less
clean, can also be used, as can other ground-state charmed hadrons 
and c-quark leptonic decays. The purities achieved are 65-90\%, but
with an overall charm tag efficiency of only a few percent.

 There are also analyses which use a `slow' $\pi$ tag (the $\pi$ in the
D$^{*}$ decay has a small p$_{T}$) in a double-tag. 
However, this tag is rather loose because there is a considerable background 
at low p$_{T}$ from fragmentation processes. 

 Several methods have been used in the determination of $\Rc$. These are:
\begin{itemize}
\item[i)] {\it Single charm-counting rate} (ALEPH, DELPHI and OPAL). 
This requires measuring the  production
 rates of the ground-state charmed hadrons ($\Dzero, \Dplus, \Ds$, as
 well as charmed baryons). Small corrections are applied for unobserved
baryonic states.
The total rate gives $\Rc$ x Prob(c$\ra$ hadrons), so if all
the ground-state charmed hadrons are detected the measurement gives $\Rc$. 
\item[ii)] {\it Inclusive/exclusive double-tag} (ALEPH, DELPHI and OPAL).
This first requires 
measurements of the production rate of $\Dstarpm$ mesons in several 
decay channels. This depends on the product 
$\Rc$ x P$_{c\ra \Dstarp}$ x BR($\Dstarp\ra\Dzero\pi^{+}$), and this
sample of $\ccbar$ (and $\bbbar$) events is used to measure
P$_{c\ra \Dstarp}$ x BR($\Dstarp\ra\Dzero\pi^{+}$), using a slow pion tag
in the opposite hemisphere. 
\item[iii)] {\it Exclusive double-tag} (ALEPH). Here, exclusively 
reconstructed $\Dstarp$, $\Dplus$ and $\Dzero$ mesons are used, giving
good purity but larger statistical errors.
\item[iv)] {\it Lifetime plus mass double-tag} (SLD). This uses the same
tagging algorithm used for $\Rb$, and achieves a purity of about 84\%.
\item[v)] {\it Single leptons} (ALEPH). This assumes a value of 
BR(c$\ra$l).

\end{itemize}

\begin{figure}[t]
\vspace*{13pt}
\begin{center}
\mbox{
\epsfig{file=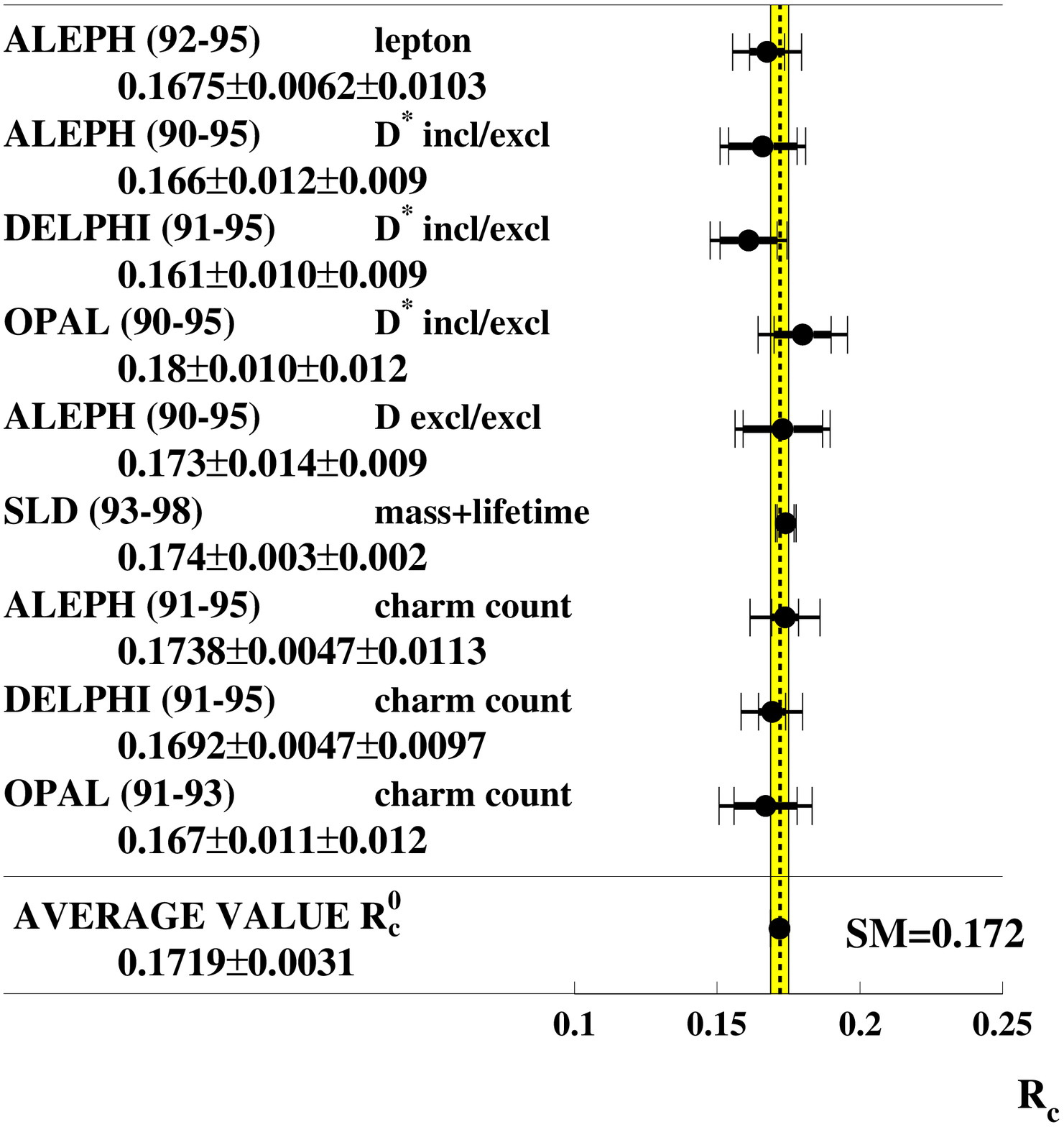,height=9cm}}
\end{center}
\caption{
Results on $\Rcc$, together with the average value for $\Rccz$, 
from a fit to all the LEP and SLC heavy flavour data.
The SM prediction 
for $\Mt$ = 174.3 GeV and $\MH$ = 150 GeV is shown as a dashed line.}
\label{fig-rc}
\end{figure}

 The LEP average value for $\Rccz$, made in Summer 1995, 
was $\Rccz$ = 0.1540 $\pm$ 0.0074. This was some 2.4 standard deviations below
the SM value of 0.172. The present value (see fig.~\ref{fig-rc}) is
$\Rccz$ = 0.1719 $\pm$ 0.0031,
and is rather close to the SM value. For the 1995 average, roughly half of 
the error 
weight came from common systematic errors between the measurements, which
relied in particular on the measurement
of Y$_{c}$ = P(c$\ra\Dstarp$) x BR($\Dstarp\ra\pi^{+}$) made at 
low ($\roots \sim$ 10 GeV) energy. The LEP data now determine this quantity 
directly, so that the present average does not depend on the use of 
low energy data. In addition techniques have been refined and more robust
analyses performed.

 The relative precision of the average value of $\Rccz$ 
is 1.8\%. The statistical component of the error is 0.0023, and that from
internal experimental systematics 0.0014. The total common systematic
error is 0.0014, with the largest components (0.0005) coming from 
both  BR($\Ds\ra\phi\pi$) and BR($\Lambda_{c}\ra$pK$^{-}\pi^{+}$).

\begin{figure}[t]
\vspace*{-1.9cm}
\vspace*{13pt}
\begin{center}
\mbox{
\epsfig{file=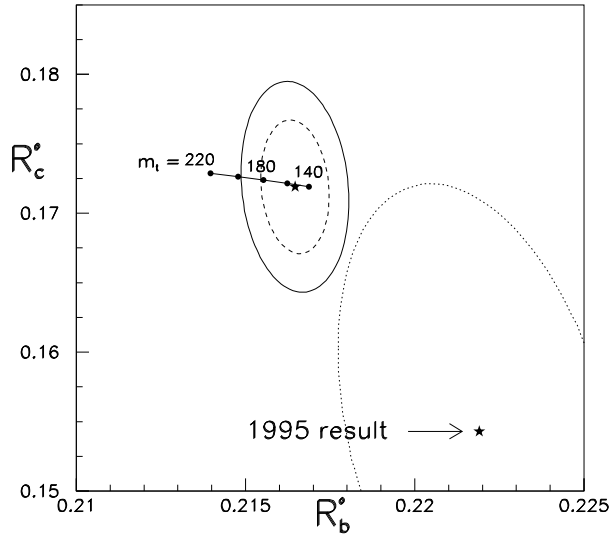,height=9cm}}
\end{center}
\caption{The 70\% and 95\% confidence level contours in 
the $\Rbbz$, $\Rccz$ plane. 
The SM prediction for various values of $\Mt$ is shown, as is
the central experimental value in 1995, together with 
its 95\% confidence level contour.
}
\label{fig-rbrc}
\end{figure}

 As discussed above, the determinations of $\Rbbz$ and $\Rccz$ are correlated,
with a correlation coefficient of -0.14.
Fig.~\ref{fig-rbrc} shows 
the 70\% and 95\% confidence level contours in 
the $\Rbbz$, $\Rccz$ plane, as well as the SM prediction for various values 
of $\Mt$.

\subsection{\bf Heavy-quark asymmetries\label{sec-hqasym}}

\begin{figure}
\vspace*{13pt}
\begin{center}
\mbox{
\epsfig{file=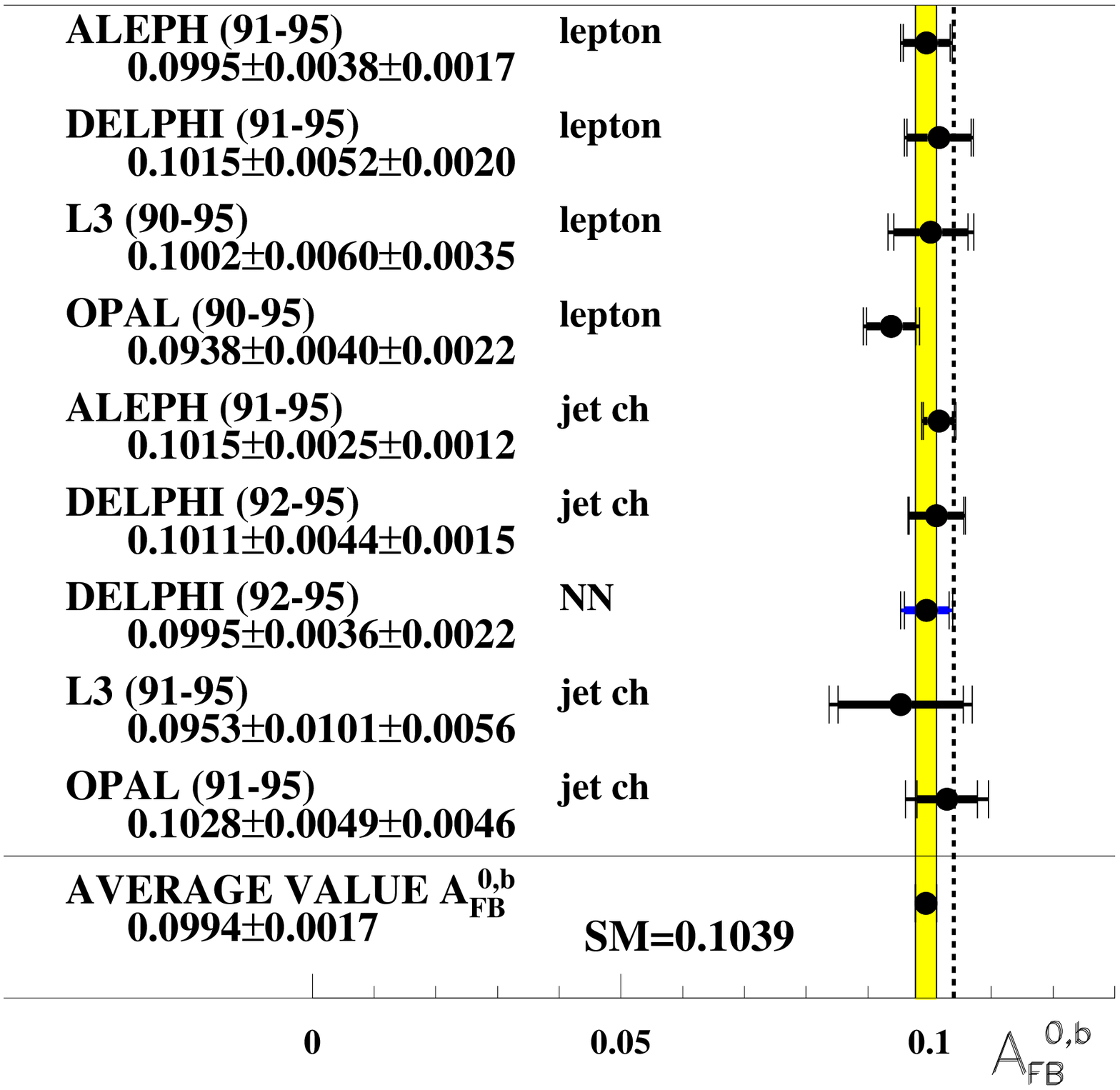,height=9cm}}
\end{center}
\caption
{
Results on $\Afbb$, together with the
average value for $\Afbzb$ from a fit to all the LEP 
heavy flavour data. 
The SM prediction 
for $\Mt$ = 174.3 GeV, $\MH$ = 150 GeV is shown as a dashed line.}
\label{fig-afbb}
\end{figure}

\begin{figure}
\vspace*{13pt}
\begin{center}
\mbox{
\epsfig{file=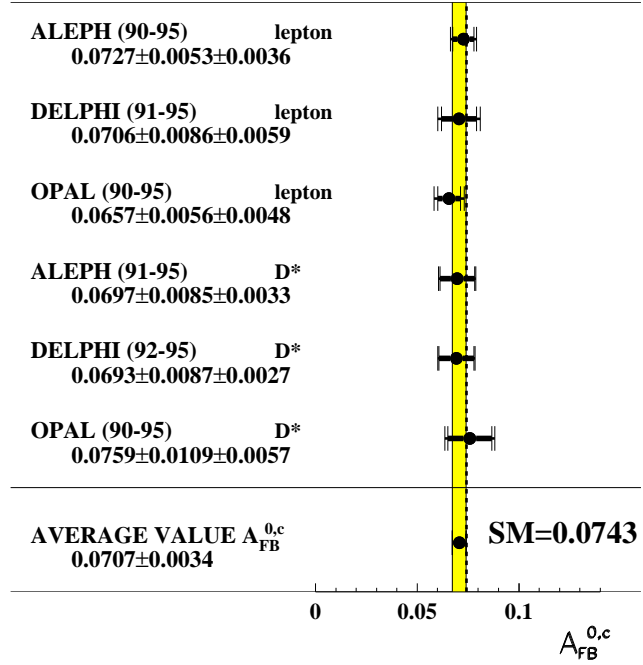,height=9cm}}
\end{center}
\caption
{ 
Results on $\Afbc$, together with the average value for $\Afbzc$
from a fit to all the LEP heavy flavour data.
The lepton results shown are still preliminary. An earlier lepton
result from L3 is included in the average but not plotted.
The SM prediction 
for $\Mt$ = 174.3 GeV, $\MH$ = 150 GeV is shown as a dashed line.}
\label{fig-afbc}
\end{figure}

 The results for $\Afbb$ and $\Afbc$ are 
given in figs.~\ref{fig-afbb} and ~\ref{fig-afbc} respectively. They are
corrected to the full experimental acceptance. The quoted values 
are corrected for QCD effects and to correspond to $\roots =$ 91.26 GeV; both
peak and off-peak data are used. 
The QCD corrections are calculated
to second-order~\cite{Catani}, and amount to 0.0063 for both b 
and c quarks~\cite{EWWG99}.
In order to obtain the pole asymmetries $\Afbzc$ and $\Afbzb$ from 
the experimentally measured results, corrections
are applied, using ZFITTER, to get to $\roots = \MZ$, for QED effects 
and for the
contributions of $\gamma$ exchange and $\gamma-\Zzero$ interference, as well
as for the b-quark mass.
These amount, in total, to additive corrections of 0.0062 for $\Afbc$ and
0.0025 for $\Afbb$. The results for  $\Afbb$ have also been corrected
for the effects of $\bbbar$ mixing. 
The methods used for the asymmetries are as follows:
\begin{itemize}
\item[i)] {\it Lepton spectra }(ALEPH, DELPHI, L3 and OPAL). The 
characteristic high transverse momentum spectrum from the heavy quarks is
exploited (sometimes in conjunction with other information) to measure
both $\Afbb$ and $\Afbc$.
\item[ii)] {\it Lifetime tag plus hemisphere charge} (ALEPH, 
DELPHI, L3 and OPAL). For $\Afbb$, and these give roughly equal precision
to the lepton results.
\item[iii)] {\it D mesons} (ALEPH for $\Afbc$, and DELPHI and OPAL for 
$\Afbb$ and $\Afbc$).
\end{itemize}
{\it Neural Network methods} have also been used for the most recent 
measurements of $\Afbb$ (ALEPH and DELPHI), incorporating much 
of the information from the above methods. 
A {\it single} and {\it double-tag} procedure is
used, as for $\Rbb$, so the method is essentially self-calibrating,
except for the effects of backgrounds and hemisphere correlations,
which are taken from simulation.
 
For both the $\Afbb$ and $\Afbc$ measurements, the systematic errors
in all the methods are smaller than the statistical errors.
For $\Afbb$ the statistical, internal
systematic and common systematic components of the errors are 0.0016, 
0.0006 and 0.0004 respectively. For $\Afbc$ the statistical, internal
systematic and common systematic components of the errors are 0.0030, 
0.0014 and 0.0009 respectively. So both these measurements are
statistics limited.

\begin{figure}[t]
\vspace*{-1.4cm}
\vspace*{13pt}
\begin{center}
\mbox{
\epsfig{file=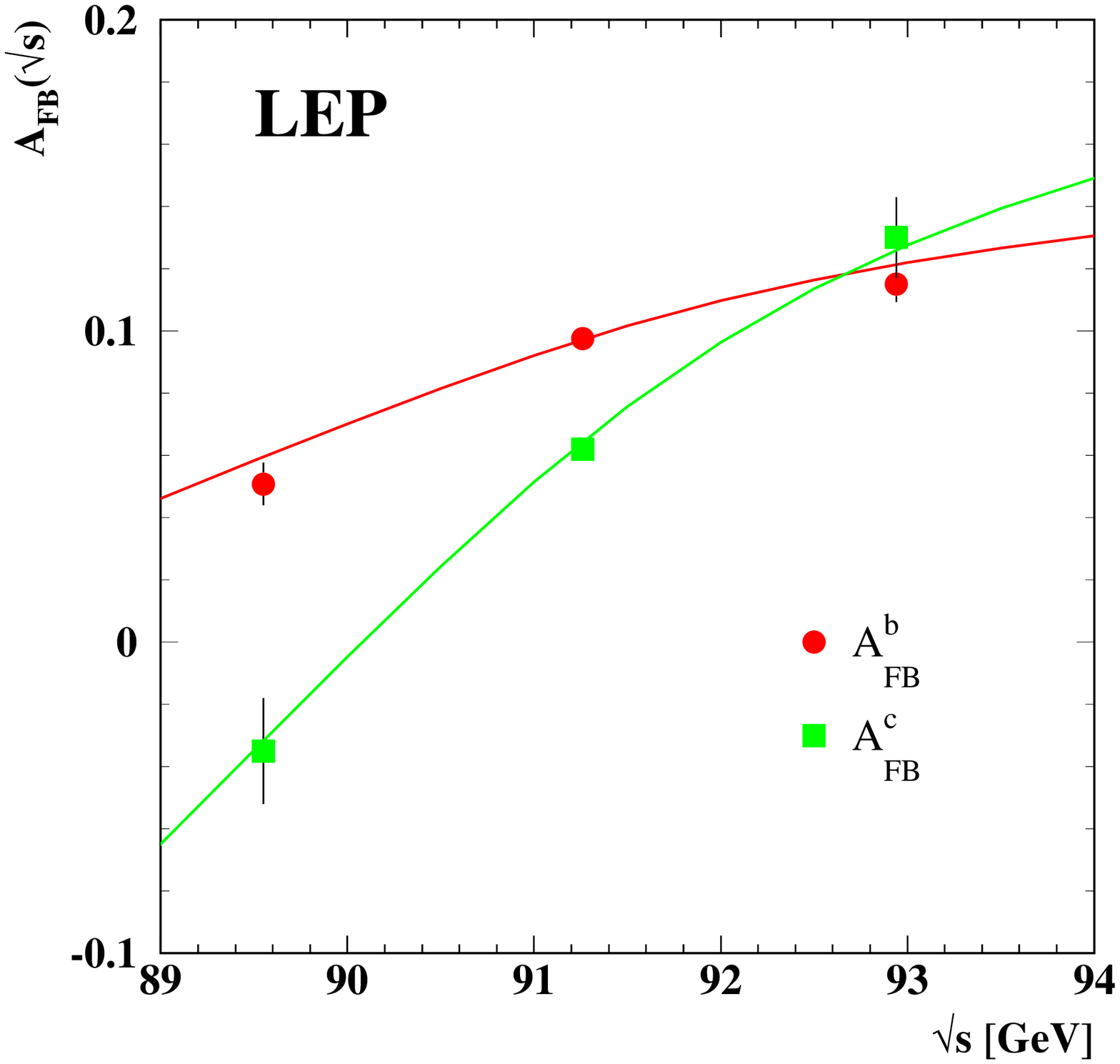,height=9cm}}
\end{center}
\caption{Energy dependence of the heavy-quark forward-backward asymmetries.}
\label{fig-afbene}
\end{figure}

 The asymmetries $\Afbzb$ and $\Afbzc$
are rather weakly correlated, and both the pole asymmetries, and their
energy dependence (see fig.~\ref{fig-afbene}), are 
compatible with the SM.

 Measurements of the heavy-quark forward-backward asymmetries, using a
longitudinally polarised beam, by the SLD Collaboration give directly
values of $\Ab$ and $\Ac$. Using lepton, kaon, D-meson and jet-charge 
plus lifetime/vertex mass tags, the values $\Ab$ = 0.922 $\pm$ 0.020 
and  $\Ac$ = 0.670 $\pm$ 0.026 are obtained~\cite{sld_tampere,EWWG01}.

\subsection{\bf Combining the heavy flavour results\label{sec-hflav_res}}

 The combination has been carried out by a LEP/SLD working 
group~\cite{EWWG01}, and details of the procedure used for the LEP experiments
can be found in ~\cite{lephf}.
Each experiment provides, for each measurement, a complete breakdown of 
the systematic errors, adjusted if necessary to agreed meanings of 
these errors. 
Direct measurements of $\Ab$ and $\Ac$ by SLD,
obtained by measuring $\Afbb$ and $\Afbc$ with a polarised beam, are also
included.
A multi-parameter fit is then performed to get the best overall
values of $\Rbbz$,$\Rccz$,$\Afbzb$, $\Afbzc$, $\Ab$ and $\Ac$, 
plus their covariance matrix. The results of a fit to both the LEP and SLD 
data are given in table~\ref{tab-hfres}. 
The effective
mixing parameter $\chibar$, and the leptonic branching 
ratios b$\rightarrow\ell$, b$\rightarrow$c$\rightarrow\bar\ell$  
and c$\rightarrow\ell$, are 
also included in the fit. It should be noted that the $\chit$/df is very
small, leading to a probability close to 100$\%$. This is, of course, rather
unlikely. However, it does indicate that the errors on the combined 
heavy flavour results are probably not underestimated.

\begin{table}[htbp]
\caption[]{Results of fits to the LEP and SLD heavy flavour data, plus
the correlation matrix. 
The $\chit$/df of the average is 47/(105-14), a probability
of greater than 99$\%$.}\label{tab-hfres}
\begin{indented}
\lineup
\item[] 
\begin{tabular}{lllrrrrrr}\br
quantity& value & error & $\Rbbz$ 
& $\Rccz$ & $\Afbzb$ & $\Afbzc$ &$\cAb$ & $\cAc$ \\
\br
$\Rbbz$   &0.21646& 0.00065& 1.000&--0.14&--0.08&  0.05&--0.07&  0.04\\
$\Rccz$   &0.1719& 0.0031  & & 1.000&  0.04&--0.03&  0.03&--0.05\\
$\Afbzb$  &0.0994& 0.0017  &  &  & 1.000&  0.16&  0.02& 0.00\\
$\Afbzc$  &0.0707& 0.0034  &  & &  & 1.000&--0.01&  0.02\\
$\cAb$    &0.922 & 0.020   & &  &  & & 1.000&  0.13\\
$\cAc$    &0.670 & 0.026   &  & & &  &  & 1.000\\
 \br
\end{tabular}
\end{indented}
\end{table}

\subsection{\bf Inclusive Hadron Charge Asymmetry $\avQfb$\label{sec-qfb}}

 Asymmetry measurements can also be made if the individual quark flavours
are not separated. The method involves the measurement, in $\Zzero$ hadronic 
events, of the hadronic charge asymmetry, based on
the mean difference in jet charges measured in the forward and backward
event hemispheres, $\avQfb$. The measured values of $\avQfb$ cannot be 
compared directly as some of them include detector dependent effects, 
such as acceptances and efficiencies. The results are best compared
using the values of $\swsqeffl$ extracted in each analysis, as given
in table~\ref{partab}. It can be seen that the systematic errors are
larger than the statistical errors. These are dominated
by fragmentation and decay modelling uncertainties.

\begin{table}[htbp]
\caption[]{Summary of the determinations of $\swsqeffl$ from inclusive 
hadronic charge asymmetries at LEP. For each experiment, the first 
error is statistical and the second systematic. 
}
\label{partab}

\begin{indented}
\lineup
\item[] 
\begin{tabular}{llc}
\br
Experiment & & $\swsqeffl$ \\
\br
ALEPH & 90-94 & $0.2322\pm0.0008\pm0.0011$ \\
DELPHI& 91    & $0.2345\pm0.0030\pm0.0027$ \\
L3& 91-95     & $0.2327\pm0.0012\pm0.0013$ \\
OPAL  & 90-91 & $0.2326\pm0.0012\pm0.0029$ \\
\br
Average  &           & $0.2324\pm0.0012$ \\
\br
\end{tabular}
\end{indented}
\end{table}

\subsection{\bf The coupling parameters $\cAf$\label{sec-coup_params}}

\begin{figure}[t]
\vspace*{13pt}
\begin{center}
\mbox{
     \includegraphics[width=0.48\linewidth]{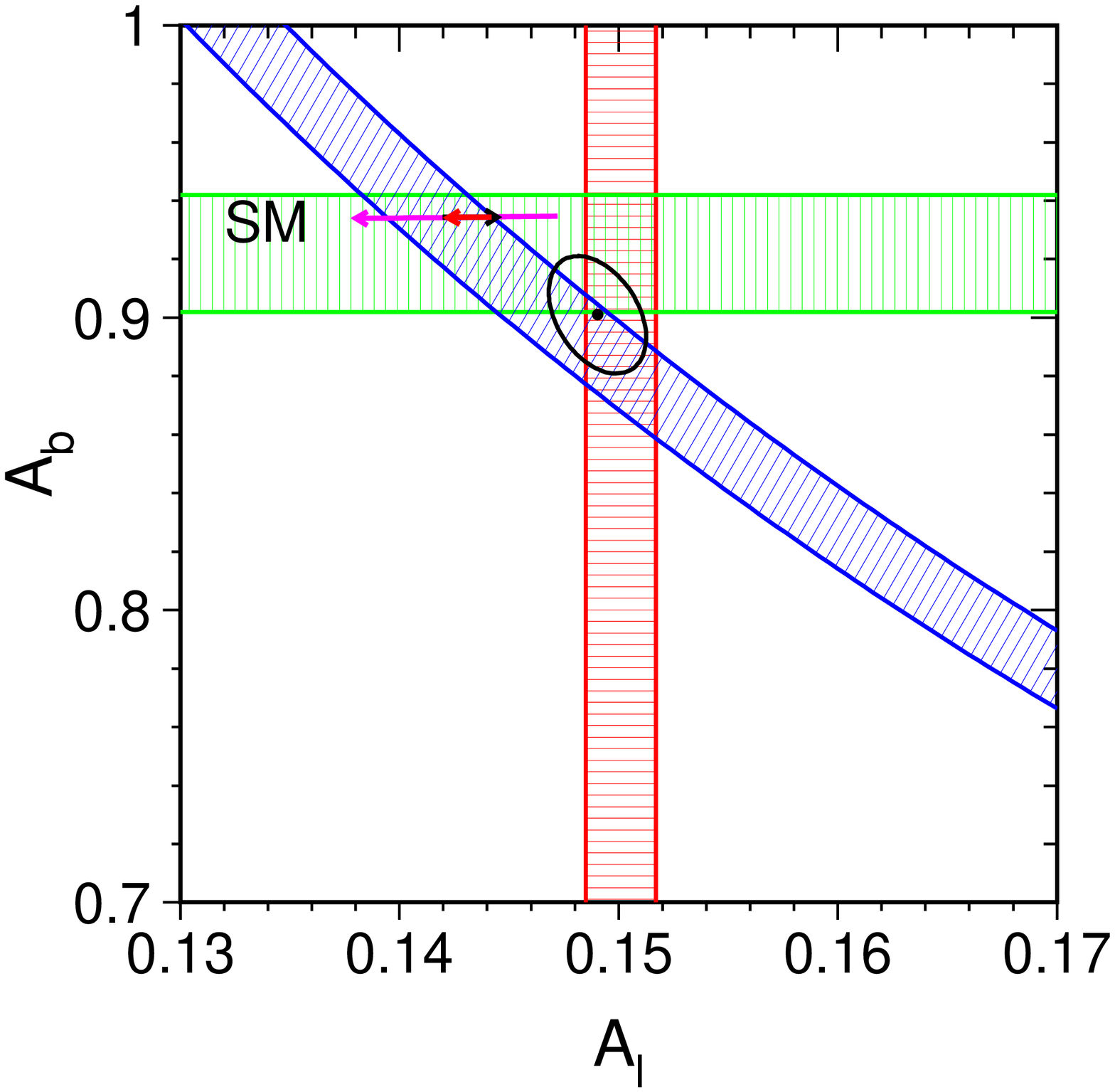}
     \includegraphics[width=0.48\linewidth]{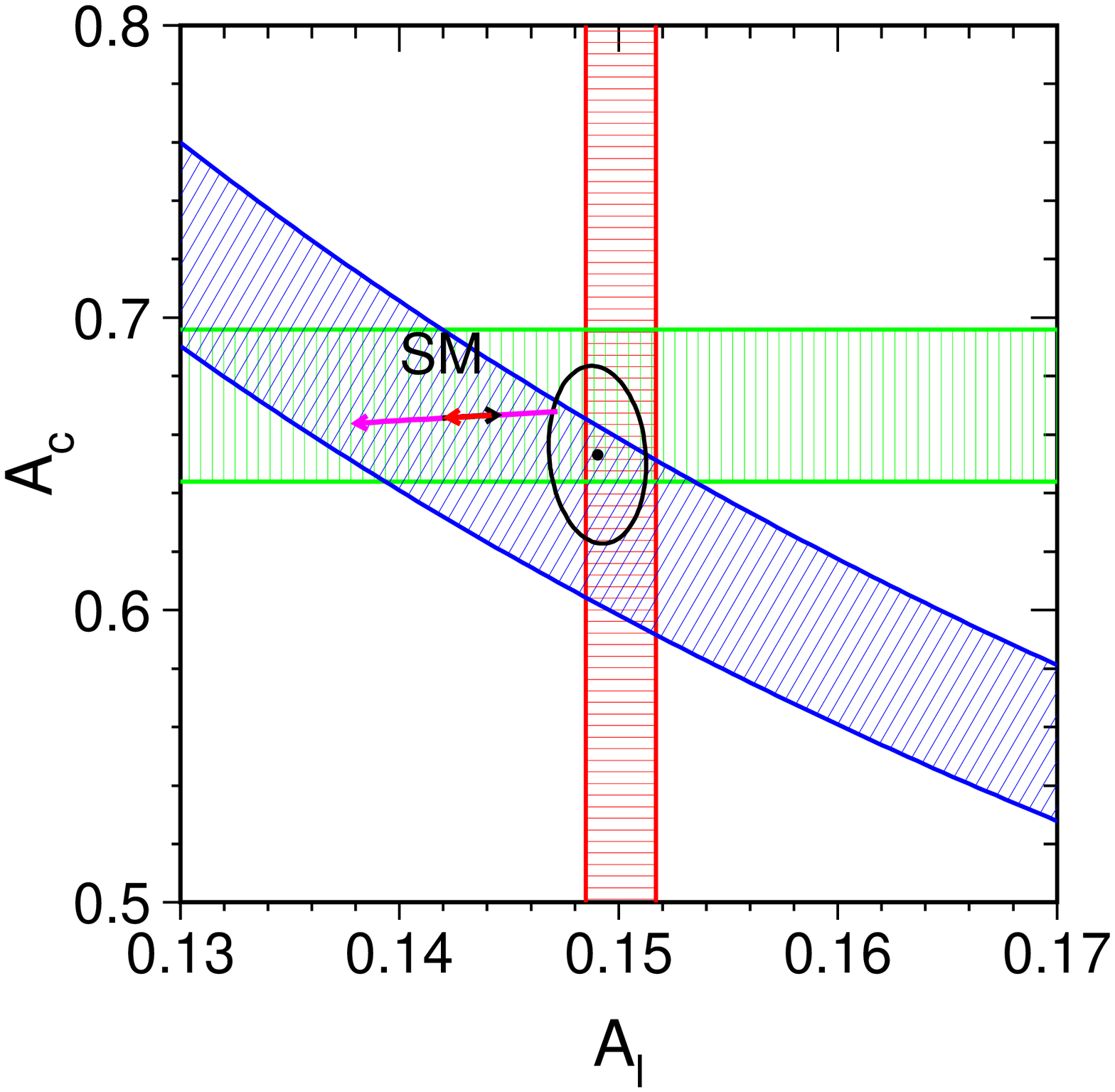}}
\end{center}
\vspace*{-0.4cm}
\caption{ Measurements of $\cAl$ from SLD+LEP (vertical band), 
$\cAb$($\cAc$) from SLD (horizontal band) and $\Afbzb$($\Afbzc$) 
from LEP (diagonal band). 
The 70$\%$ confidence level contour for the combined fit to these variables
is also shown.
The SM predictions are shown as arrows, with the left-pointing arrow showing
the variation from $\MH$ = 300$^{+700}_{-186}$ GeV and the right-pointing 
arrow showing the variation from $\Mt$ = 174.3 $\pm$ 5.1 GeV. 
There is an additional contribution from the error 
on $\dalfahad$ ($\pm$ 0.00036), which is in the same sense as the Higgs arrow
but with a size similar to that from the top-quark uncertainty. 
}
\label{fig-ab_al}
\end{figure}

 The coupling parameters $\cAf$ are obtained directly in the case of
the SLD polarisation measurements or from the $\tau$ lepton polarisation.
The forward-backward asymmetries for different fermions at LEP, 
using eqn.($\ref{eqn-z2}$), determine the product of $\cAe$ and $\cAf$.
The results for $\cAe$, determined
assuming lepton universality where appropriate, are given in table~\ref{tab-AL}.
The results for $\cAb$ and $\cAc$, both those measured directly and 
those derived
from forward-backward asymmetry measurements and assuming a value of $\cAe$,
are given in table~\ref{tab-AF-Q}. The results are displayed graphically in
fig~\ref{fig-ab_al}.

\begin{table}[htbp]
\caption[]{
  Determinations of the leptonic coupling parameter $\cAl$, assuming
  lepton universality. The cumulative averages for $\cAl$, and 
the $\chi^2$ per degree of freedom for these, are also given.  }
\label{tab-AL}
\begin{indented}
\lineup
\item[] 
\begin{tabular}{cccr}

\br
                  & $\cAl$            & Cumulative Average & $\chi^2$/df\\

\mr
$\Afbzl$           & $0.1512\pm0.0042$ &                    &               \\
$\ptau(\cos\theta)$& $0.1465\pm0.0033$ & $0.1482\pm 0.0026$ &  0.8/1        \\
\mr
$\cAl$ (SLD)& $0.1513\pm0.0021$ & $0.1501\pm 0.0016$ &  1.6/2          \\
\br

\end{tabular}
\end{indented}
\end{table}

\begin{table}[htbp]
\caption[]{
  Determination of the quark coupling parameters $\cAb$ and $\cAc$
  from LEP data alone (using the LEP average for $\cAl$), from SLD
  data alone, and from LEP+SLD data (using the LEP+SLD average for
  $\cAl$), assuming lepton universality.  }
\label{tab-AF-Q}
\begin{indented}
\lineup
\item[] 
\begin{tabular}{cccr}

\br
         &    LEP                  & SLD  & LEP+SLD  \\
         & ($\cAl=0.1482\pm0.0026$)&      & ($\cAl=0.1501\pm0.0016$) \\
\mr
$\cAb$   & $0.893\pm0.022$    & $0.922 \pm 0.020$  & $0.901\pm0.013$   \\
$\cAc$   & $0.634\pm0.033$    & $0.670 \pm 0.026$  & $0.653\pm0.020$  \\
\br
\end{tabular}
\end{indented}
\end{table}

\begin{figure}[t]
\vspace*{13pt}
\begin{center}
\mbox{
\epsfig{file=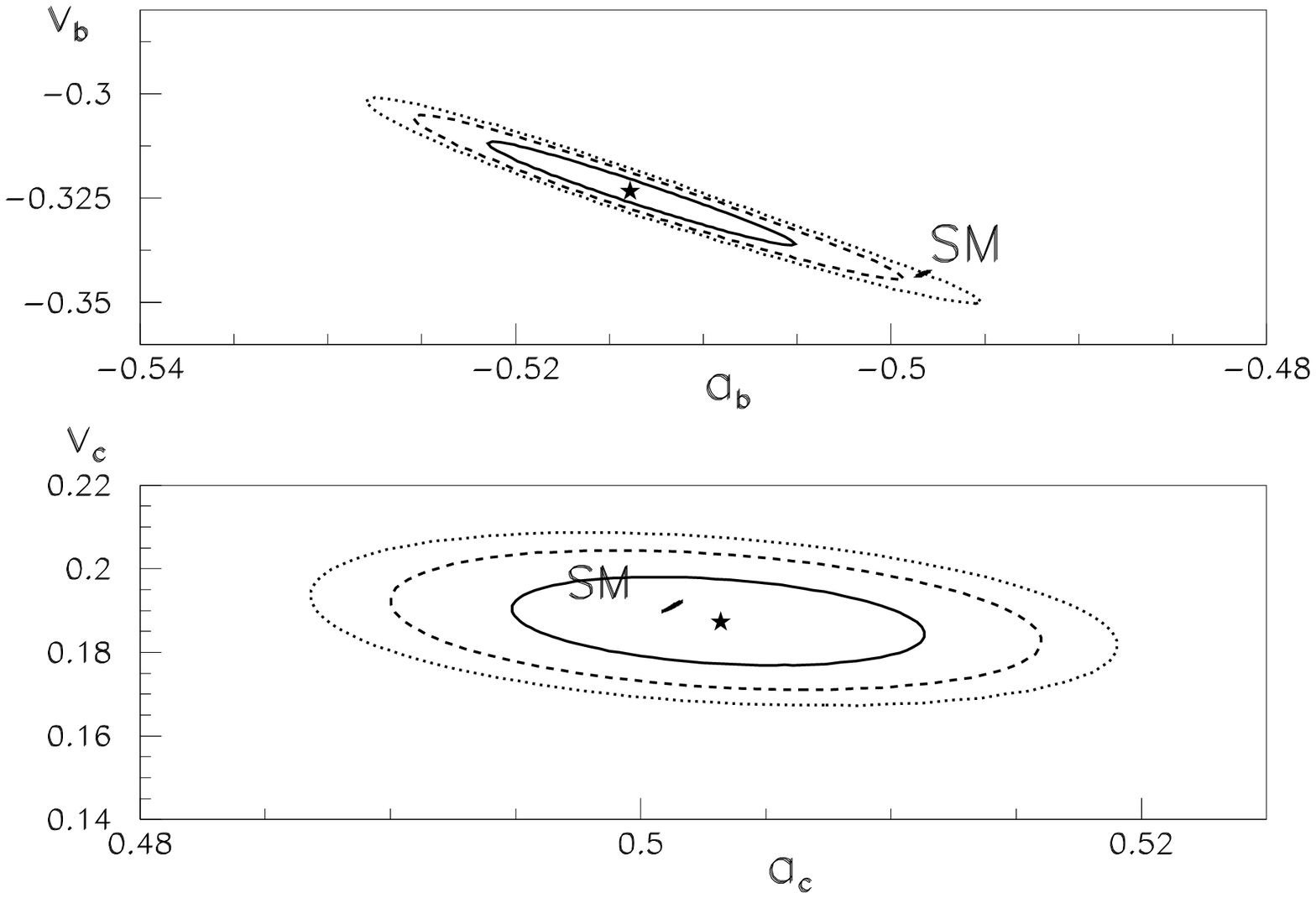,height=9cm}}
\end{center}
\caption{Results of a fit to the b and c quark vector and axial-vector 
couplings. The contours are for the 70, 95 and 99\% confidence limits.
The SM predictions are also shown.  }
\label{fig-coupvabc}
\end{figure}

 It can be seen that the SLD values of $\cAb$ and $\cAc$ are in good agreement
with the SM predictions of 0.935 and 0.668, 
which are essentially independent of $\Mt$ and $\MH$.
However, the values of $\cAb$ and $\cAc$, extracted from $\Afbzb$ and $\Afbzc$
and the measured values of $\cAe$,
are somewhat below the SM predictions. When combined with the SLD results,
which for $\cAb$ is slightly below the SM prediction, the values for $\cAb$ and
$\cAc$ are respectively 2.6 and 0.8 standard
deviations below the SM predictions.


\subsection{\bf Extraction of heavy-quark couplings\label{sec-hflav_coup}}

 An alternative approach in trying to understand the possible implications
of the heavy flavour results is to extract the individual quark 
couplings\cite{pbr_coup}. The 
measurements used are $\Rbbz = \Gb/\Ghad$ (which, using $\Ghad$ from the
lineshape, gives $\Vb^{2}$+ $\Abcoup^{2}$), $\Rccz$ 
($\Vc^{2}$ +  $\Accoup^{2}$), $\cAe$ from LEP/SLD ($\Ve/\Ae$),
$\Afbzb$ ($\Vb/\Abcoup$, $\Ve/\Ae$), $\Ab$ ($\Vb/\Abcoup$), 
 $\Afbzc$ ($\Vc/\Accoup$, $\Ve/\Ae$) and $\Ac$ ($\Vc/\Accoup$). The constraint 
$\alphasmz$ = 0.118 $\pm$ 0.002 is imposed 
(although the results are rather insensitive to this, as discussed below),
and lepton universality is assumed. 

 The signs of the b- and c-quark couplings are uniquely determined from the
LEP data. From the sign of the measured value of $\Afbzb$ (i.e. positive),
it follows that $\Vb$ and $\Abcoup$ have the same sign.
The behaviour of the energy dependence of A$_{FB}^{b}$ away from the Z-pole
depends on the
product Q$_{\rm e}$Q$_{\rm b}\Ae\Abcoup$ of the electric charges and 
axial-vector couplings of the electron and b-quark.  From the data 
shown in fig.~\ref{fig-afbene}
it can thus be deduced that $\Abcoup$ is negative, and thus $\Vb$ is also
negative. Similar considerations show that $\Vc$ and $\Accoup$ are both
positive.
The results for the vector and axial-vector couplings, of both b and c quarks,
are shown in  table ~\ref{tab-hfva} 
and fig.~\ref{fig-coupvabc}. Also shown are
the SM predictions corresponding to $\Mt=174.3\pm5.1$~GeV and $\HI$. 
Note that there
is a very strong anti-correlation between $\Vb$ and $\Abcoup$.
As discussed above, the signs and magnitudes of all the couplings
have been determined. These confirm the SM quantum number assignments.
Of course, they are measured to good precision, so the results are
sensitive to small deviations from the simplest predictions.

\begin{table}[htb]
\caption[]{Results, plus correlation matrix, of a fit to the vector and 
axial-vector couplings of b and c quarks. The $\chit$ probability for the
fit is 11\%. }
\label{tab-hfva}
\begin{indented}
\lineup
\item[] 
\begin{tabular}{ccrrrr}\br
parameter & fitted value  & $\Vb$ &$\Abcoup$ &$\Vc$ &$\Accoup$  \\
\br
$\Vb$     & $-0.3233\pm0.0079$ & 1.00 & -0.97 & -0.19 & 0.06  \\ 
$\Abcoup$ & $-0.5139\pm0.0052$ &      & 1.00  & 0.18  & -0.03 \\
$\Vc$     & $0.1873\pm0.0068$  &      &       & 1.00  & -0.29   \\
$\Accoup$ & $0.5032\pm0.0053$  &      &       &       & 1.00 \\
\br
\end{tabular}
\end{indented}
\end{table}

\begin{figure}[t]
\vspace*{13pt}
\begin{center}
\mbox{
\epsfig{file=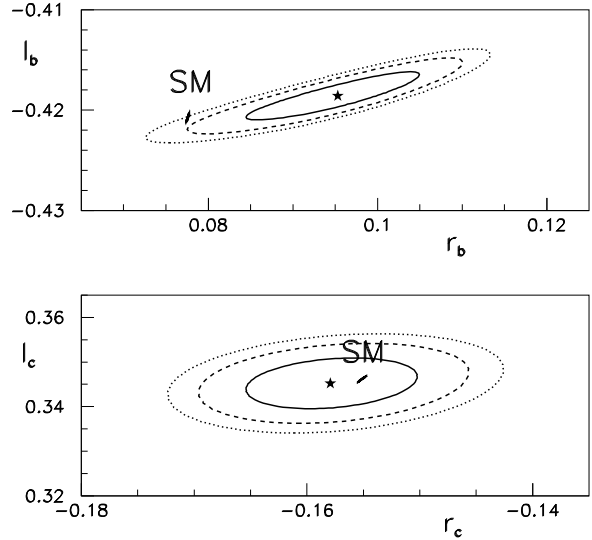,height=9cm}}
\end{center}
\caption{Results of a fit to the b and c quark left-handed and 
right-handed couplings.
The contours are for the 70, 95 and 99\% confidence limits.  }
\label{fig-couplrbc}
\end{figure}

 The b-quark couplings can also be expressed in terms of the
left-handed  $\ell_{b}$ = ($\Vb$ + $\Abcoup$)/2 and  
right-handed r$_{b}$ = ($\Vb$ - $\Abcoup$)/2 couplings.
The results are shown in fig.~\ref{fig-couplrbc}. The corresponding results
for the c-quark are also shown. The b-quark couplings are not in particularly
good agreement with the SM predictions, with the largest discrepancy
being for the right-handed coupling, r$_{b}$.

 The fitted values of $\Vb$ and $\Abcoup$ (or  $\ell_{b}$ and  r$_{b}$)
give a value of $\Rbbz$ greater than the SM value,
and a value of $\cAb$ (or $\Afbzb$) less than the SM value. In that
sense the b-quark data are mutually consistent with the observed deviations
from the ~SM.
The point in the SM band giving the smallest $\chit$
to the fitted data values corresponds to $\Mt$ = 169.2 GeV and
$\MH$ = 114 GeV. The $\chit$ probability for compatibility to this point
is 2.8\%.

 It is worthwhile therefore exploring further this possible discrepancy.
In the above fits the assumed value of $\alphasmz$ was taken to
be 0.118 $\pm$ 0.002. If a central value of 0.116 is used, then the leptonic
couplings are unchanged and the shifts in the b- and c-quark couplings
are less than 0.0002. Hence the results are not very 
sensitive to $\alphasmz$.
This is to be expected since the ratios $\Rbbz$ and $\Rccz$ are, by
construction, rather insensitive to $\alphasmz$.

 The results from the SLD Collaboration on $\ALR$, $\Ab$ and
$\Ac$~\cite{sld_tampere,EWWG01} require a precise determination of
the degree of polarisation of the
electron beam. It can be noted that the values of $\cAe$ (from  $\ALR$),
$\Ab$ (from $\Alrfbb$, see eqn.\ref{eqn-alrfb}) are above
and below the SM predictions respectively. Since, in both cases, what is
measured is
proportional to the product of the polarisation and the required parameter,
the measurements cannot both be reconciled with the SM simply by a change in the
value of the electron polarisation. It is worth stressing that the uncertainty
on $\ALR$ due to the  polarisation is about 0.5\%.
This is to be compared to the overall
statistical component of the error of about 1.3\%.

 Measurements of $\Afbzb$ determine the product of $\cAe$ and $\Ab$.
Thus the value of $\Ab$ extracted depends critically on that of $\cAe$.
In the standard fits given above the information on $\cAe$ comes from all of
the data, and the fitted value is $\cAe$ ~= 0.1501 $\pm$ 0.0016. Most of
the information comes from the measurements of $\ALR$, 
the $\tau$-polarisation
and $\Afbzl$. In the SM the value of $\cAe$ increases for
increasing $\Mt$ and decreasing $\MH$. However, as $\Mt$ is now well
constrained, the main variation is from  $\MH$. As can be seen from
fig.~\ref{fig-leptuniv}, the lepton coupling data favour a light Higgs.
Within the ranges $\tI$ and $\HI$, the closest SM value 
is 0.1485, which
corresponds to $\Mt$ = 179.4 GeV and $\MH$ = 114 GeV. 
The values of $\Vb$ and  $\Abcoup$ extracted, when this value
for $\cAe$ is imposed for the measurement of $\Afbzb$, are given in
table~\ref{T:vabvalues}. Also given are the $\chit$ probabilities that
the results are compatible with this SM point. If $\Afbzb$ 
and $\Afbzc$ are
removed from the fit, then the  probability increases to 38\%.

 In summary, the fit to the couplings gives a satisfactory $\chit$, and the
fitted vector and axial-vector couplings are reasonably compatible with
the SM values. The largest contribution to the fit $\chit$ comes from
the measurement of $\Afbzb$.

\begin{table}[htb]
\caption[]{ Values of $\Vb$ and $\Abcoup$ for different assumptions
about  the use of $\Afbzb$.
The SM values used correspond to $\MH$ = 114 GeV and $\Mt$ = 169.2 GeV,
except for the second line where $\Mt$ = 179.4 GeV is used for
consistency with the value of $\cAe$.}
\label{T:vabvalues}
\begin{indented}
\lineup
\item[] 
\begin{tabular}{cccc}\br
 conditions on $\Afbzb$  & $\Vb$  & $\Abcoup$ & $\chit$
 prob. for SM \\
\mr
 none  & -0.3233 $\pm$ 0.0079 & -0.5139 $\pm$ 0.0052 & 2.8\% \\
 $\cAe$ = 0.1485   & -0.3246 $\pm$ 0.0072 & -0.5130 $\pm$ 0.0048 & 2.3\% \\
 remove $\Afbzb$($\Afbzc$) & 
-0.3356 $\pm$ 0.0128 & -0.5058 $\pm$ 0.0088 & 38\% \\
\br
\end{tabular}
\end{indented}
\end{table}
 
\boldmath
\subsection{${\mathbf{f\overline{f}}}$ \bf production at LEP 2}
\label{sec-FF}
\unboldmath
 

 The above results have all come from data collected at energies at, or
close to, the $\Zzero$ peak ({\it LEP 1 phase}). Data have also been collected 
at various centre-of-mass energies from 130 to 209 GeV, from 1995 until 2000, 
when LEP was closed. The part of the programme with centre-of-mass energies
above the W-boson pair-production threshold (161 GeV) is called 
the {\it LEP 2 phase}. In total, an integrated luminosity of more 
than 700 pb$^{-1}$ per experiment was collected; well beyond the initially 
expected luminosity of 500 pb$^{-1}$ per experiment.

 The reaction $\eeff$ has also been extensively studied at LEP 2 energies. For
energies well above $\roots \simeq \MZ$, the contribution from the Z propagator
(see equation~\ref{eqn-zbwig}) is much reduced, as the distance in $\roots$
from the Z-pole is many factors of the Z width. 
However, the main difference in the analysis of LEP 2 $\ff$ data
is that there is a significant probablility
that an initial state photon (see fig.~\ref{fig-mumu}), or photons, 
are emitted, leaving the energy of 
the remaining $\ee$ system ($\rootsp$) close to that of the Z resonance. 
Since the Z cross-section is large, there is a relatively large probability
for the process of {\it radiative return} to the Z. To study the physics of 
the {\it direct} (or {\it non-radiative}) process, it is generally required to 
have $\rootsp/\roots >$ 0.85.

\begin{figure}[htp]
\vspace*{13pt}
 \begin{center}
   \mbox{
     \includegraphics[width=0.48\linewidth]{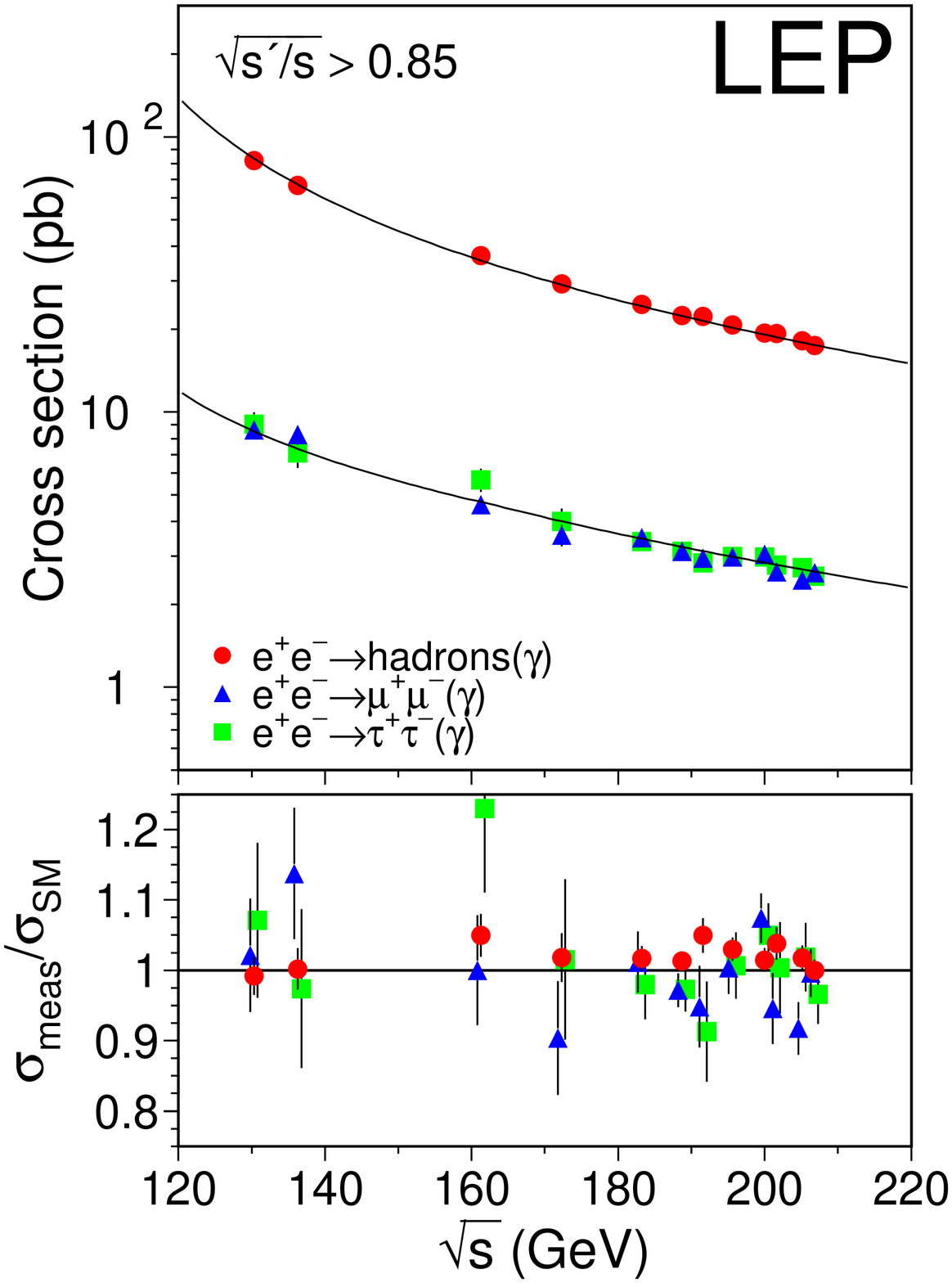}
     \includegraphics[width=0.48\linewidth]{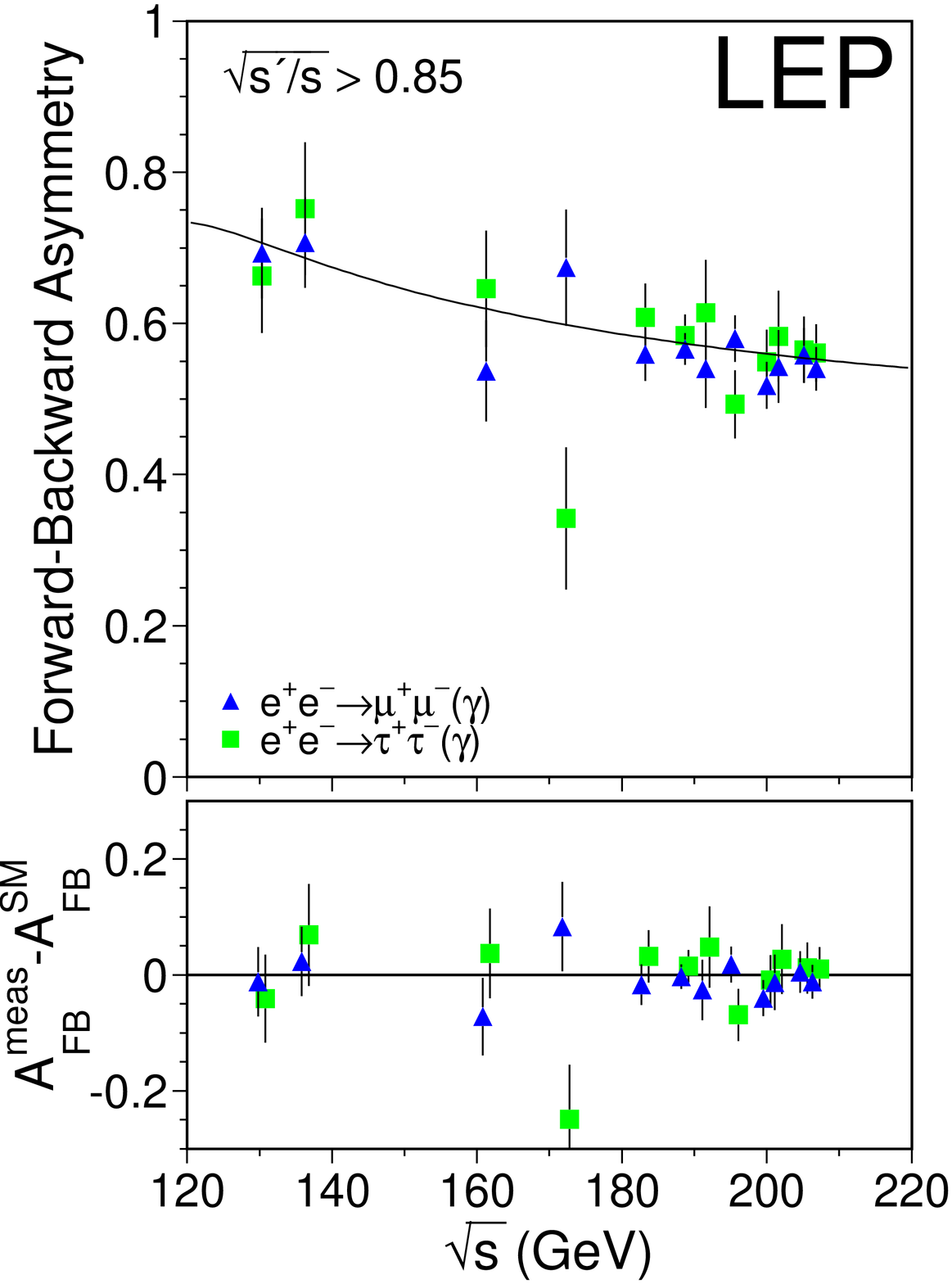}
     }
\end{center}
 \caption{Combined preliminary LEP cross-sections for
 $\qqb$, $\mumu$ and $\tautau$ final states, and forward-backward asymmetries
 for $\mumu$ and $\tautau$ final states, as a function of \CoM\
 energy. The predictions of the Standard Model,
 computed with ZFITTER, are shown as curves.
 The lower plots show the ratio of the data divided by the predictions
for the cross-sections, and the difference between the measurements and
 the predictions for the asymmetries. 
}
\label{fig-lep2-ff-xs-afb-lep}
\end{figure}

 The cross-sections and forward-backward asymmetries for the combined LEP 2
data~\cite{EWWG01}, are shown in fig.~\ref{fig-lep2-ff-xs-afb-lep}, 
together with the Standard Model predictions. 
It can be seen that the data are in reasonably
good agreement with these predictions. This shows that the use of the SM
in the calculation of some of the small corrections used in the Z lineshape
analyses is well justified. In particular, the $\gamma$-Z interference term 
for the $\qqb$ final states is poorly known from the Z-pole data, and is fixed 
to the SM value (see sect. \ref{sec-lep_comb}). 
This interference term is
highly correlated with the $\Zzero$ mass parameter and, if left free in the
fit, leads to a much reduced precision on $\MZ$. The data from LEP 2, and also
from the TRISTAN accelerator operating at around $\roots \simeq$ 61 GeV,
can be used to significantly limit the size of 
the hadronic $\gamma$-Z interference.
Although detailed fits have not yet been performed, the error on $\MZ$ 
with this interference term free should be about 2.3 MeV,
compared to that of 2.1 MeV obtained when the SM constraint is imposed.
Heavy flavour production
rates and asymmetries have also been studied at LEP 2. Again the results
are compatible with the SM.

 The $\ff$ data can also set very stringent limits
on many models containing physics beyond the Standard Model. These models
include additional heavy Z vector bosons, lepto-quarks, R-parity violating
supersymmetry, models of gravity in extra dimensions as well as contact 
interaction models which parameterise
new physics in terms of the left- and right-handed components of the
initial and final-state fermions. There is no evidence in the data for
the existence of any of these effects, and so limits are obtained on 
the masses or scales
below which such effects can be ruled out. For example, the limit on a
hypothetical heavy Z-boson, having the same couplings as the $\Zzero$
(sequential Z-boson), can be ruled out for masses up to 1.9 TeV.

 In the absence of new physics beyond the SM the $\ffb$ data on cross-sections
and asymmetries can be used the test the running of the electromagnetic
coupling constant $\alpha$. The OPAL Collaboration find a 
value 1/$\alpha$(191 GeV) 
= 126.2 $\pm$ 2.2~\cite{opal-alpha}, in agreement with the SM 
expectation of 127.9.

\subsection{\bf The Drell-Yan process \label{subsec-drellyan}}

\begin{figure}[t]
\vspace*{13pt}
\begin{center}
\mbox{
\epsfig{file=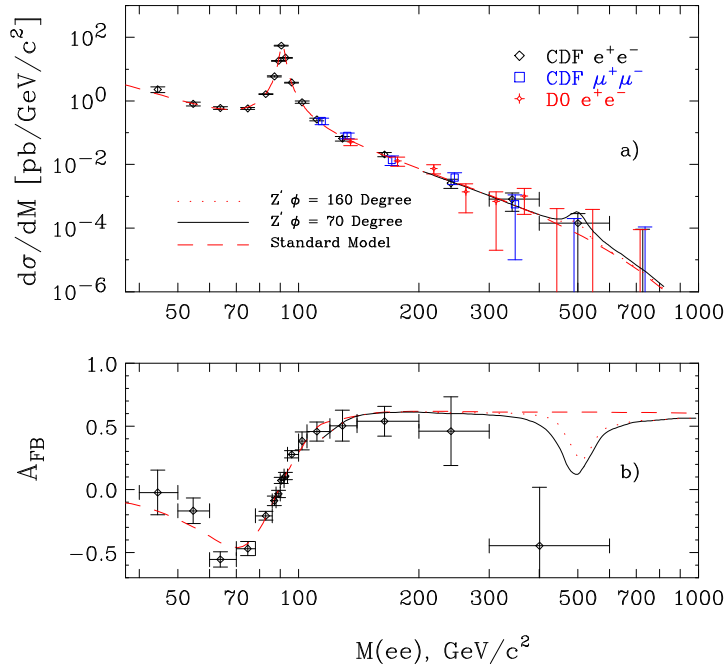,height=12cm}}
\end{center}
\vspace*{-5.0cm}
\caption{Dilepton mass spectrum from the Drell-Yan process from the CDF
and D0 experiments at the Tevatron for Run 1.}
\label{fig-drell-yan}
\end{figure}

 A process analogous to the $\eeff$ interaction which has been studied at LEP 
is $\qpqbll$, the Drell-Yan process~\cite{drellyan}. 
This is studied at the $\ppb$ Tevatron Collider, at Fermilab in the USA, 
for both electrons and muons in the final state. 
The interest for electroweak physics
is in the region where the $\lept$ pair has a large invariant mass. 
Measurements~\cite{cdf-dy} of the invariant mass distributions,
and the forward-backward 
asymmetries, are shown in fig.~\ref{fig-drell-yan}. 
The behaviour of A$_{FB}$ around
the Z resonance is in agreement with the SM predictions.
The data at invariant masses above the Z are used
to test the validity of the SM, and to search for physics beyond it. 
The invariant mass range explored goes well beyond that studied directly 
at LEP. 
Also shown in fig.~\ref{fig-drell-yan} are the predictions obtained if there 
was an additional Z${'}$ resonance with a mass of 500 GeV. The data are, 
however, in good agreement with the SM predictions alone.

\section{ The W boson \label{sec-wmass}} 

In the on-shell renormalisation scheme $\swsqa = \swsq$, so that precise
measurements of the W and Z masses give directly the weak mixing angle.
An accurate measurement of the W-boson mass gives a rather precise indirect 
estimate of the Higgs boson mass in the SM, from electroweak 
radiative corrections.
The W-boson decays weakly into either 
a quark-antiquark pair or a lepton and its corresponding neutrino.
The partial leptonic decay width is given by~\cite{rosner94}
\begin{equation}
    \Gamma(\Wtoenu) = \frac{\GF \MW^{3} }{ 6\pi\sqrt{2}}  
( 1 + \delta^{sm}) = 227.0 \pm 0.3 \hspace*{0.1cm} \MeV, \ \ 
\end{equation}
where the error is dominated by the present uncertainty in $\MW$ (see below).
If the values of $\GF$ and $\MW$ are used to determine the SM value
of $\Gamma(\Wtoenu)$, then the electroweak
corrections $\delta^{sm}$ are small ($\simeq$ -0.35 $\%$),
because the bulk of the corrections are absorbed in $\GF$ and $\MW$.
The partial width to $\qqb$ final states, for massless quarks, is given by
\begin{equation}
    \Gamma(\Wtoqqb) = f_{QCD} \Gamma(\Wtoenu) \mid V_{ij} \mid^{2}.  \ \ 
\end{equation}
where f$_{QCD}$ = 3(1 + $\alphasmw/\pi$ 
+ 1.409($\alphasmw/\pi$)$^{2}$ + ...)
 is a QCD colour correction factor (similar to eqn.~\ref{eqn-zgf})
and V$_{ij}$  is the 
Cabibbo-Kobayashi-Maskawa~\cite{cabibbo,km}(CKM) matrix element.
The total width $\GW$ in the SM is given by
\begin{equation}\label{eqn-gwsm}
    \GW = ( 3 + 2 f_{QCD} ) \Gamma(\Wtoenu) = 2.0986 \pm 0.0028 
\hspace*{0.1cm} \GeV,  \ \
\end{equation}
where the uncertainty from $\alphasmw$ (=0.121 $\pm$ 0.002)
is 1.0 MeV, and that from $\MW$ is 2.6 MeV.

The main $\qqb$ decay modes are $\udb$ and $\csb$.The $\qqb$ branching ratio
thus gives mainly constraints on the matrix elements V$_{ud}$ 
and V$_{cs}$. Since the former is well known from other measurements, the
$\qqb$ mode can be used to give V$_{cs}$.
The decay branching ratios (as measured at LEP~\cite{EWWG01,mor_eweak_01}) 
are given in table \ref{tab-w_branch}. In the combination procedure,
common systematic errors (e.g. from the 4-jet QCD background) are taken into
account. The data allow sensitive tests of the validity 
of {\it lepton universality} of the {\it weak charged-current},
at a level of better than 3$\%$:
\begin{eqnarray}\label{cc_univ}
 B(\Wtomnu) / B(\Wtoenu) = 1.000 \pm 0.021, \nonumber \\
 B(\Wtotnu) / B(\Wtoenu) = 1.052 \pm 0.029, \nonumber \\
 B(\Wtotnu) / B(\Wtomnu) = 1.052 \pm 0.028.  
\end{eqnarray}
Assuming lepton universality,  
B($\Wtolnu$) = 10.69 $\pm$ 0.06 (stat) $\pm$ 0.07 (syst)$\%$, compatible 
with the SM value of 10.82$\%$.

 The decay branching ratios have also been extracted at hadron colliders
by measuring the ratio 
\begin{equation}\label{ratiowz}
    R = \frac{ \sigma( \ppb \rightarrow W \rightarrow \enu) }
             { \sigma( \ppb \rightarrow Z \rightarrow \ee)} ,  \\ 
\end{equation} 
which can be written as
\begin{equation}\label{ratiowz1}
     R  = \frac { \sigma ( \ppb \rightarrow W + ..) }
               { \sigma ( \ppb \rightarrow Z + ..) }
         \frac{ \GZ}{\Gamma(Z\rightarrow \ee)}
         \frac{\Gamma(\Wtoenu)}{\GW} \ \ .
\end{equation} 
Using the measured value of the Z leptonic branching fraction from LEP,
and the SM theoretical calculation of the ratios of the W and Z 
cross-sections ($\simeq$ 3.3), the CDF and D0 Tevatron measurements 
give~\cite{lancaster99} BR($\Wtoenu$) = (10.43 $\pm$  0.25)$\%$.
In this, the total systematic uncertainty is 0.23$\%$, with 0.19$\%$
coming from the QED uncertainties in the acceptance calculations 
and in the $\sigma_{W}$/$\sigma_{Z}$ ratio.

 The combined LEP and Tevatron value is BR($\Wtoenu$) 
= (10.66 $\pm$  0.09)$\%$. In terms of the CKM matrix elements
\begin{equation}\label{breckm}
 \frac{1}{BR(\Wtoenu)} = 3 + f_{QCD} \sum_{ij}  \mid V_{ij} \mid^{2} ,
\end{equation}
where the sum is over i=(u,c) and j=(d,s,b). This gives
\begin{equation}\label{vsumig2}
 \sum_{ij}  \mid V_{ij} \mid^{2}  = 2.044 \pm 0.024,
\end{equation}
where an error of 0.001 comes from $\delta\alphasmw$ and the rest is 
from $\delta$BR($\Wtoenu$). Using the experimental value for the
sum of all elements except $\mid V_{cs} \mid^{2}$, 
namely 1.0477 $\pm$ 0.0074~\cite{pdg2001}, the value
\begin{equation}\label{vcs}
 \mid V_{cs} \mid = 0.998 \pm 0.013
\end{equation}
can be extracted. In this, the uncertainty from the measured W branching
fraction is $\pm$ 0.013, the input CKM uncertainty is $\pm$ 0.004,
and that from $\alphasmw$ is negligible.

 Alternatively, the combined leptonic branching ratio from LEP and the
Tevatron, together with the SM value of $\Gamma(\Wtoenu)$, can be
used to make an {\it indirect} measurement of the W-boson
width of $\GW$ = 2.130 $\pm$ 0.017 GeV. This value is 1.8$\sigma$ from
the SM prediction given in eqn.(\ref{eqn-gwsm}).

\subsection{\bf Mass and width of the W boson \label{sec-wmassandwidth}}

 The measurements of the mass of the W boson, $\MW$, have been made at
proton-antiproton colliders (by UA2~\cite{MWUA2} at CERN  and 
CDF~\cite{MWCDF} and D0~\cite{MWD0} at the Tevatron) and at LEP
(by ALEPH~\cite{MWALEPH}, DELPHI~\cite{MWDELPHI}, L3~\cite{MWL3} and
OPAL~\cite{MWOPAL}, and references therein), with updates 
reported in~\cite{EWWG01}.

\subsubsection{\bf Mass and width of the W boson from hadron 
colliders\label{sec-wmass_tev}}

\begin{figure}[htbp]
\vspace*{13pt}
\vspace*{-1.5cm}
\begin{center}
\mbox{
\epsfig{file=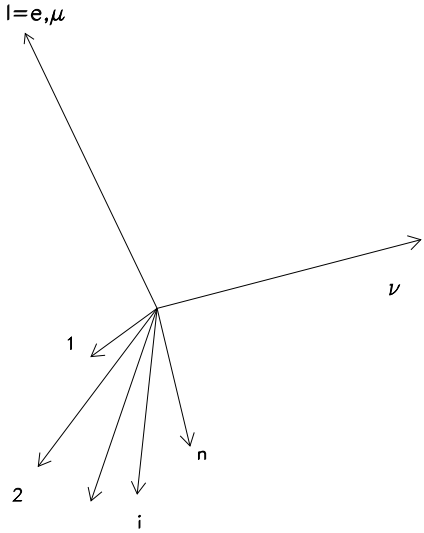,height=8cm}}
\end{center}
\vspace*{-0.7cm}
 \caption{Kinematics of W decay in the plane transverse to the beam direction.
The recoil energy vector of the n detected hadrons 
is ${\bf u}$ = $\sum_{i=1}^{n}$ {\bf $E_{T}^{i}$}.
}
\label{fig-upicture}
\end{figure}

 For measurements of $\MW$ at hadron colliders the purely hadronic W decay 
mode suffers from too much background and only the electron and muon
leptonic decays have been used. The W-bosons are produced by the
reaction 
$\qpqbd \rightarrow$ W $\rightarrow$ e($\mu$) + $\nu_{e}$($\nu_{\mu}$). 
The event topology selected is an isolated electron or muon, plus the
residual hadronic system from the collision. The neutrino is not detected and
can give a sizeable missing transverse energy/momentum.
Since a large fraction of the longitudinal
energy and momentum escapes detection in the forward regions of detectors
in, or close to, the beam-pipe, only the plane perpendicular to the beam-axis
can be used to impose energy-momentum constraints. Measurement is made of
the lepton energy/momentum, plus that of the recoil jet (${\bf u}$); see
fig.~\ref{fig-upicture}. From these quantities the 
transverse mass, $\mtrans$, is constructed
\begin{eqnarray}\label{mtrans}
\mtrans^2 = 2 p_{T}^{\ell} p_{T}^{\nu} ( 1 - \cos \Delta \phi ) ,
\end{eqnarray} 
where p$_{T}^{\ell}$ and p$_{T}^{\nu}$ are the charged lepton and neutrino
transverse momenta and $\Delta \phi$ is their azimuthal separation.
The neutrino component is reconstructed from measurement of the 
recoil ${\bf u}$, and the understanding of this recoil system is crucial 
to the analysis.  The Z boson is produced by the process
$\qpqb \rightarrow$ Z $\rightarrow$ $\ee$ ($\mumu$).
A study of Z boson production and leptonic decays is 
of great importance, both in calibrating the energy/momentum scale
and in understanding the p$_{T}$ production spectrum of heavy bosons.
The leptonic decays of the J/$\Psi$ boson are also useful in the scale
calibrations.
Another important systematic uncertainty is the imprecision in the knowledge
of the parton density functions (PDFs) in the incident proton and 
antiproton. The lepton transverse energy spectra have also be used to 
determine $\MW$, but these of course are correlated to $\mtrans$.

\begin{figure}[htbp]
\vspace*{13pt}
\begin{center}
\mbox{
\epsfig{file=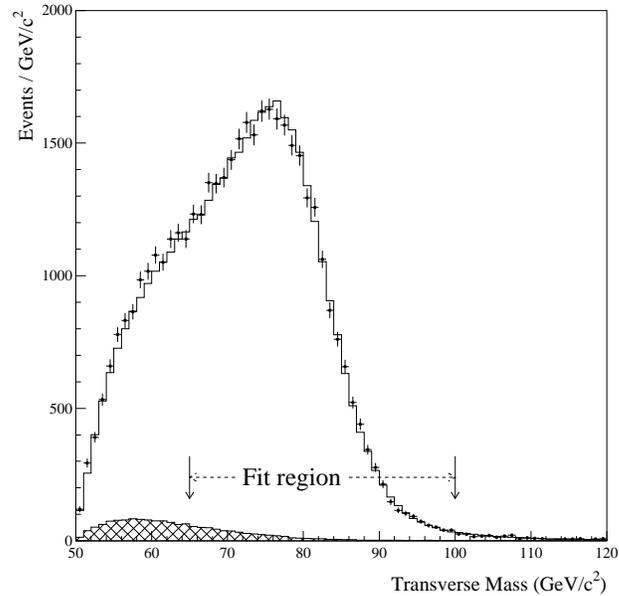,height=9cm}}
\end{center}
\vspace*{-2.3cm}
\caption
{Transverse mass distribution from the CDF experiment for the decay
$\Wtoenu$. The points represent the data and the histogram
shows the Monte Carlo simulation of the signal and background.
The region used to fit $\MW$ is shown.
}
\label{fig-cdf_mt_e}
\end{figure}

 All the data from the Tevatron Run 1, which finished in 1995 and yielded
an integrated luminosity of about 110 pb$^{-1}$, have been analysed.
An example of a transverse mass distribution, from the $\Wtoenu$ decay,
is given in fig.~\ref{fig-cdf_mt_e}. 
The results from the Tevatron CDF and D0 experiments, and the earlier
UA2 experiment at CERN are given
in table \ref{tab-mw_hadcol}(from ~\cite{Gerber}), 
together with the average value. In combining 
the Tevatron data, a 25 MeV common systematic error is used. This covers
common uncertainties in the PDFs, W-width and QED corrections.

 The W width is extracted by making a likelihood fit to the large transverse
mass part of the $\mtrans$ spectrum. This region is
rather insensitive to detector resolution effects, which fall off in an 
approximately Gaussian manner, but is sensitive to the W width. The 
combined CDF result from the electron and muon channels~\cite{cdf_wwidth},
using the range $\mtrans \geqsim$ 120 GeV, 
is $\GW$ = 2.05 $\pm$ 0.10 $\pm$ 0.08 GeV. For D0 (see ~\cite{Gerber}) the
range 90 $\leqsim \mtrans \leqsim$ 200 GeV is used for the electron channel,
giving $\GW$ = 2.23 $\pm$ 0.14 $\pm$ 0.09 GeV. Assuming a common systematic
uncertainty of 50 MeV, a combination of the Tevatron
results gives $\GW$= 2.11 $\pm$ 0.11 GeV.


\begin{table}[htbp]
\caption[]{ W boson branching ratios from measurements at LEP.
For the $\qqb$ mode lepton universality is assumed.
}
\label{tab-w_branch}

\begin{indented}
\lineup
\item[] 
\begin{tabular}{lc}
\br decay mode   & branching ratio $\%$        \\ \br
$\qqb$        & $67.92 \pm 0.27$    \\
$\enu $    & $10.54 \pm 0.17$   \\
$\mnu $    & $10.54 \pm 0.16$    \\
$\tnu $    & $11.09 \pm 0.22$    \\
\br
\end{tabular}
\end{indented}
\end{table}

\begin{table}[htbp]
\caption[]{ W mass measurements from hadron colliders.
}
\label{tab-mw_hadcol}

\begin{indented}
\lineup
\item[] 
\begin{tabular}{llc}
\br experiment   & decay modes used    & $\MW$ (GeV)       \\ \br
 UA2             & $\Wtoenu$,$\Wtomnu$ &$80.360 \pm 0.370$ \\
 CDF             & $\Wtoenu$,$\Wtomnu$ &$80.433 \pm 0.079$ \\
 D0              & $\Wtoenu$           &$80.483 \pm 0.084$ \\ \br
 average         &                     &$80.454 \pm 0.060$ \\ 
\br
\end{tabular}
\end{indented}
\end{table}


\vspace*{0.6cm}

\subsubsection{\bf Mass and width of the W boson at LEP\label{sec-wmass_lep}}

\begin{figure}[htbp]
\vspace*{13pt}
\begin{center}
\mbox{
\epsfig{file=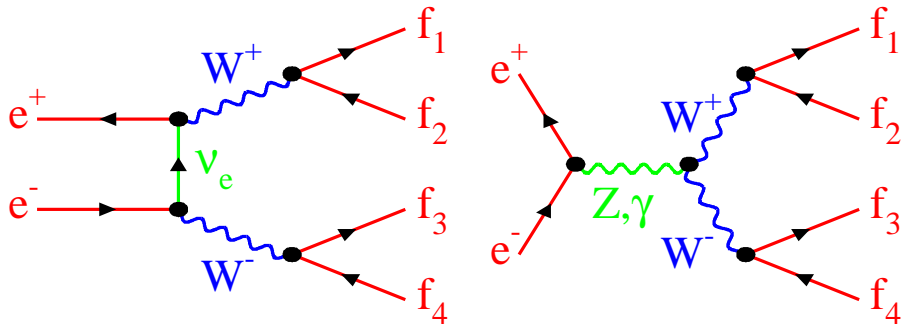,height=8cm}}
\end{center}
\caption
{The three lowest-order charged current (CC03) diagrams for $\eeww$. 
}
\label{fig-cc03}
\end{figure}

 One of the main purposes behind increasing the energy for LEP 2 above
the W-pair threshold (around 161 GeV) was
to study the production and decay properties of the W boson. The 
lowest-order diagrams for producing W pairs are shown in fig.~\ref{fig-cc03}.
These diagrams are collectively known as CC03 diagrams, as the three diagrams
are charged-current interactions. They consist of t-channel neutrino exchange
and s-channel exchange of a photon or Z boson. In the Standard Model there 
are large cancellations between these diagrams, and thus
measurement of the production cross-section of $\eeww$ is a very sensitive 
test of the SM.

\begin{figure}[htbp]
\vspace*{13pt}
\begin{center}
\mbox{
\epsfig{file=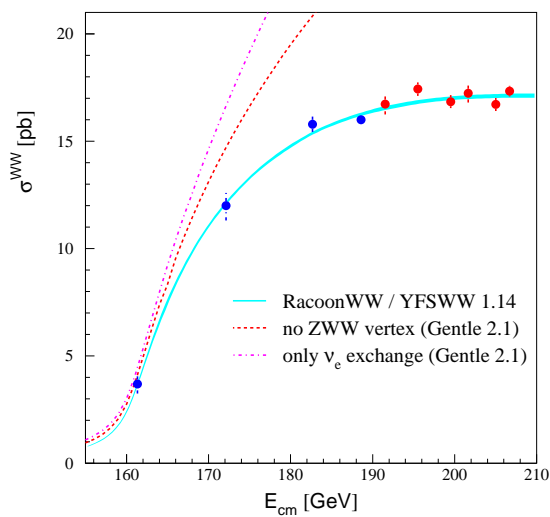,height=8cm}}
\end{center}
\caption {
Cross-section for the CC03 process $\eeww$, as a function of cms energy.
}
\label{fig-ww_xsec_w01}
\end{figure}

 The reaction $\eeww$ can be extracted relatively cleanly at LEP and all the
decay modes of the W can be used. Thus the samples analysed consist of
fully hadronic final states (46~\%), semileptonic final states (44~\%)
and purely leptonic final states (10~\%). The purely leptonic and 
semileptonic final states can be selected relatively cleanly from background,
but there is a potentially sizeable background from $\eeqqg$ in the 
fully hadronic final state. The cross-sections for these CC03 processes have
been measured and combined by the four LEP experiments. Small corrections 
have to be applied for other 4-fermion final states with the same topologies.
The results are shown in fig.~\ref{fig-ww_xsec_w01}. It can be seen that
the data are compatible with the SM predictions, which at the highest
energies have an uncertainty of about 0.5\%. The results clearly demonstrate 
the existence of the triple-gauge boson couplings.

 The first data taken in the LEP 2 phase consisted of a dedicated run 
at $\roots$ = 161 GeV; just above the W-pair threshold. The cross-section
in this region is very sensitive to $\MW$. Hence, from the measurement
of the cross-section, $\MW$ may be extracted; assuming the validity of
the Standard Model.
The indirect determination of $\MW$ by this method is given in 
table~\ref{tab-mw_lep}. The contribution of the 2\% uncertainty in the SM
predictions gives a 34 MeV component to the systematic error.

 The threshold method is valid in the SM. A comparison of the LEP data,
averaged over all cms energy values, and taking into account common
systematic uncertainties, gives a 
value of R$_{\rm WW}$ = 0.998 $\pm$ 0.009 for the ratio of the measured
cross-sections to the SM predictions. These are from the 
YFSWW generator~\cite{YFSWW}, which employs a double-pole approximation
approach and has $\cal{O}(\alpha)$ electroweak corrections. An alternative
model, RACOONWW~\cite{RACOONWW}, gives a very compatible value,
namely R$_{\rm WW}$ = 1.000 $\pm$ 0.009. These results show that the data
are consistent with the SM up to the highest available energies.

 A much more accurate determination of $\MW$ has been made using {\it direct
reconstruction} of the W mass in the reaction $\eeww$. The purely leptonic
mode yields a rather large error, so emphasis is placed on the 
semi-leptonic ($\qqb\lnu$) and fully hadronic ($\qqb\qqb$) final states.
For the $\qqb\qqb$ analysis events are selected which contain four (or more)
jets. Again $\eeqqg$ is the largest background. In a four-jet final state
there are three possible combinations with which the jets can be paired
into two W bosons.
However, in practice only one, or sometimes two, are compatible with the 
previously known value of $\MW$. 
The experimental resolution on the invariant mass of
a pair of jets is typically 5-10 GeV. Some fraction of the jet energy is
generally not detected. The resolution can be greatly 
improved (to $\simeq$ 1 GeV) by using a
kinematic fit in which energy-momentum constraints (4C-fit) are imposed. 
This requires
a rather precise knowledge of the LEP beam energy, and the uncertainty 
on this is one of the main systematic errors.

  For the $\qqb\lnu$ channel the assignment of the jets to the parent W boson
is straightforward. The lepton is well measured in the case of electron and
muon decays; however, there is a missing neutrino. In this case the kinematic
fit gives only one constraint, except if, as is often the case, it is assumed
that the mass of the W from the $\qqb$ and $\lnu$ systems are equal; in which
case it becomes a 2C-fit. For the $\qqb\tnu$ channel the leptonic part is
not constrained, and all the information on $\MW$ comes from the $\qqb$ system,
which is again scaled to the LEP beam energy. The W-boson width, $\GW$, is
also extracted in these fits, and has only a small correlation with $\MW$.

 A variety of different sophisticated techniques are used
to extract $\MW$ from these event samples. The most straightforward
method is the {\it convolution method}, in which the theoretical Breit-Wigner
form~\footnote{An s-dependent Breit-Wigner, similar in form to that for 
the Z in eqn.(\ref{eqn-zbwig}), is used.} for the invariant mass of the
W decay products is convoluted with the experimental
resolution on this quantity. Both the W mass and width can be determined.
The method must be `calibrated' by fitting Monte Carlo samples of known
mass (or masses). The linearity of the method with respect to the determination
of $\MW$, and also the validity of the assigned statistical uncertainty, 
must both be studied. 
An alternative is the {\it Monte Carlo reweighting method}
in which Monte Carlo events are reweighted such
that they correspond to a series of different generated masses.
The value of the W mass which best fits the data is found.
The distribution which is fitted is the invariant mass
of the W-decay products, plus possibly other distributions which are sensitive
to $\MW$. In both these methods it is usual to force the mass of the decay
products from the two W-pairs in the kinematic fit to be equal.
An {\it ideogram method} has also been used in which all jet pairings are
taken, and the equal mass constraint is not imposed. A likelihood function is
then constructed for each event, as a function of $\MW$, and these likelihoods
are combined to give the best value from all the events. 

A particular problem is the $\eeqqg$ background in $\qqb\qqb$. This
background must either be removed by suitable selection cuts, or taken
into account by weighting the events in such a way that the probabilility
that they are signal or background events is taken into account,  
The jet fragmentation process is
potentially a common systematic uncertainty between channels and between the
different LEP experiments. The effects of initial state radiation (ISR)
are also a common uncertainty. Considerable effort has been made to 
understand these
processes. In the fully hadronic final states there is the additional
potential problem of `cross-talk' between the decay products of the two W's.
This can arise through {\it colour reconnection}, as, for example, 
`strings' can be formed between the quarks and gluons from the different
W's. These effects can only be computed in QCD at the perturbative level,
where they are small. However, it is in the region of soft particles that
the effects are expected to be more pronounced. This is in the domain
of Monte Carlo models, and several options for models exist. 
These can yield shifts in $\MW$ up to about 100 MeV, but the
bulk of the models give shifts $\leqsim$~50 MeV. 
Ideally it is desirable to find measurable quantities, which are insensitive 
to $\MW$, but which are sensitive to the parameters of the model.
The measurements of these parameters, within the context of the model,
can then be used to `calibrate' the model by finding the shifts in the
values of $\MW$ corresponding to the measured model parameter range.
One promising possibility is to measure the particle 
flow (i.e. the density of particles) between jets coming
from different W-bosons, and compare it with that measured between jets 
from the same W. From such measurements it is hoped that the colour 
reconnection parameters in the various models can be constrained.
In practice these attempts
are complicated by the pairing ambiguities in jets, and the fact that the
4 jets in the final state are generally not coplanar.

 There can also be {\it Bose Einstein correlations (BEC)} between the
decay products of the two W's. Such correlations are well known in 
multiparticle final states and lead to an enhancement in the invariant
mass distribution of like-sign bosons near threshold. In principle such
correlations can arise between, for example, pairs of pions from 
different W-bosons. Studies of differences in the hadronic systems of 
hadrons from jets in a semi-leptonic W decay (where there can only
be BEC between the W decay products) and those in fully-hadronic
decays (which might also have BEC between decay products of different W's)
are being carried out. There are also several options in the Monte Carlo 
simulation models of BEC. 
However, it is worth stressing that, in terms of the underlying quantum
mechanical effects, all the existing models are somewhat {\it ad hoc}.
The technique used is to modify the
four-momenta of pairs of particles, and to compensate the overall 
energy-momentum conservation either {\it locally} or {\it globally}. 

 The {\it colour reconnection} and {\it BEC} can both lead to `cross-talk'
effects, and thus an apparent shift in $\MW$ and $\GW$. They are collectively
referred to as {\it final state interaction (FSI)} effects. They constitute
one of the largest systematic uncertainties in the $\qqb\qqb$ final
state. Investigations have shown that all the experiments are equally
sensitive to these effects, and so the same common systematic uncertainties
for these effects are used in making a LEP combination.

\begin{figure}[htp]
\vspace*{13pt}
 \begin{center}
   \mbox{
     \includegraphics[width=0.485\linewidth]{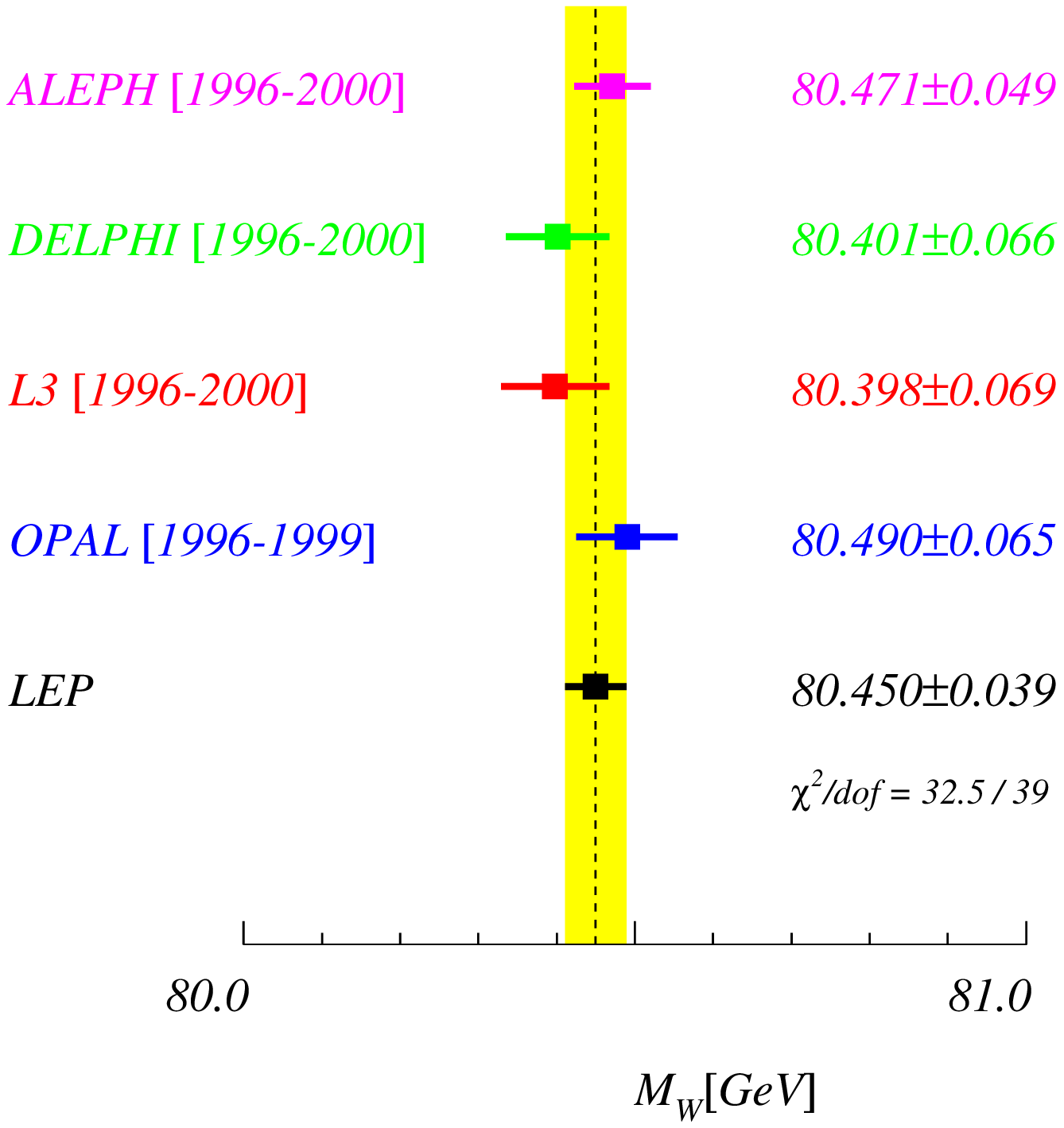}
     \includegraphics[width=0.485\linewidth]{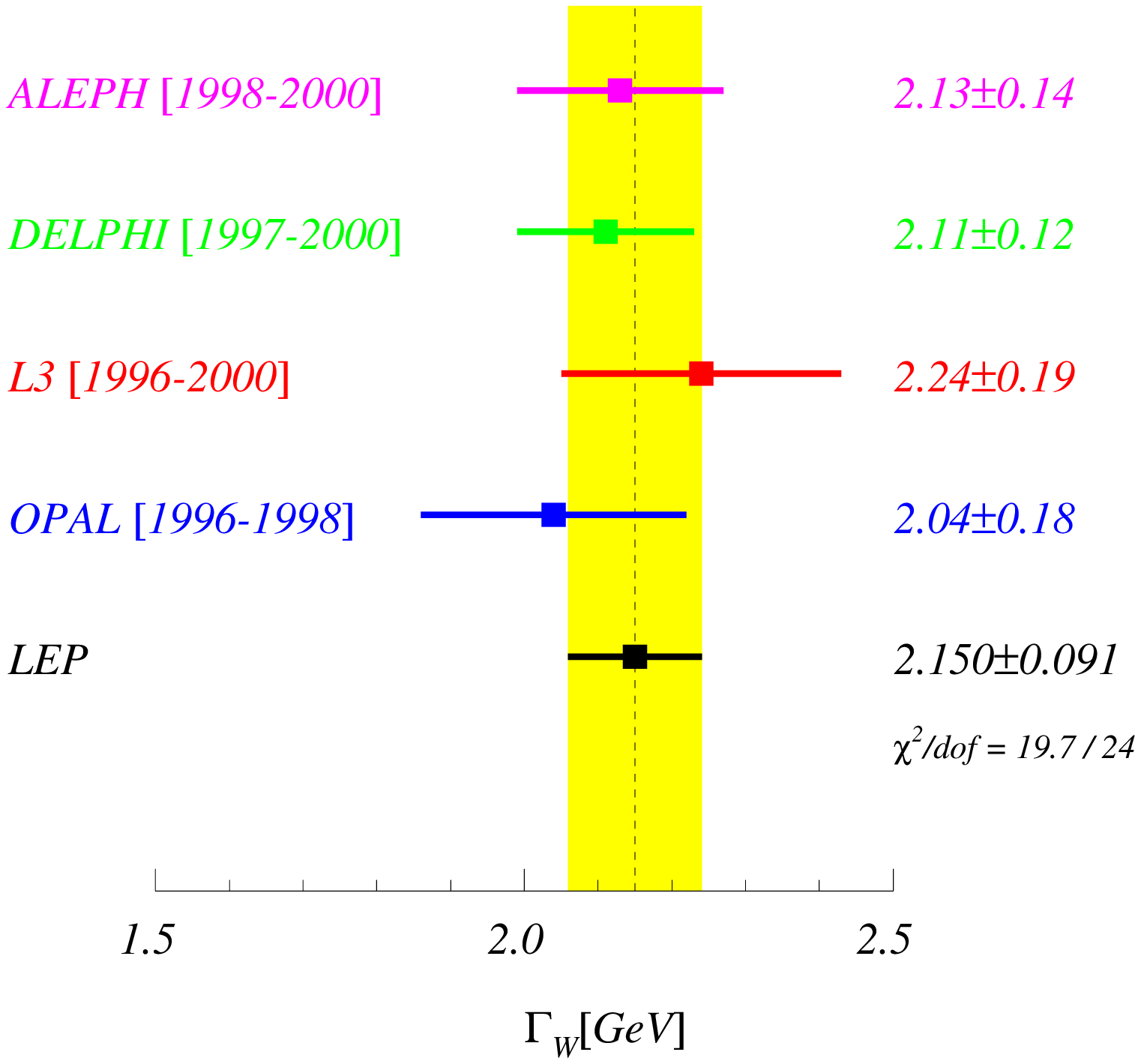}
     }
\end{center}
 \caption{Mass and width of the W boson as measured by the LEP experiments.
The combined LEP average is also shown.
}
\label{fig-mw-gw-lep}
\end{figure}

The measurements of the W mass and width from the LEP experiments are
combined, taking into account the common systematic errors. For the mass, 
the main errors common to both the $\qqb\lnu$ and $\qqb\qqb$ are the
fragmentation uncertainty (17 MeV), initial and final state radiation (8MeV),
and the LEP beam energy (17 MeV; see below). The colour reconnection and Bose
Einstein correlation errors are currently assigned to 
be 40 and 25 MeV respectively for the $\qqb\qqb$ channel. 
These get reduced to 11 and 7 MeV in the combined
result, as the $\qqb\qqb$ carries less weight in the combination. The weight
of the $\qqb\qqb$ channel is 27$\%$. This is
essentially due to the relatively large size of these effects, since both
channels would give roughly the same weight without these FSI errors.
For $\GW$, the colour reconnection and Bose Einstein correlation errors are 
taken to be 65 and 35 MeV repectively.
The results for the mass and width for the four LEP experiments are
shown in fig.~\ref{fig-mw-gw-lep}. The combined LEP results for $\MW$
from the threshold method, and for the different decay channels for the
direct reconstruction method, are given in table~\ref{tab-mw_lep}. 
The overall LEP combined results for $\MW$ and $\GW$ are
\begin{eqnarray}\label{mwgw}
\MW = 80.450 \pm 0.039  \hspace*{0.2cm} {\rm GeV} \nonumber\\
\GW = 2.150 \pm 0.091 \hspace*{0.2cm} {\rm GeV}.
\end{eqnarray} 
For $\GW$, the combined statistical and systematic errors 
are 0.068 and 0.060 GeV respectively.

 The difference between the fully hadronic and semi-leptonic mass 
measurements can also be extracted. A significant non-zero value
for $\Delta\MW$ could be indicative of large FSI effects in 
the $\qqb\qqb$ channel. The FSI errors are neglected in making this
difference, and this gives $\Delta\MW(\qqb\qqb -\qqb\lnu)$ = +9 $\pm$ 44 MeV,
well compatible with zero. However, it should be noted that this
difference is not precise enough to be sensitive to shifts of the
size of the FSI errors currently assigned. Further efforts are being
made in devising track selection procedures which are less sensitive
to FSI effects; for example, by removing or de-weighting low momentum
tracks and/or tracks at  wide angles to the W jet-axes.

 The LEP beam energy in the LEP 2 range has an uncertainty of about 20 MeV.
This arises because the method of resonant depolarisation only works up to
about 60 GeV beam energies. Hence extrapolations need to be made, and this
significantly increases the uncertainty. The main method is to use 16 NMR
devices which were installed around the ring and to use the flux-loop method,
which measures the total field seen by the beam, as a check of the linearity
of the extrapolation.

\subsection{\bf World average values of the W boson mass and width
\label{sec-wmass_aver}}

\begin{figure}[htbp]
\vspace*{13pt}
\begin{center}
\mbox{
\epsfig{file=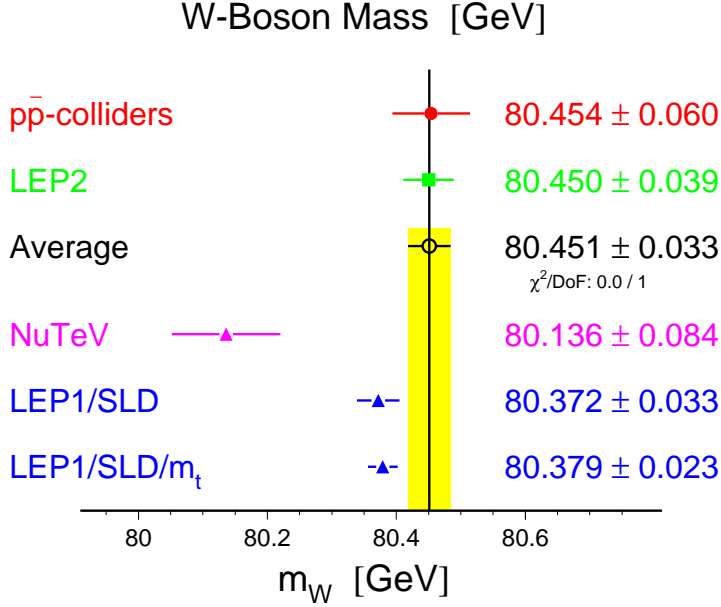,height=8cm}}
\end{center}
\caption {
Comparison of direct and indirect determinations of $\MW$.
}
\label{fig-mw_comp}
\end{figure}

The combined Tevatron and LEP measurements of $\MW$ and $\GW$ give
\begin{eqnarray}\label{mwgw_world}
\MW = 80.451 \pm 0.033  \hspace*{0.2cm} {\rm GeV} , \nonumber\\
\GW = 2.134 \pm 0.069 \hspace*{0.2cm} {\rm GeV}.
\end{eqnarray} 
This direct $\MW$ result is compared to more indirect determinations 
in fig.~\ref{fig-mw_comp}. It can be seen that the NuTeV value, and to
a lesser extent the indirect electroweak fit results (these are both discussed
below), favour lower values of $\MW$.
The result for $\GW$ can be used to make an estimate of the W mass,
by utilising the SM relationship (\ref{eqn-gwsm}). 
This gives $\MW$ = 80.90 $\pm$ 0.87 GeV, which is
consistent with the direct measurement, but has a sizeable error.


\begin{table}[htbp]
\caption[]{ W mass measurements from LEP. The components of the errors
from statistics and systematics are shown, with the contribution of
the LEP energy uncertainty shown separately. The last two results have
a correlation coefficient of 0.28.
}
\label{tab-mw_lep}

\begin{indented}
\lineup
\item[] 
\begin{tabular}{lcccc}
\br method     & $\MW$ (GeV) & stat. error. & syst. error & LEP error  \\ \br
 threshold     & 80.40   & 0.20         & 0.07        & 0.03       \\
 $\qqb\lnu$    & 80.448  & 0.033        & 0.028       & 0.017      \\
 $\qqb\qqb$    & 80.457  & 0.030        & 0.054       & 0.017      \\
\br
\end{tabular}
\end{indented}
\end{table}

\section{\bf The running of $\alpha$}
\label{sec-alfaem}

 The fine structure constant is known at q$^{2} \simeq$ 0 with the
impressive relative precision of 4x10$^{-9}$. However, what is important for 
the interpretation
of heavy gauge boson results is the value at a scale $\MZ$, $\alphamz$. 
The running of $\alpha$(s) is given by eqn.(\ref{alfa1}).
The running component $\Delta \alpha(s)$ = $ - \Pi_{\gamma\gamma} (s) $, 
where  $\Pi_{\gamma\gamma}$ is the photon self-energy. 
At s = $\MZ^{2}$ the leptonic contribution $\Delta\alpha_{lept}$ can be
computed analytically, and is known to third-order. 
The top-quark contribution, $\Delta\alpha_{top}$, is well known if $\Mt$ is 
specified ($\simeq$-~0.00007). 
The remaining hadronic part $\Delta \alpha_{had}^{(5)}$
cannot be calculated entirely from QCD because of ambiguities in defining
the light quark masses  m$_{u}$ and m$_{d}$, and also the inherent 
non-perturbative nature of the problem at small energy scales.
The largest uncertainties come from this term. Instead, use is made of 
the data on
\begin{equation}\label{alfa2}
R_{had}(s) = \frac{\sigma (\eehad) }{\sigma(\eemumu)},
\end{equation}
from which one can compute
\begin{equation}\label{eqn-alfa3}
{\rm Re} \Pi_{\gamma\gamma} (s) = \frac{\alpha {\rm s} }{ 3 \pi} P
\int \frac{ R_{had}(s^{'})} { s^{'}(s^{'} - s ) } ds{'} \ \ .
\end{equation}

 Most of the sensitivity is to $R_{had}$ at low values of $\roots$,
below about 10 GeV. In practice, 
there are difficulties in evaluating the integral:- 
\begin{itemize}
\item [a)] from the resonant structure in the data ($\rho$,$\omega$,J/$\psi$,
$\Upsilon$),
\item [b)] some data are not very accurate (e.g. large systematic errors)
 and old (not always enough information is given),
\item [c)] somewhat arbitrary choices need to be made in the use
 of the data (e.g. the form of local parameterisation, interpolation or fit,
how to deal with inconsistent data and how to cross thresholds etc.).
\end{itemize}

\begin{figure}[htbp]
\vspace*{13pt}
\begin{center}
\mbox{
\epsfig{file=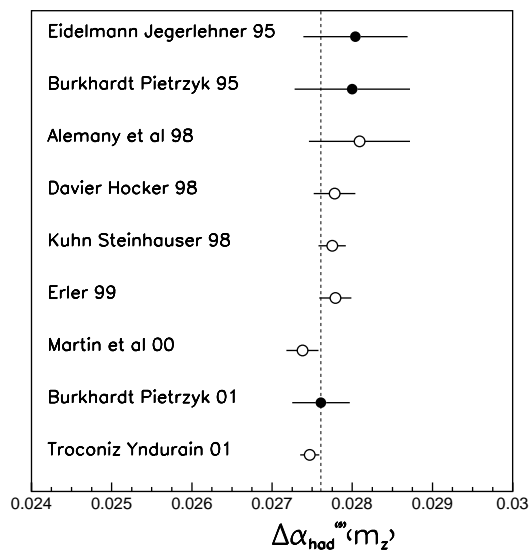,height=8cm}}
\end{center}
\caption {
Recent estimates of $\Delta \alpha^{(5)}_{had}(\MZ)$. Those which
rely mostly on experimental data are shown as solid circles, while
the more theory-driven estimates are shown as open circles. 
}
\label{fig-alfahad}
\end{figure}

\begin{figure}[htbp]
\vspace*{13pt}
\begin{center}
\mbox{
\epsfig{file=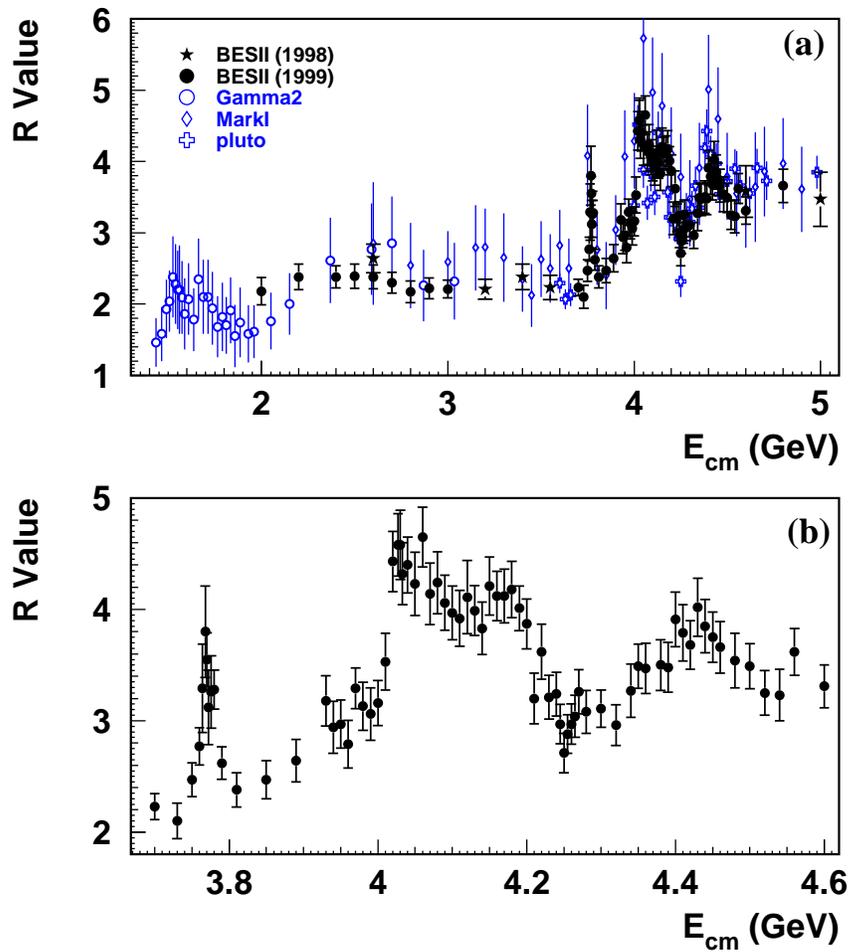,width=0.98\linewidth}}
\end{center}
\caption {
Measurements of R by BES and other experiments: (a) over the cms energy
range 1 to 5 GeV and (b) around the charm threshold.
}
\label{fig-besr}
\end{figure}


 Fig.\ref{fig-alfahad} shows some of the more recent 
determinations~\cite{ej95,bp95,alem98,davier98,kuhn98,erler99,mart00,bolek01,
troc01},
starting with the previous value used by the LEP Electroweak Working Group,
namely $\Delta \alpha^{(5)}_{had}(\MZ)$ = 0.02804 $\pm$ 0.00065. 
Only the most recent of these use the new data from
Novosibirsk around the $\rho$ resonance and the BES-2 data for the
very important region from 2-5 GeV; see fig.\ref{fig-besr}.
 For the fits below, the value used is~\cite{bolek01}
\footnote{References to the data used,
and to previous work, can be found here.} 

\begin{equation}\label{alfa3}
  \Delta \alpha^{(5)}_{had}(\MZ) = 0.02761 \pm 0.00036.
\end{equation}
This incorporates the recent data, and experimental data are used 
below 12 GeV and third-order perturbative QCD used above.

Some other determinations are more theory-driven, 
and use perturbative QCD to low $\sqrt{s}$, justifying this by the
success of extracting $\alphasmz$ from $\tau$-decays.
As an example, a recent more theory-driven determination, which also
includes results from BES, 
of $\Delta \alpha^{(5)}_{had}(\MZ)$ = 0.02747 $\pm$ 0.00012~\cite{troc01},
is also used below. 

\subsection{\bf muon (g-2)}

 The {\it anomolous magnetic moment} of the muon, which is defined as
\begin{equation}\label{alfa4}
  a_{\mu} =  \frac{ g_{\mu} -2} {2},
\end{equation}
has been measured at Brookhaven by experiment E821~\cite{bnle821} to 
remarkable precision, giving the updated average value
\begin{equation}\label{alfa5}
  a_{\mu}({\rm expt}) =  (11659203 \pm 15) 10^{-10}. 
\end{equation}
At the time the E821 result was announced the SM estimate was such that
\begin{equation}\label{alfa6}
  a_{\mu}({\rm expt}) - a_{\mu}({\rm SM}) =  (43 \pm 16) 10^{-10},
\end{equation}
a difference of 2.6 standard deviations. The SM computation is rather similar
to that for $\alphamz$, with contributions from QED, weak effects and also
hadronic effects. Since then, a sign error in the computation of part of the 
so-called light by light term, which is part of the hadronic term and
about 5.6 10$^{-10}$ in magnitude, was found\cite{knecht}.
With this corrected
\begin{equation}\label{alfa7}
  a_{\mu}({\rm expt}) - a_{\mu}({\rm SM}) =  (26 \pm 16) 10^{-10},
\end{equation}
so the difference is reduced to 1.6 standard deviations. This has significantly
reduced the flow of papers interpreting the difference.

\section{\bf Other electroweak measurements
\label{sec-sm_other}}

\subsection{\bf The top quark 
\label{sec-sm_other_top}}

 The discovery of the top quark at the Fermilab Tevatron, by the
CDF~\cite{cdfdisc} and D0~\cite{d0disc} Collaborations,
is clearly of fundamental importance in the field of electroweak physics.
The parameter which is most important in the context of this review
is the mass of the top-quark, $\Mt$. Since the top-quark is so massive
it decays very quickly on the time-scale of the hadronisation processes.
Hence the interpretation of the top-quark mass is more straightforward
than that of the lighter quark masses. The main decay mode 
is t $\rightarrow$ Wb, so the event classification of the produced $\ttb$
pair depends on the decay modes of the W bosons.
Both CDF and D0 measure the mass
in both the dilepton and lepton plus jets channels, with the latter being
the more accurate. The results for the top-quark pole mass
are $\Mt$ = 172.1  $\pm$ 7.1~GeV for CDF~\cite{cdfmtop} and 
$\Mt$ = 176.1  $\pm$ 6.6 GeV for D0~\cite{d0mtop}. Typical systematic 
uncertainties on these measurements are $\simeq$ 5 GeV.
A combination of the measured values gives~\cite{topmass}
\begin{equation}\label{eqn-mt1}
              \Mt = 174.3  \pm 5.1 \ \ \GeV.
\end{equation}

\subsection{\bf Atomic parity violation 
\label{sec-sm_other_apv}}
Measurements of Atomic Parity 
Violation (APV) in Cesium~\cite{wood97,bennett99} are 
also used to give information on the weak neutral current.
The nuclear spin-independent weak interaction of an electron with the
nucleus is of the form H $\propto \GF\rho(r)$Q$_{W}$,
where $\rho$(r) is the nuclear density.
The quantity measured is the {\it weak charge} 
\begin{equation}\label{eqn-apv1}
Q_{W}(Z,N) = -2 [ C_{1u}( 2 Z + N ) + C_{1d}( Z+2N ) ],
\end{equation}
where Z and N are the number of protons and neutrons in the nucleus 
and C$_{1q}$ = 2$\Ae\Vq$ is defined in terms of the electron axial-vector
and quark vector couplings at a scale q$^{2}~\simeq$~0. So, to leading
order,
\begin{eqnarray}\label{eqn-apv2}
 C_{1u} = \rho ( -\frac{1}{2} + \frac{4}{3}\swsqa ), \nonumber\\
 C_{1d} = \rho (  \frac{1}{2} - \frac{2}{3}\swsqa ).
\end{eqnarray}

 Measurements have also been made for Thallium, but there are outstanding 
questions on theoretical corrections~\cite{derevianko,kozlov,dzuba}, which
have only recently been addressed for Cesium. 
Using the new evaluation of the 6s-7s transition, 
the corrected experimental result for Cesium (Z=55,N=78) is~\cite{dzuba}
\begin{equation}\label{eqn-apv3}
\Qwcs =  - 72.39 \pm 0.29 (expt) \pm 0.51 (theory)  = - 72.39 \pm 0.59.
\end{equation}

\subsection{\bf Neutrino neutral to charged current ratio
\label{sec-sm_other_nc}}

 Measurements of the neutral-current (NC) to charged-current (CC) ratio in deep
inelastic $\nu$($\bar{\nu}$)-nucleon scattering can be used to make
a determination of the weak mixing angle. Note that this is a t-channel 
process, in contrast to $\eeff$, which is s-channel (for f$\neq$e). 

 The tree-level Lagrangian for weak neutral current neutrino-quark scattering is
\begin{equation}\label{eqn-nccc1}
{\cal {L}} = - \frac{\GF \rho_{0}}{\sqrt{2}}
 [ \bar{\nu}\gamma^{\mu}(1-\gamma^{5})\nu]
[\epsilon^{q}_{L}\bar{q}\gamma_{\mu}(1-\gamma^{5})q +
\epsilon^{q}_{R}\bar{q}\gamma_{\mu}(1+\gamma^{5})q ],
\end{equation}  
where any deviations from $\rho_{0}$ = 1 describe non-standard sources
of SU(2) breaking. $\epsilon^{q}_{L,R}$ are the left- and right-handed
chiral quark couplings, which contain a term ~-Q$_{q}\swsqa$. For the weak
charged current the corresponding couplings 
are $\epsilon^{q}_{L}$ = t$^{3}_{q}$ and  $\epsilon^{q}_{R}$ = 0.
Thus measurement of the NC/CC ratio gives values
of $\rho_{0}$ and $\swsqa$. In the context of the SM, this measurement
of $\swsqa$ is equivalent to a measurement of $\MW$, but outside the
SM it provides measurements of the weak couplings of light-quarks at
a momentum scale far below $\MZ$.

 The NC/CC ratio for scattering off an isoscalar nuclear target is
\begin{eqnarray}\label{eqn-nccc2}
R^{\nu(\bar{\nu})} \equiv \frac{ \sigma( \nu(\bar{\nu})N \rightarrow
 \nu(\bar{\nu})X )}
{\sigma( \nu(\bar{\nu})N \rightarrow \ell^{-(+)}X )}
 \nonumber\\
= g_{L}^{2} + r^{(-1)}g_{R}^{2},
\end{eqnarray}
  where r = $\sigma$( $\bar{\nu}$N $\rightarrow \ell^{+}$X)/
$\sigma$( $\nu$N $\rightarrow \ell^{-}$X) $\simeq$ 0.5 and
\begin{eqnarray}\label{eqn-nccc3}
g_{L}^{2} = (\epsilon_{L}^{u})^{2} + (\epsilon_{L}^{d})^{2} 
= \frac{1}{2} - \swsqa + \frac{5}{9}\swsqsq 
 \nonumber\\
g_{R}^{2} = (\epsilon_{R}^{u})^{2} + (\epsilon_{R}^{d})^{2}
= \frac{5}{9}\swsqsq.
\end{eqnarray}
To reduce the dependence on the quark-density functions, and 
the contribution of strange and other sea-quarks, 
the combination (Paschos and Wolfenstein~\cite{paschos})
\begin{equation}\label{eqn-nccc4}
 R^{PW} = \frac{ R^{\nu} - rR^{\bar{\nu}} }{1-r} 
= g_{L}^{2} - g_{R}^{2} = \frac{1}{2} - \swsqa
\end{equation}
is used. The strange sea-quark distribution is studied in dimuon CC events,
which are produced from the s$\rightarrow$c reaction.

The most recent, and most precise,
result comes from the NuTeV Collaboration who used both  neutrino and
anti-neutrino beams to 
give~\cite{NuTeV-final}\footnote{This result updates a preliminary result 
from the NuTeV and CCFR experiments of $\swsqa = \swsq \equiv 
0.2255 \pm 0.0021$.}
\begin{eqnarray}\label{eqn-nccc5}
\swsqa = 1-\MW^2/\MZ^2
 \nonumber\\
\equiv 0.2277\pm0.0016 -0.00022\frac{\Mt^2-(175~\GeV)^2}{(50~\GeV)^2}
+0.00032\ln\frac{\MH}{150~\GeV},
\end{eqnarray} 
where the explicit dependence on $\Mt$ and $\MH$ is given. This dependence,
which arises from the electroweak radiative corrections to the result,
is used in the electroweak fits discussed below. The total uncertainty
of 0.0016 has statistical and systematic components of 0.0013 and
0.0009 respectively. The largest systematic uncertainty comes from charm
production and the knowledge of the strange sea. The uncertainty in the
fraction of the $\nu_{e}$, $\bar{\nu_{e}}$ component of the beam is
another important systematic.

This result can be converted into an indirect measurement of the W mass; 
$\MW$ = 80.136 $\pm$ 0.084 GeV, using
$\MZ$ = 91.1875 GeV, $\Mt$ = 175 GeV and $\MH$ = 150 GeV.
Fig.~\ref{fig-mw_comp} shows a comparison
of the extracted $\MW$ value with the direct measurements. 
The difference between the world average direct result and that
of NuTeV is 300 $\pm$ 92 MeV, almost 3.3 standard deviations apart.
 
 Results are also given for a two-parameter fit to the R$^{\nu}$ 
and  R$^{\bar{\nu}}$ measurements, namely
\begin{equation}\label{eqn-nccc6}
                  \rho_{0} = 0.9983 \pm 0.0040, 
\hspace*{1.0cm}   \swsqa = 0.2265 \pm 0.0031, 
\end{equation}
with a correlation coefficient of 0.85. Alternatively, the results can
be expressed as
\begin{equation}\label{eqn-nccc7}
                (g^{eff}_{L})^{2} = 0.3005 \pm 0.0014, 
\hspace*{1.0cm} (g^{eff}_{R})^{2} = 0.0310 \pm 0.0011,
\end{equation}
with a correlation of -0.02.
These are to be compared to the predicted SM values, from a fit to 
other data, of
(g$^{eff}_{L})^{2}$ = 0.3042 and  (g$^{eff}_{R})^{2}$ = 0.0301.
That is, the measured (g$^{eff}_{L})^{2}$ is lower than the SM expectation
by 2.5$\sigma$, whereas (g$^{eff}_{R})^{2}$ is compatible with the SM.

\section{\bf Constraints and tests of the Standard Model
\label{sec-sm_tests}}
 The electroweak data described in the previous sections can be used
both to test the self-consistency of the Standard Model and make estimates
of the SM parameters.

\subsection{\bf Measurements of the weak mixing angle
\label{sec-sm_angle}}
\begin{figure}[htbp]
\vspace*{13pt}
\begin{center}
\mbox{
\epsfig{file=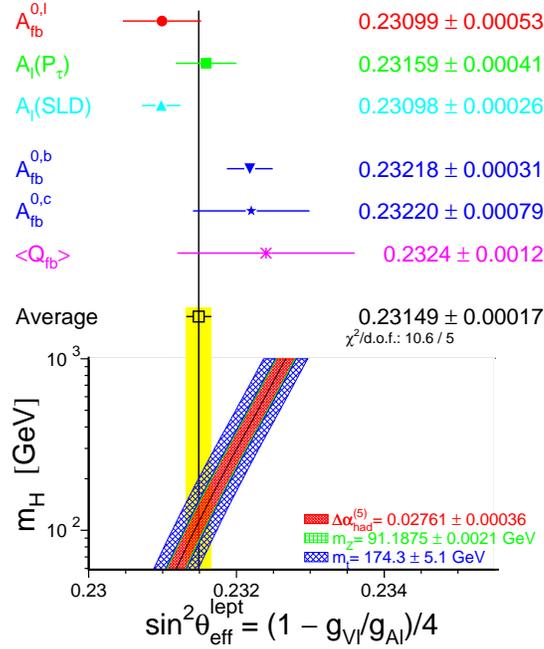,height=9cm}}
\end{center}
\caption {
Comparison of the determinations of $\swsqeffl$. Also shown is the
SM expectation as a function of $\MH$, with the bands showing the
uncertainties from other SM parameters.
}
\label{fig-seff2}
\end{figure}

 The various asymmetries discussed in section \ref{sec-Zdata} 
determine the ratios of the
vector to axial-vector couplings of the Z to one or more fermions.
From eqn.(\ref{eqn-z3}) it can be seen that the measurements thus
determine $\swsqeffff$. The value used for reference and comparison
is that for leptons, $\swsqeffl$. 
Fig.~\ref{fig-seff2} shows a comparison of the various determinations.
For the lepton asymmetries ($\Afbzl$, $\cAl(\ptau)$, $\cAl$(SLD)), $\swsqeffl$
is computed using only the assumption of lepton universality.
For the heavy quark measurements ($\Afbzb$,$\Afbzc$) a small
correction to get from  $\swsqeffff$ to $\swsqeffl$ is applied, computed
from the SM. The overall $\chit$ gives a probability of 5.9$\%$ for
agreement of the results; the main contributions to
the $\chit$ coming from the SLD $\ALR$ measurement and from $\Afbzb$.
As can also be seen from fig.~\ref{fig-seff2}, the $\ALR$ value favours a light
value of $\MH$, whereas $\Afbzb$ favours a rather heavy value. The other
measurements using quarks also favour a rather heavy Higgs mass, but
they are less precise. It can be noted that if $\cAe$ is computed
from $\Afbzb$ and the measured value of $\Ab$, using equation~(\ref{eqn-z2}), 
then the value $\swsqeffl$ = 0.23194 $\pm$ 0.00050 is extracted. This
is slightly closer to the values extracted from leptonic asymmetries.

\subsection{\bf Global electroweak fits
\label{sec-sm_fits}}

 The electroweak data discussed in this report are used in 
electroweak fits to test
the validity of the Standard Model (see section \ref{sec-SM}),
and to estimate the mass of the important
missing ingredient of the SM, the Higgs boson mass. Electroweak
quantities are sensitive through propagator and vertex correction terms,
which enter (mainly) as $\Mt^{2}$ and ln($\MH$). The relative
sensitivity is thus much higher for indirect determinations of $\Mt$
than $\MH$, but $\MH$ can be significantly constrained,
especially if $\Mt$ is specified.
As already discussed, this led to a prediction of the top-quark mass in
advance of its discovery. The present data are now precise
enough to significantly constrain the Higgs mass. It is, of course, of
considerable interest to compare the limits of direct searches for the Higgs
with those from the electroweak fits.

 The results of the direct searches for the SM Higgs boson at LEP 
are briefly discussed in sect.~\ref{sec-direct}. The 95$\%$ c.l. lower limit on
the Higgs mass is 114.1 GeV, with the possibility of a signal around
115 GeV with a significance of about 2 standard deviations.
These values are compared in this section to the values
for $\MH$ coming from electroweak fits.

 In the electroweak fits the values of the well known `constants' $\GF$
= (1.16637 $\pm$ 0.00001) 10$^{-5}$ GeV$^{-2}$~\cite{pdg2001}, 
$\MZ$ and $\alphamz$ are used. 
The fits then give values of $\Mt$, $\MH$ and $\alphasmz$. 
The SM computations
are provided by the semi-analytic programs ZFITTER and TOPAZ0, which 
contain a large amount of theoretical input, and have been thoroughly tested.
  
 The quantities which are used in the fits are:-
\begin{itemize}
\item[1)] the results of the 5-parameter Z lineshape fits 
(i.e. assuming lepton universality), from table~\ref{tab-five}
\item[2)] the 6-parameter heavy flavour fits,  from table~\ref{tab-hfres}
\item[3)] the combined LEP tau-polarisation result, from eqn.~(\ref{eqn-ptau})
\item[4)] the result for $\cAe$ from the SLD experiment, 
from eqn.~(\ref{eqn-alr})
\item[5)] the inclusive hadron charge-asymmetry $\avQfb$,
from table~\ref{partab}
\item[6)] the world average values of $\MW$ and $\GW$, 
from eqn.~(\ref{mwgw_world})
\item[7)] the combined Tevatron value of the top-quark mass, 
from eqn.~(\ref{eqn-mt1})
\item[8)] the NuTeV result for $\swsqa$, from eqn.~(\ref{eqn-nccc5})
\item[9)] the atomic physics parity violation parameter $\Qwcs$ for Cesium,
from eqn.~(\ref{eqn-apv3}).
\end{itemize}
The full set of correlations between these parameters is taken into account
in the fits.

 In order to test the consistency of the data with respect to the Standard
Model a fit is first made to all the Z-pole data (from LEP and SLD).
This is fit 1 in tab.\ref{tab-ewfits}, and gives fitted values for $\Mt$ and
$\MW$ which are reasonably compatible with the directly measured values 
of $\Mt$ = 174.3 $\pm$ 5.1 GeV and $\MW$ = 80.451 $\pm$ 0.033 GeV respectively.

 In fit 2 in tab.\ref{tab-ewfits}, the direct measurement of $\Mt$ is also
included. The value of $\MW$ derived from this fit, which
doesn't include any direct measurements of $\MW$ or $\GW$, is in reasonable
agreement with the directly measured value of $\MW$.
In order to get the best indirect estimate for $\Mt$, a fit is made
to the Z-pole data plus the direct measurements of $\MW$ and $\GW$. 
This is fit 3 in tab.\ref{tab-ewfits}, and the value of $\Mt$ derived 
from this fit is again in good agreement with the directly measured value.
Note that the $\chisq$ probabilities of all these fits are reasonably good.

\begin{figure}[htbp]
\vspace*{13pt}
\begin{center}
\mbox{
\epsfig{file=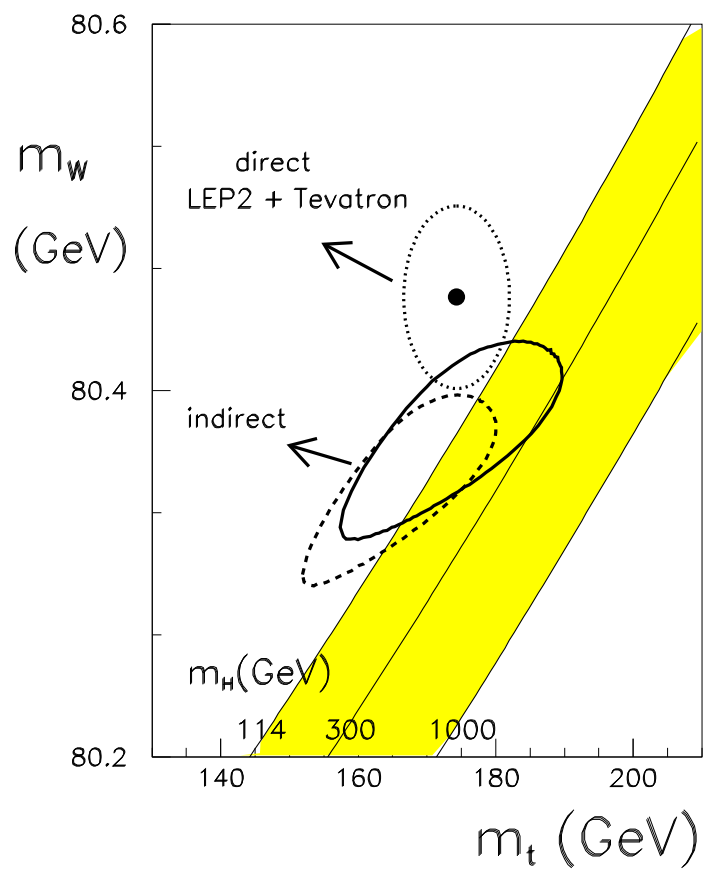,height=13cm}}
\end{center}
\caption {
Direct and indirect determinations of $\Mt$ and $\MW$.
The contours shown are for a confidence level of 70$\%$.
The central value of the direct measurements is shown as
a solid point and the contour by a dotted line.
The indirect determination which is shown as a solid line is for
all high q$^{2}$ electroweak measurements, 
except the direct measurements of $\Mt$ and $\MW$.
The results obtained when the NuTeV and APV results are also
included in the indirect determination is shown as a dashed line.
}
\label{fig-mtvmw}
\end{figure}

One of the most stringent tests of the SM, first proposed in~\cite{pbr_lp95}, is
to compare the {\it direct} measurements of $\Mt$ and $\MW$ against 
the  {\it indirect} determinations. This is shown in fig.~\ref{fig-mtvmw}.
Each point in the $\Mt$-$\MW$ plane corresponds to a unique Higgs mass,
and contours of fixed values of $\MH$ are also shown. 
The indirect determination corresponding to the solid line includes
only the high q$^{2}$ data. It can be seen that the direct and indirect
70$\%$ c.l. contours have only a small overlap. If the NuTeV and 
APV results are included then the 70$\%$ c.l. contour for the indirect
determination (dashed line) does not overlap with that from the direct 
measurement.
If the central values were to remain the same, then improved precision 
could indicate a breakdown of the SM. 
It can also be seen in fig.~\ref{fig-mtvmw} that, in all cases, 
the data favour a low Higgs mass. 
Furthermore, the region in the $\Mt$-$\MW$ plane where both the direct
and indirect measurements are situated is just that region expected 
in many SUSY models.

\begin{table}[htbp]
\caption[]{
Results of the electroweak fits to high q$^{2}$ data (see text). The value
$\alphasmz$ = 0.119 $\pm$ 0.003 is obtained in these fits.
}\label{tab-ewfits}
\begin{indented}
\lineup
\item[] 
\begin{tabular}{llll}\br
quantity      & fit 1                  & fit 2                  
& fit 3 \\ \br
$\Mt$(GeV)    & 170.6 $^{11.4}_{-9.0}$ & 173.6 $\pm$ 4.6     
& 180.8 $^{10.9}_{-8.5}$   \\
$\MH$(GeV)    & 82 $^{+109}_{-41}$     & 100 $^{+64}_{-41}$  
& 121 $^{+166}_{-65}$      \\
$\MW$(GeV)    & 80.372 $\pm$  0.033    & 80.379 $\pm$  0.023 
& 80.411 $\pm$  0.023      \\
$\chisq$/df   & 15.3 / 10  (12$\%$)    & 15.4 / 11 (17$\%$)          
& 18.5 / 12 (10$\%$)       \\
\br
\end{tabular}
\end{indented}
\end{table}

\begin{figure}[htbp]
\vspace*{13pt}
\begin{center}
\mbox{
\epsfig{file=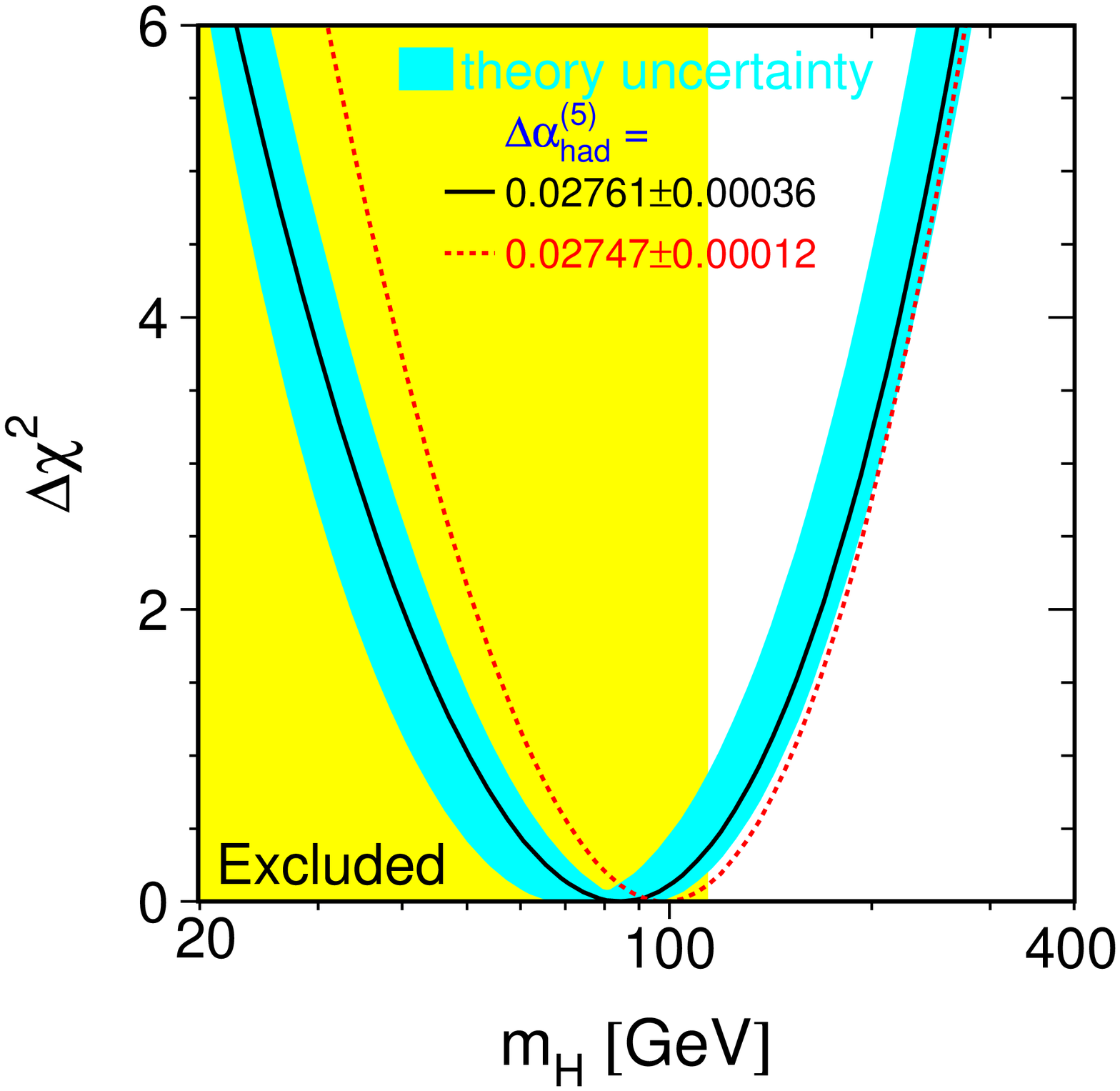,height=9cm}}
\end{center}
\caption {$\Delta\chisq$ as a function of $\MH$ for the global fit to all
electroweak data. The shaded region is excluded by direct searches at LEP.
The band shows an estimate of the theoretical uncertainty.
}
\label{fig-blueband}
\end{figure}

 The fits given in tab.\ref{tab-ewfits} only include electroweak 
data corresponding to a scale q$^{2} \simeq \MZ^{2}$. 
That is, items 8) and 9) in the list of quantities
above, which correspond to a scale much below $\MZ^{2}$, are not included.
The fitted parameters from a {\it global} fit to all the electroweak data,
that is items 1)-9) in the above list,
are given in tab.\ref{tab-global}. The $\chisq$ with respect to the
minimum value, as a function of the Higgs mass $\MH$, is shown 
in fig.~\ref{fig-blueband}.
The central value for $\MH$ is slightly below the lower limit for direct
searches at LEP (114 GeV at the 95$\%$ c.l.). Taking into account the
theoretical uncertainty (shown as a band), the one-sided 95$\%$ c.l. 
is $\MH \leq$ 196 GeV. If the value 
$\Delta \alpha^{(5)}_{had}$(s) = 0.02747 $\pm$ 0.00012~\cite{troc01} is 
used, then the upper limit becomes 199 GeV.

\begin{table}[htbp]
\caption[]{
Results of the global electroweak fit.
The $\chisq$/df is 29/15, a probability of 1.7$\%$.
The value for $\MW$ is that derived from the fit.}\label{tab-global}
\begin{indented}
\lineup
\item[] 
\begin{tabular}{lll}\br
quantity      & fitted value & error \\ \br
$\Mt$(GeV)    & 174.7    & 4.4      \\
$\alfas$      & 0.118    & 0.003    \\
$\MH$(GeV)    & 85       & $^{+54}_{-34}$  \\
\br
$\MW$(GeV)    & 80.394   & 0.018    \\
\br
\end{tabular}
\end{indented}
\end{table}

\begin{figure}[htbp]
\vspace*{13pt}
\begin{center}
\mbox{
\epsfig{file=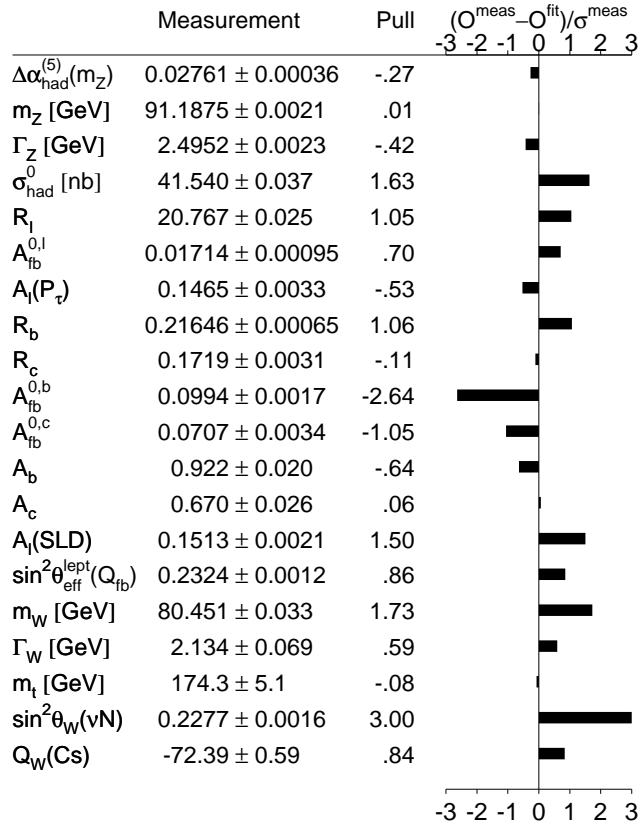,height=13cm}}
\end{center}
\caption {
Pull distribution for the global electroweak fit.
}
\label{fig-pulls}
\end{figure}

 The pull values of the fitted quantities are shown in 
fig.~\ref{fig-pulls}. The pull for an observable O$_{i}$ is defined 
as (O$_{i}^{\rm meas}$ - O$_{i}^{\rm fit}$)/$\sigma_{i}^{\rm meas}$.
It can be seen that the largest pulls are
3.00 for the NuTeV result and -2.64 for $\Afbzb$.
The rms of the pull values is 1.28. This quantity has a statistical
uncertainty of 0.17, and so is reasonably compatible with unity.

\begin{figure}[htbp]
\vspace*{13pt}
\begin{center}
\mbox{
\epsfig{file=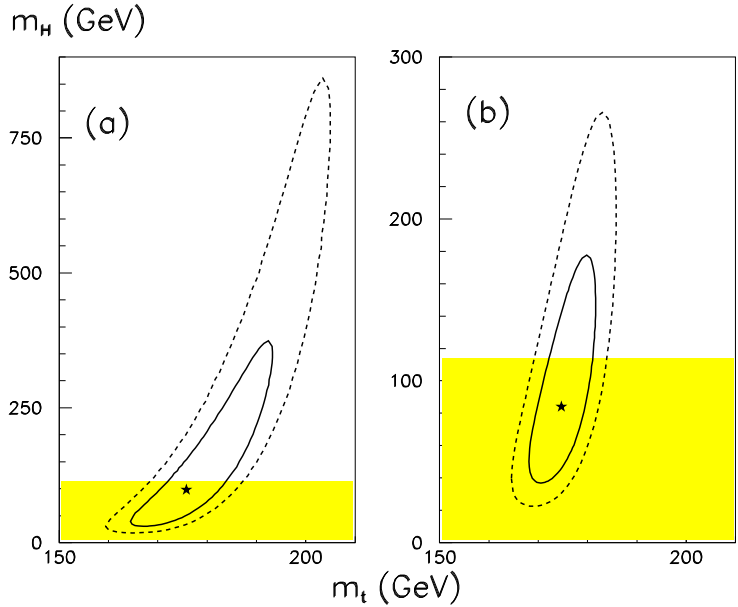,height=9cm}}
\end{center}
\caption { Fitted values of $\Mt$ and $\MH$ from a fit to a) all electroweak
data except $\Mt$ and b) all  electroweak data, 
together with the 70 and 95$\%$ confidence level contours. The shaded region
is that excluded by the direct Higgs search at LEP.
}
\label{fig-mtmh}
\end{figure}

 The 70 and 95$\%$ confidence level contours for the parameters 
$\Mt$ and $\MH$ are shown in fig.~\ref{fig-mtmh}. It can be seen that
there is a strong correlation ($\simeq$ 0.7) between $\Mt$ and $\MH$.
The importance of using the direct measurement of $\Mt$
in improving the constraints on $\MH$ can also be seen.

 Although the global electroweak fits which have been carried out in recent
years have always favoured a light Higgs mass, the central
fitted values, and the upper limits, have undergone some important changes. 
For example, at the time of the ICHEP 2000 Conference in Osaka~\cite{EWWG00}, 
the fitted value was $\MH$ = 60 $^{+52}_{-29}$ GeV, 
with a $\chisq$ probability of 14$\%$. The main input changes in the fits
since then are the updates on the measurement of $\MW$, the heavy flavour data, 
the NuTeV NC/CC ratio, the inclusion of $\Qwcs$ 
and the new value of $\alphamz$. The effect of the inclusion of each
of these separately with respect to the Osaka data is as follows. 
The updated $\MW$ result alone changes the Osaka $\MH$ value by -6 GeV, 
while the new heavy flavour data change it by +3 GeV. The inclusion
of the new NuTeV NC/CC ratio changes the Osaka $\MH$ value by +3 GeV
(but the $\chisq$ increases significantly), whereas
the inclusion of $\Qwcs$ increases $\MH$ by +2 GeV.
However, the largest effect comes from the new $\alphamz$, which with
the Osaka data gives $\MH$ = 88 GeV, i.e. a change of +28 GeV.
Thus the overall change of +25 GeV with respect to the Osaka input data 
comes largely from $\alphamz$. Note that the correlation between $\alphamz$
and $\MH$ is about -0.5, so an increase in the value of $\alphamz$ used
leads to a decrease in the fitted $\MH$ value.

As an indication of the importance of the external constraint on $\alphamz$,
a fit to all electroweak data, but without the $\alphamz$ constraint, 
gives $\MH$ = 25 $^{+39}_{-11}$ GeV. $\alphamz$ is strongly correlated
with $\Mt$ and $\MH$, with correlation coefficients of 0.52 and -0.80
respectively.

\begin{figure}[htbp]
\vspace*{13pt}
\begin{center}
\mbox{
\epsfig{file=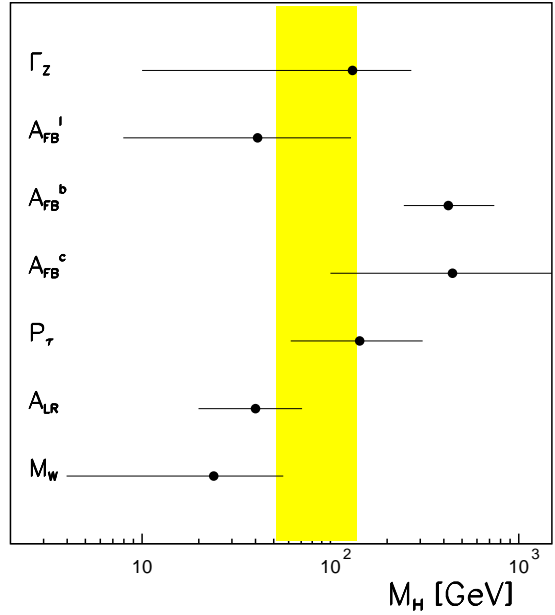,height=9cm}}
\end{center}
\caption { Sensitivity of individual electroweak measurements
to the Higgs mass, $\MH$. The error bars shown correspond to
one standard deviation. The one standard deviation results of the fit to
all the data is shown as a vertical band.
}
\label{fig-higgs-sens}
\end{figure}

 The various electroweak quantities have different sensitivities to $\MH$.
This can be seen in fig.\ref{fig-higgs-sens}, where the results from
the seven most sensitive individual measurements, M$_{i}$, are displayed. 
In each of these fits only the measurement in question is used and
the values  $\MZ$ = 91.1875 GeV, $\Mt$ = 174.3 GeV, 
$\alphasmz$ = 0.118 and $\Delta \alpha^{(5)}_{had}(\MZ)$ = 0.02761
are imposed. If these quantities were allowed to vary in the fit
then their correlations with $\MH$ would be different for each M$_{i}$. 
Thus fig.\ref{fig-higgs-sens} displays only part of the influence
of the observables M$_{i}$ in the overall fit. It is of interest to
quantify which measured quantities are currently the most sensitive 
to $\MH$. To see this the measured quantities are set to the SM values,
corresponding to $\MH$ = 300 GeV, and the above single measurement fits
repeated. The sensitivities, in decreasing order of sensitivity, are
$\MW$ (0.19), $\ALR$ (0.20), $\Afbzb$ (0.24), $\GZ$ (0.27), 
$\ptau$ (0.33), $\Afbzl$ (0.45) and $\Afbzc$ (0.65), where the quantities
in parentheses are the fit errors on log$_{10}$($\MH$).

 What conclusions can be drawn about $\MH$ from the above results ? There
are two, or more, points of view. Firstly, one can note that the 
overall $\chisq$ of 29/15 df has a probability of 1.7$\%$. 
Although this is somewhat on the low side,
we expect some statistical fluctuations, even
if all the measurements are reliable and taken at face value.

\begin{table}[htbp]
\caption[]{
Sensitivity of fitted $\MH$ value and upper limit to input data. The NuTeV
measurement is not included in these fits.
}\label{tab-sensitivity}
\begin{indented}
\lineup
\item[] 
\begin{tabular}{lcc}\br
    fit  & $\MH$ GeV &   95$\%$ c.l. (GeV) \\
\br
standard                                  &  81$^{+49}_{-32}$ & 174 \\
if exclude A$_{LR}$                       & 108$^{+66}_{-43}$ & 233 \\
if exclude $\Afbzb$ and $\Afbzc$          &  43$^{+32}_{-18}$ & 106 \\
if exclude A$_{LR}$,$\Afbzb$ and $\Afbzc$ &  48$^{+44}_{-23}$ & 135 \\
if scale errors                           &  75$^{+58}_{-35}$ & 189 \\
\br
\end{tabular}
\end{indented}
\end{table}

 An alternative approach is to conclude that the large 
overall $\chisq$ is either indicating a breakdown of the SM or that
one, or more, measurements contributing to the $\chisq$ are, at some level,
incorrect. 
 
  The largest contribution to the overall $\chisq$ (9 out of
29) comes from the NuTeV NC/CC ratio. If this measurement is
removed from the fit the $\chisq$ reduces to 20 (probability of 14$\%$) and
the fitted value of $\MH$ is not significantly changed: it is reduced
by about 3 GeV. The NuTeV measurement is of great interest,
in that it might be an indication of physics beyond the Standard Model.
However, the present accuracy, measured in terms of the equivalent 
uncertainty on value of $\MW$, is such
that the inclusion of the result does not significantly change 
the extracted $\MH$ value, but it does
increase substantially the $\chisq$. To investigate the effects of the
other measurements, the NuTeV result is omitted in the following considerations.

 For the remaining  measurements it is mainly the quantities most sensitive
to $\MH$ which give the largest contributions to $\chisq$. 
Namely, $\Afbzb$, $\Afbzc$, $\ALR$, $\MW$, $\Afbzl$, $\GZ$ and $\ptau$ 
contribute 15 to total $\chisq$ of 20, so maybe
one should scale these errors by their $\sqrt{\chisq/(df-1))}$ = 1.6 ?
Tab. \ref{tab-sensitivity} gives
the results for the central values and 95$\%$ c.l. upper limits, without
consideration of the theoretical uncertainty, for a series of fits. 
It can be seen that the
95$\%$ c.l. upper limit increases to 233 GeV if the $\ALR$ measurement is
excluded from the fit, but decreases to 106 GeV if $\Afbzb$ and $\Afbzc$
are both excluded.
Excluding both  $\ALR$ and the heavy flavour asymmetries $\Afbzb$ and $\Afbzc$
still favours a light Higgs. If the errors of the
Higgs sensitive quantities are scaled, as discussed above, 
then the central value does not
change much, but the 95$\%$ c.l. upper limit increases to 189 GeV.

\subsection{\bf Direct Higgs search and limits and electroweak fits}
\label{sec-direct}

 In the above discussion neither the lower limit on the direct search for
the SM Higgs boson at LEP 2, nor the possible observation of a signal, 
are taken into account in the electroweak fits. A signal, with a mass of about
115 GeV and compatible with being from the SM Higgs boson, has been found 
at LEP, with a statistical significance of about 2 standard 
deviations~\cite{gianotti}. 

 In the experimental analysis, the main channel searched for 
is $\ee \rightarrow$ ZH, followed by the decay H $\rightarrow \bb$.
This is a threshold process, so the searches for this Z$\bb$ topology in
the highest cms energy runs at LEP 2 ($\roots$ up to 209 GeV)
were the most important. 
In the experimental analysis events are classified in terms of the 
mass $\MH$ of a potential Higss boson and a variable ${\cal G}$ quantifying the
Higgs-like nature of the event. This likelihood variable ${\cal G}$ is 
constructed to be large for a Higgs-like topology and small for a background 
topology. A likelihood ratio ${\cal L}_{s+b}$/${\cal L}_{b}$ is constructed, 
as a function of $\MH$, where b is the hypothesis of background only, and
s+b that for a Higgs signal plus background. This test statistic amounts 
essentially to the difference in the $\chisq$ between the two hypotheses, 
and shows a possible Higgs signal at a mass around 115 GeV~\cite{gianotti}.

 In the analysis the probablility is calculated, as a function of $\MH$,
that the background could fluctuate to the level seen in the data, 
and whether the excess is compatible with the expected production rate 
of the SM Higgs boson. This latter point is well defined, since the 
production rate is known if $\MH$ is specified. 
Note however that, in addition to the possible signal at 115 GeV, there
are other signals of smaller significance. Furthermore, nothing can be said
about the region beyond the kinematic range of LEP. Indeed, further Higgs
bosons could exist at higher masses, as expected in SUSY models. In this case
the SM would be shown to be invalid. 

 In electroweak fits, it is assumed that there is just one Higgs boson 
of unknown mass. The Higgs mass which gives the
best fit to all the data is found and, provided the 
overall probablility that the SM fits the ensemble of electroweak data 
is satisfactory, the value or limits found are the best (indirect) 
estimates of $\MH$, within the context of the SM. 
Thus the statistical question addressed in the electroweak fits is rather 
different to that in the direct search, and so the information is 
difficult to combine. 
However, it should be noted that the electroweak fits are certainly 
compatible with a Higgs of mass 115 GeV.
 
 The lower limit on $\MH$ from the direct search at LEP 2 is about 114 GeV,
at the 95$\%$ confidence level. Since the production cross-section for 
the  Higgs boson falls off rapidly with increasing Higgs mass, the limit 
for lower Higgs masses is much more stringent.
This lower limit can be represented essentially as a `brick wall' in the
$\chisq$ of the electroweak fit, in that we know that the SM Higgs cannot be 
significantly below the lower limit, or else it would have been observed.

\begin{figure}[t]
\vspace*{13pt}
\begin{center}
\mbox{
     \includegraphics[width=0.48\linewidth]{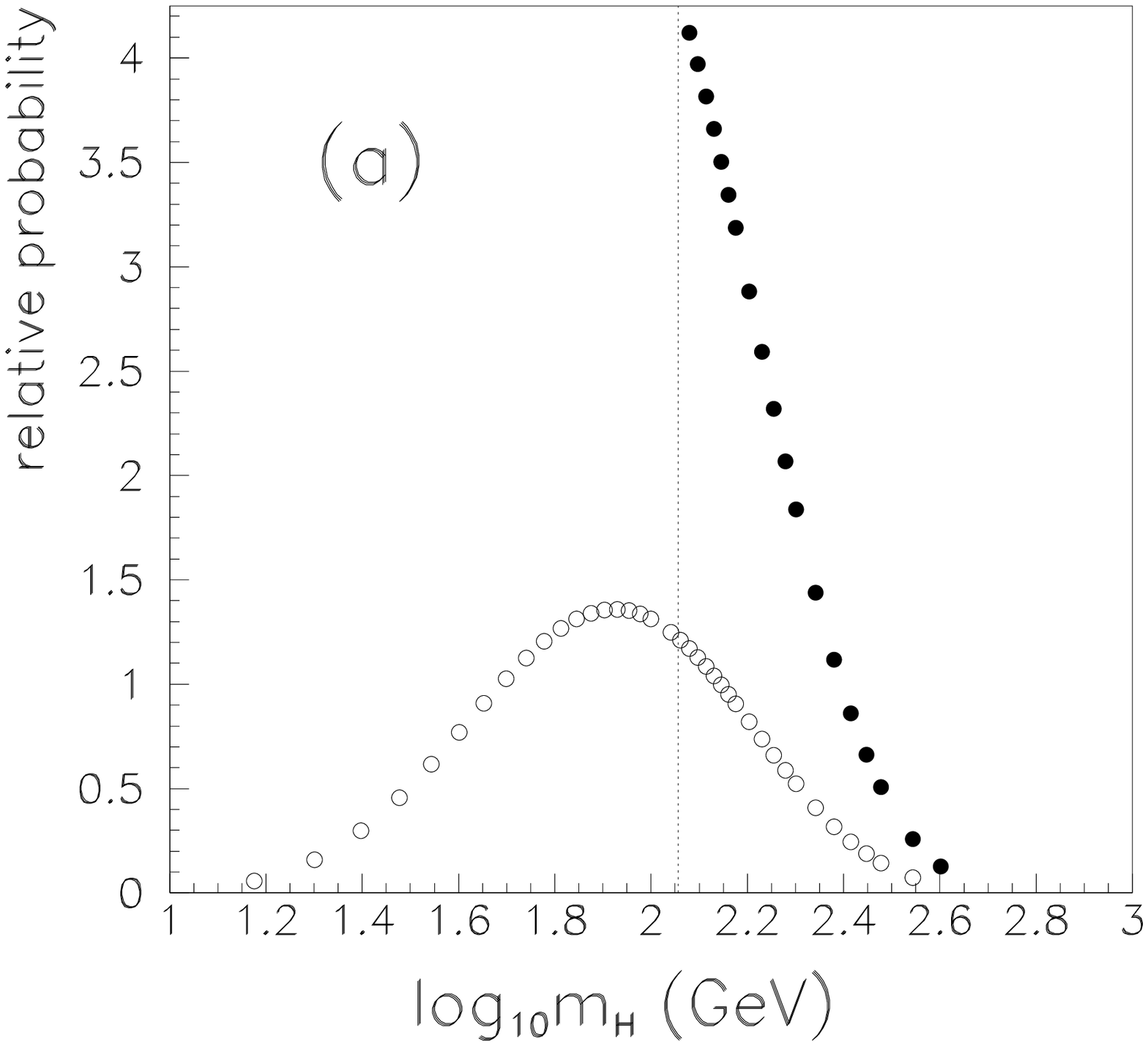}
     \includegraphics[width=0.48\linewidth]{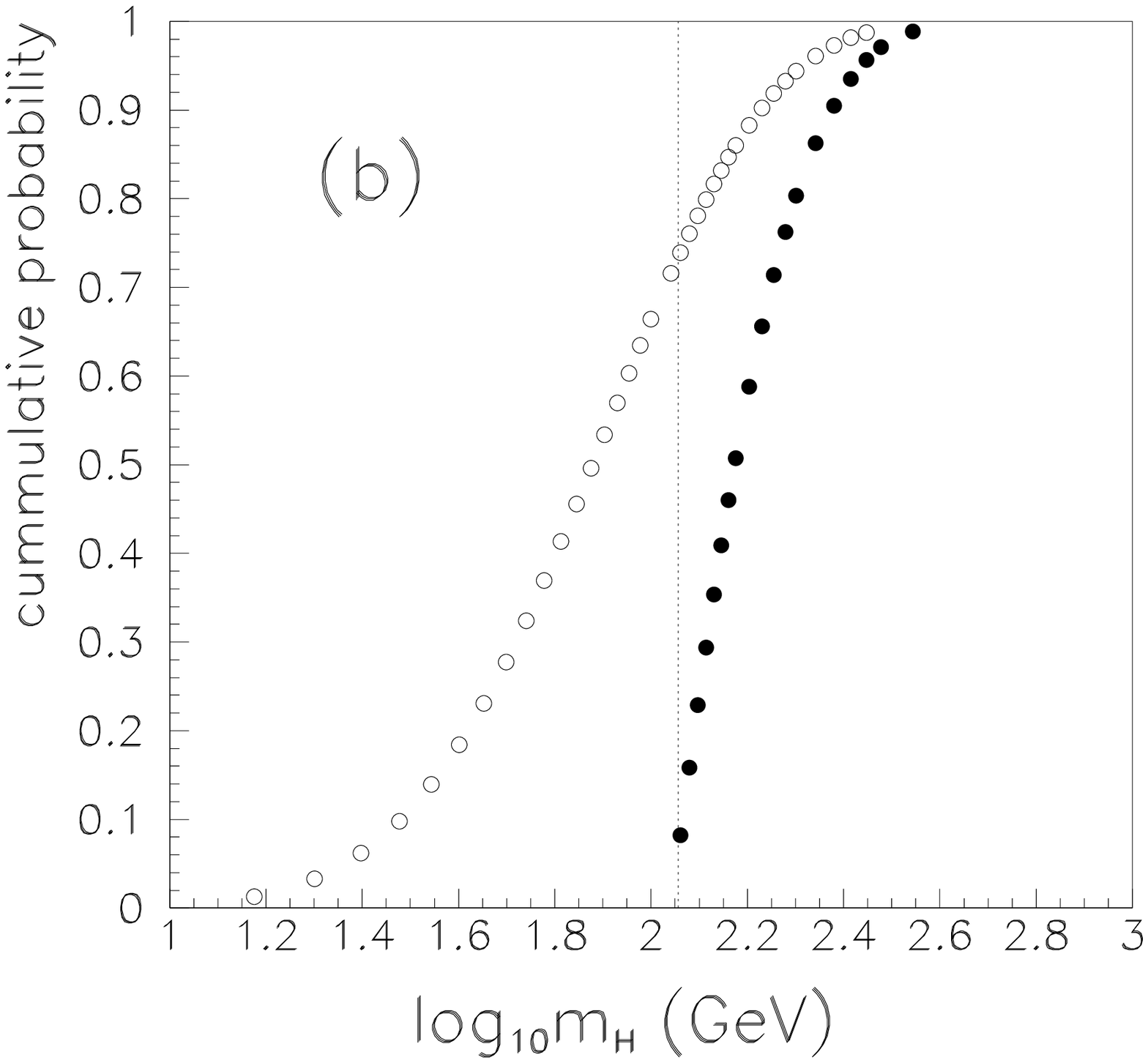}}
\end{center}
\caption{ Relative probabilty (a) and cumulative 
probabilty (b), as a function of log$_{10}$($\MH$),
for the cases where no lower limit is imposed (solid circles) and
where a lower direct limit of 114 GeV is imposed. This limit is shown as a
dashed line.
}
\label{fig-higgsprob}
\end{figure}

 The non observation of the SM Higgs can be incorporated into the electroweak
fits, provided additional assumptions are made. It is first assumed that the
{\it prior} probability for the Higgs boson is uniform as a function of
log($\MH$). The $\chisq$ probabilities for the electroweak fits to all data
are then computed,
as a function of $\MH$,  between some very low mass value (in practice 
3 GeV is used here) and 1000 GeV, where the theory breaks down.
The differential or relative probablity distribution, as a function 
of log($\MH$), is then constructed, such that the overall probablity that 
the Higgs is in this mass range is unity. This is because, in the absence 
of further information, the SM Higgs should exist in this range.
The experimental lower limit is taken into account by computing the 
probability again, this time from a lower limit of 114 GeV up to 1000 GeV.
This probability is then normalised to unity, as we are assuming now that
the SM Higgs is in this mass range. The distributions of the probabilities
obtained, with and without the use of the lower limit, are shown 
in fig.\ref{fig-higgsprob}. From the cumulative probability distribution
a 95$\%$ 
confidence level upper limit can be extracted. 
This is about 215 GeV for the case where no lower direct limit 
is imposed, and rises to about 275 GeV when the direct limit is imposed.
This increase in the limit is just a consequence of restributing the
unit probability into a more restricted area.

 In summary, from the considerations in this and previous sections,
the best estimate for the  Higgs mass is that it is relatively
light. However, the data are not fully compatible, so some caution should
be made in drawing conclusions.

\subsection{\bf Further considerations of the NuTeV result}\label{sec-further}

 As discussed in section \ref{sec-sm_other_nc}, the main discrepancy from
the NuTeV data with respect to the SM is in the value 
of g$_{L}^{2}$, which is about 1$\%$ below the SM prediction.
A recent review of both the theoretical corrections needed for the 
measurement, and for possible interpretations in terms of physics beyond
the SM, can be found in ~\cite{davidson}. The NuTeV analysis assumes
that $s = \bar{s}$, for the strange-sea. It is suggested in ~\cite{davidson}
that if this equality is violated, such 
that $s - \bar{s} \simeq$ 0.002, as obtained from neutrino dimuon data,
then a good fraction of the anomaly can be explained. However, NuTeV 
dimuon data give $s - \bar{s} \simeq$ -0.0027 $\pm$ 0.0013. That is, 
the measured NuTeV asymmetry has the opposite sign, and using this value
would increase the significance of the anomaly.

 It is interesting to note that the NuTeV data, when interpreted in terms
of a deviation of the neutrino NC rate, gives
$\rho^{2}$ = 0.9884 $\pm$ 0.0042~\cite{mcfarland}. That is, 
the $\nu$ couplings are (1.16 $\pm$ 0.42)$\%$ less than 
their SM values. From the results given in
section \ref{sec-lep_comb}, it can be computed that the ratio of the 
invisible width of the Z-boson to the SM value (i.e. for N$_{\nu}$ = 3) is
0.9947 $\pm$ 0.0028, which is (0.53 $\pm$ 0.28)$\%$ less than the SM value.
It worth pointing out that the energy scales probed are very different;
with q$^{2} \simeq$ -20 GeV$^{2}$ (t-channel) for the neutrino beam, 
and q$^{2} \simeq$ 8300 GeV$^{2}$ (s-channel) at LEP. 
The current precision of the
data is not sufficient to draw any firm conclusion on a possible violation
of the SM for the neutrino NC couplings.

 Various possibilities in terms of physics beyond the SM, which might
explain the  NuTeV anomaly (if it is taken to be real) are discussed 
in ~\cite{davidson}.
In general it is found that models which preserve a fair degree of 
symmetry have difficulty in explaining the results. 
SUSY models, for example, give effects which are too small in magnitude and of
the wrong sign. Only models in which the couplings are much more {\it ad hoc}
can be tuned to fit the data, but these tend to have a large number of
parameters.

\subsection{\bf Quasi model-independent variables}

 The information content of all of the precision electroweak quantities
can, to a very good approximation, be described by four quasi model-independent
quantities $\eone$,$\etwo$,$\ethree$ and $\eb$~\cite{epsilon}, 
plus $\alphasmz$. 

 The $\epsilon$'s are defined as follows:
\begin{eqnarray} 
 \eone & = & \Drho  \\
\etwo  & = & \czt\Drho + \frac{ \szt\Drw}{( \czt - \szt)} - 2\szt\Dkap \\ 
 \ethree & = & \czt\Drho + ( \czt - \szt )\Dkap  
\end{eqnarray}
where the axial-vector and vector $\Zzero$-lepton couplings 
are (see eqn.~\ref{eqn-z3})
\begin{equation}
 \Alll = - \frac{1}{2} ( 1 + \Drho/2 ) 
\end{equation}
and
\begin{equation}
 \Vlll/\Alll =  1 - 4\swsqeffl =  1 - 4( 1 + \Dkap )\szt \ \ , 
\end{equation}
with
\begin{equation}
 \szt\czt = \frac{\pi\alphamz}{ \roottwo \GF\MZ^{2} } .
\end{equation}
Numerically, $\szt$ = 0.23118. The quantity $\Drw$ is given by $\MW$
\begin{equation}
 ( 1 - \frac{\MWS}{\MZS})\frac{\MWS}{\MZS} = 
\frac{\pi\alphamz}{\roottwo\GF\MZS(1 - \Drw) } .
\end{equation}
 Inserting the value of $\szt$, the expression for $\ethree$ becomes
\begin{equation}
 \ethree =  0.77\Drho + 0.54\Dkap \ \ . 
\end{equation}

In this approach it is assumed that new physics, beyond the SM, enters
through gauge-boson propagator (vacuum polarisation) functions
and/or vertex corrections to
the $\Zzero\lept$ vertex, assuming lepton universality. The 
effects arising through propagator contributions are often
called {\it oblique} electroweak corrections.

Note that the measurements of the total or partial $\Zzero$ widths, which 
constrain $\Drho$, contribute to both $\eone$,$\etwo$ and $\ethree$, whereas
measurements of $\seff$, from forward-backward or other asymmetry
measurements, contribute only to $\etwo$ and $\ethree$. 
The direct measurement of the W mass determines $\Drw$, and this only
enters in the variable  $\etwo$.

 Data on the $\Zzero\qqb$ vertices can also introduced if it is
further assumed that all additional deviations from the SM are contained in
the vacuum polarisation functions and/or the $\Zzero\bbbar$ vertex.
Further, $\alphasmz$ must also be introduced to describe the QCD corrections.
A new parameter $\eb$ is needed to describe the $\Zzero\bbbar$ vertex
and is defined such that
\begin{equation}
 \Abcoup = \Alll ( 1 + \eb ) 
\end{equation}
and
\begin{equation}
  \frac{\Vb}{\Abcoup}  = \frac{ 1 - \frac{4}{3}\swsqeffl + \eb } { 1 + \eb } .
\end{equation}
The determination of $\eb$ comes essentially from the measurement of $\Rb$.
As noted previously, the forward-backward asymmetries for heavy-quarks
depend mainly on $\cAe$. For the other quarks, small corrections are
made in order to obtain $\Drho$ and $\Dkap$  from  the extracted values of the
quark vector and axial-vector couplings. These corrections are computed
in the SM. Corrections are made for QCD effects where necessary,
using the ZFITTER package. With these additions, all the LEP and SLD 
measurements can be included in the fits.
 
 In the SM, the main dependence on $\Mt$ is either quadratic or logarithmic,
whereas the dependence on $\MH$ is only logarithmic. The parameter $\eone$ 
depends on $\Mt^2$ and ln($\MH$), $\etwo$ depends
on ln($\Mt$) , $\ethree$ depends on ln($\MH$) and ln($\Mt$) 
and $\eb$ depends on $\Mt^2$. The SM contributions to the $\Zzero\bbbar$ vertex
come from  t-W-b loops.

 Of course, the usefulness of these variables, or the closely related
S,T and U variables~\cite{stu}\footnote{These are related approximately
as follows: $\eone$ = $\alpha$T ,
$\etwo = - \alpha\rm{U}/4\seff$ 
and $\ethree = \alpha\rm{S}/4\seff$. Other roughly equivalent formalisms 
exist in the literature.}, is in studying physics beyond the SM. 
The aim is thus to reduce the errors on the $\epsilon$'s to be as
small as possible, so as to be as sensitive as possible to such physics.

\begin{figure}[htbp]
\vspace*{13pt}
\begin{center}
\mbox{
\epsfig{file=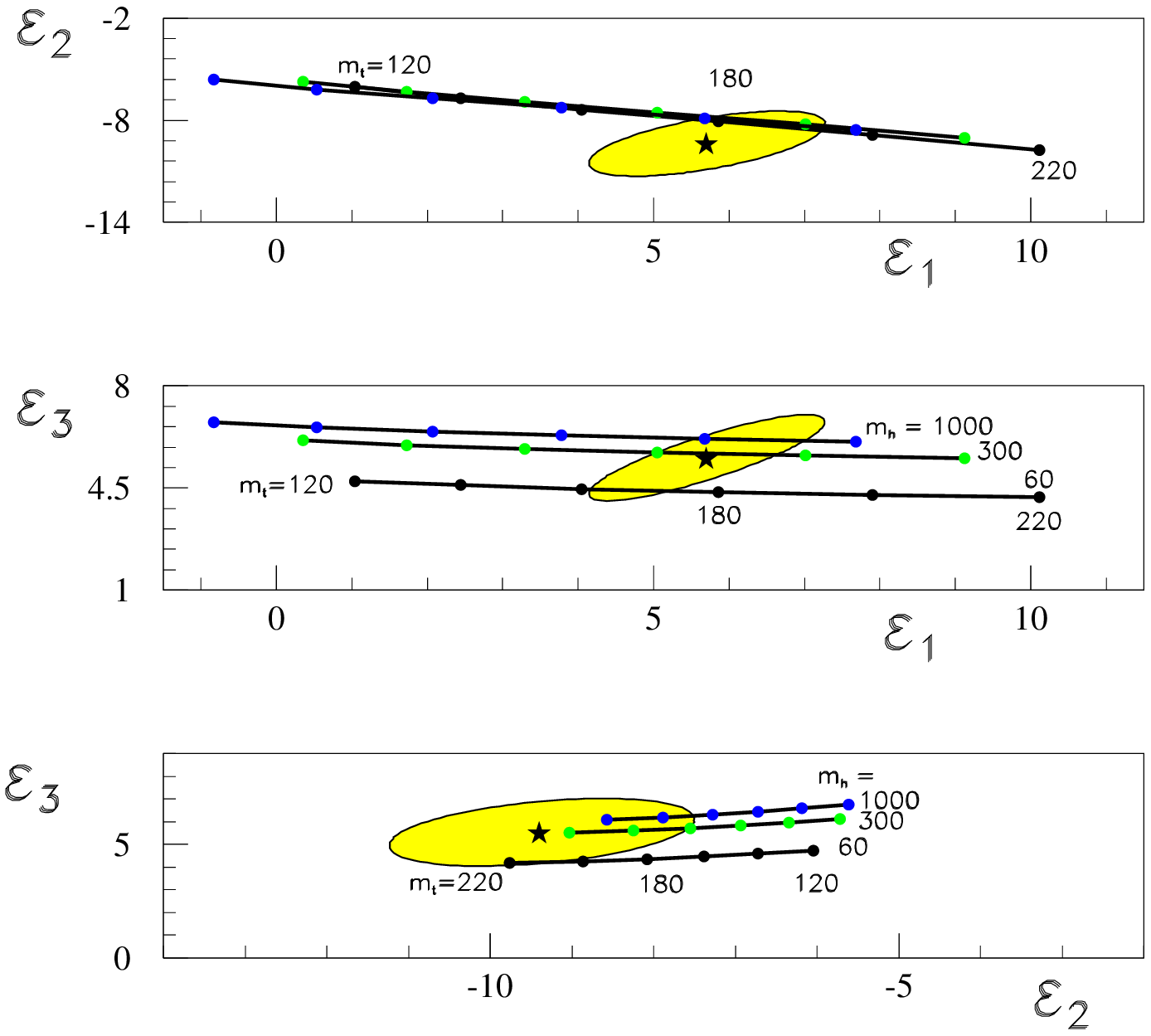,height=15cm}}
\end{center}
\caption
{ Results of a fit to $\eone$, $\etwo$ and $\ethree$, in units
of 10$^{-3}$, showing 
the 70\% confidence level contours and the expectations in the SM,
for different values of $\Mt$ and $\MH$. 
For $\eone$ versus $\etwo$ there is little sensitivity to $\MH$.}
\label{fig-eps123}
\end{figure}

 The result of a fit to the electroweak data is given in table~\ref{tab-eps}
\footnote{An indication of how the experimental
situation has evolved  can be seen by comparing the fits here with those
in ~\cite{pbrzpc,pbrjmpa}.}.
The contour plots for the 70\% confidence levels for $\eone$, $\etwo$
and $\ethree$ are also shown in fig.~\ref{fig-eps123}.
The direct top-quark mass measurement and the NuTeV result are not included,
the latter because the result is expressed as a function of $\Mt$ and $\MH$.
The value of $\alphamz$ is constrained according to eqn.(\ref{alfa3}).
The value of $\alphasmz$ is fitted from the data. If the constraint
$\alphasmz$ = 0.118 $\pm$ 0.002 is imposed, then the central values of the
$\epsilon$ variables are shifted by only a small amount.
The results are, as expected, compatible with the SM expectations at the
values of $\Mt$ and $\MH$ discussed in previous sections 
(see fig.~\ref{fig-eps123}). 
From table~\ref{tab-eps} it can be seen that the value of
$\eb$ is strongly correlated to that of $\alphasmz$.
In the SM, $\eb$ = -5.8 10$^{-3}$ for $\Mt$ = 175 GeV, and is essentially
independent of $\MH$.

 The variable $\eone$ is sensitive to new physics which violates the 
custodial weak isospin symmetry, whereas $\ethree$ is an isospin
symmetric observable. 
The $\epsilon$ (or S,T,U) variables have been used to place severe constraints
on some {\it technicolour} theories. In these theories (see e.g.~\cite{farhi})
an alternative approach to symmetry breaking is advocated, in which the
symmetry breaking arises from strongly interacting technicolour effects.
Technicolour is analoguos to QCD, but with an expected characteristic scale
of $\Lambda_{TC} \simeq$ 500 GeV, compared to the QCD scale
$\Lambda \simeq$ 200 MeV. There should also be a spectrum of
techniparticles (techni$\pi$, techni$\rho$ etc.), none of which have so
far been observed. The most reliable predictions are for the 
variable $\ethree$. For the simplest cases~\cite{stu}, the deviation
from the SM value of $\ethree$ is given in terms of the number of
technicolours $\NTC$ as
\begin{eqnarray}\label{eqn-techcol}
\Delta\ethree \simeq \left\{\begin{array}{ll} 
                   0.0035 + 0.0009( \NTC  -4 ) & \\
                   0.013  + 0.003 \ ( \NTC  -4 ), & \\                   
                  \end{array}
                  \right.
\end{eqnarray} 
where the first line is for a doublet of $\NTF$ = 2 technifermions, 
and the second line for a full technigeneration ($\NTF$ = 8). An alternative
formulation~\cite{stu} gives $\Delta\ethree \simeq 0.0024(\NTF/2)(\NTC/3)$,
giving values close to those of eqn.(\ref{eqn-techcol}).

 The experimental limit on $\Delta\ethree = \ethree$(meas) - $\ethree$(SM) 
is $\Delta\ethree \leq$ 1.5 10$^{-3}$, at the 95$\%$ c.l. Thus, within the
context of this technicolour model, a full technigeneration can be excluded, 
and a single doublet of technifermions is almost excluded. 
However, in developments of the technicolour ideas in  
{\it walking technicolour}~\cite{walktc}, and other variants, these 
limits can be evaded, since $\Delta\ethree$ can be smaller or even
negative. Note that, in general, QCD-like models have difficulties in 
preventing flavour-changing neutral-currents or in giving too fast a rate 
for proton decay.

 An alternative scenario for new physics is the existence of a fourth 
generation of fermions, which are heavy on the scale of $\MZ$. These
would contribute through $\ffb$ loop corrections to the propagators. 
A complete heavy degenerate fourth generation would
contribute $\Delta\ethree$ = 0.0017, and so is at the limit of exclusion
at the 95$\%$ c.l.

 The effect of SUSY depends on the mass spectrum of the SUSY particles.
In the `heavy' limit, where all the s-particles are rather massive,
then the MSSM predictions tend to reproduce the SM results corresponding
to a light Higgs in the region $\MH \simeq$ 100 GeV (see~\cite{barbieri}).
This scenario is, of course, compatible with the electroweak data.

\begin{table}[htb]
\caption[]{Results of the fit to the $\epsilon$ variables to all electroweak
data, with the exception of the direct measurement of $\Mt$. The correlation
matrix is also given. The $\chit$ for the fit is 16/10 df,
a probability of 11\%.
}\label{tab-eps}
\begin{center}
\renewcommand{\arraystretch}{1.1}
\begin{tabular}{ccrrrrr}\br
parameter & fitted value  & $\eone$ &$\etwo$ &$\ethree$ 
&$\eb$ &$\alphasmz$ \\
\br
$\eone$ & $(5.7\pm1.0) 10^{-3}$       & 1.00 & 0.62 & 0.86 &  0.00 & -0.37 \\
$\etwo$ & $(-9.4\pm1.2) 10^{-3}$      &      & 1.00 & 0.40 &  0.00 & -0.26 \\
$\ethree$ & $(5.5\pm1.0) 10^{-3}$     &      &      & 1.00 &  0.02 & -0.27 \\
$\eb$ & $(-4.2\pm1.6) 10^{-3}$        &      &      &      &  1.00 & -0.64 \\
$\alphasmz$ & $0.115\pm0.004$ &    &      &      &       &  1.00 \\
\br
\end{tabular}
\end{center}
\end{table}

\section{\bf Future prospects
\label{sec-future}}

\begin{table}[htbp]
\caption[]{
Effect of expected improved precision on $\MH$.  
}\label{tab-whatif}
\begin{indented}
\lineup
\item[] 
\begin{tabular}{ccc}\br
    fit  & $\MH$ GeV &   $\chisq$/df \\ \br
now: $\delta\Mt = \pm$ 5.1 GeV, $\delta\MW = \pm$ 33 MeV 
& 85$^{+54}_{-34}$ & 29 / 15 \\
if $\delta\Mt = \pm$ 2.0 GeV, $\delta\MW = \pm$ 33 MeV 
& 83$^{+38}_{-28}$ & 29 / 15 \\
if $\delta\Mt = \pm$ 5.1 GeV, $\delta\MW = \pm$ 15 MeV 
& 67$^{+40}_{-27}$ & 34 / 15 \\
if $\delta\Mt = \pm$ 2.0 GeV, $\delta\MW = \pm$ 15 MeV 
& 50$^{+21}_{-16}$ & 35 / 15 \\
if $\delta\Mt = \pm$ 1.0 GeV, $\delta\MW = \pm$ 10 MeV 
& 35$^{+12}_{-10}$ & 38 / 15 \\

\br
\end{tabular}
\end{indented}
\end{table}

If the Higgs mass were
$\MH$ = 115 $\pm$ 1 GeV, as the direct search at LEP 2 might 
indicate, then if this constraint is introduced into the SM fit using 
the present measurements, the 
top-quark mass increases by 2.0 GeV, due to the correlation with $\MH$.
The error is reduced from $\delta\Mt = \pm$ 4.4 to $\pm$ 3.1 GeV. 
The $\chisq$ probability increases slightly to 2.2$\%$,
and the other quantities are largely unchanged, since the standard
fit is compatible with this Higgs mass.

 Some improvements are expected, in the relatively near future, on the precision
of the W boson and top quark masses. For $\MW$, the final LEP value is still
awaited and Run 2, at the Tevatron, should produce a much improved precision
compared to Run 1. The same holds for the top quark mass. It is difficult to
make a precise estimate of the improvements, but to give an indication of what
the impact would be, $\delta\Mt = \pm$ 2.0 GeV and $\delta\MW = \pm$ 15 MeV
are taken. The effect of these on the global electroweak fit is 
given in tab.\ref{tab-whatif}, where it is
assumed that the present central values remain unchanged. 
It can be seen that improved precision in the measurements of 
both $\MW$ and $\Mt$ is needed. If this were achieved,
and if the central values remain unchanged, then the 95$\%$ c.l. upper
limit for $\MH$ would be 87 GeV, plus the theory uncertainty. 
This would be incompatible with the direct search limits, and thus 
would indicate a breakdown of the SM. If the limits given in the last line
of tab.\ref{tab-whatif} could be achieved, namely 
$\delta\Mt = \pm$ 1.0 GeV and $\delta\MW = \pm$ 10 MeV, this would give a very
precise central value for $\MH$, and the 95$\%$ c.l. upper limit 
would reduce to 56 GeV.

\section{\bf Summary
\label{sec-summary}}

 Most of the LEP 1 data, apart from some of the heavy flavour data, are now
finalised and published. For the $\Zzero$-fermion couplings
at a scale q$^{2} \simeq \MZ^{2}$, those of
the b-quark, and to a lesser extent those of the $\tau$-lepton, are the least
consistent with SM expectations. More precise data are needed to see if there
is indeed an inconsistency for the third fermion generation. 

 The measurements from the NuTeV experiment might also indicate an inconsistency
in the SM. It is thus important that this situation is resolved by improved
experiments.
 
 The Higgs mass appears to be relatively light, but
some of the data are rather inconsistent. More work on $\alphamz$ is needed,
as this quantity has a strong influence on the extracted $\MH$ value. 
The precision on both $\MW$ and $\Mt$ needs to be improved at the Tevatron 
in order to significantly improve the accuracy on the Higgs mass.
This is important, particularly if the central values remain essentially
unchanged. In this case, there would be clear incompatibilities in
the Standard Model.


\vspace*{3.0cm}

\ack

 I would like to thank members of the LEP Electroweak Working Group, 
past and present, for their contributions in the development of this field.
In particular, thanks go to the present convenor, Martin Grunewald, and
also to Dave Charlton and Gerald Myatt for their help.

\end{document}